\documentclass[11pt,a4paper]{article}
\pdfoutput=1
\usepackage{jheppub}

\usepackage[dvipsnames]{xcolor}

\usepackage[numbers,sort&compress]{natbib}
\usepackage{subcaption}
\usepackage{enumerate}
\usepackage{xifthen} % if-then-else statements
\usepackage{tikz}
\usetikzlibrary{patterns,calc}
\usepackage{multirow} % Table formatting
\usepackage{mathbbol}
\usepackage{amsthm}
\usepackage{mathtools}
\usepackage{slashed}

\theoremstyle{lemma}

\theoremstyle{lemma}

\theoremstyle{definition}

\renewcommand*{\le}{\left}
\newcommand*{\ri}{\right}

\renewcommand*{\a}{\alpha}

\newcommand*{\g}{\gamma}
\renewcommand*{\d}{\delta}
\newcommand*{\e}{\epsilon}
\newcommand*{\ve}{\varepsilon}
\newcommand*{\h}{\eta}
\newcommand*{\q}{\theta}
\renewcommand*{\l}{\lambda}
\newcommand*{\m}{\mu}
\newcommand*{\n}{\nu}
\renewcommand*{\r}{\rho}
\newcommand*{\s}{\sigma}

\renewcommand*{\k}{\kappa}
\newcommand*{\z}{\zeta}
\newcommand*{\y}{\psi}

\newcommand*{\G}{\Gamma}

\newcommand*{\Om}{\mathcal{O}_m}
\newcommand*{\Omv}{\langle \mathcal{O}_m\rangle}
\newcommand*{\Opsi}{\mathcal{O}_{\psi}}
\newcommand*{\Opsiv}{\langle \mathcal{O}_{\psi}\rangle}

\newcommand*{\p}{\partial}
\newcommand*{\vev}[1]{\langle#1\rangle}
\newcommand{\id}{\mathbb{1}} % Identity matrix
\newcommand*{\cA}{\mathcal{A}}

\newcommand*{\cC}{\mathcal{C}}
\newcommand*{\cD}{\mathcal{D}}

\newcommand*{\cL}{\mathcal{L}}

\newcommand*{\N}{\mathcal{N}}
\newcommand*{\cO}{\mathcal{O}}
\newcommand*{\cZ}{\mathcal{Z}}
\newcommand*{\beq}{\begin{equation}}
\newcommand*{\eeq}{\end{equation}}
\newcommand*{\bea}{\begin{eqnarray}}
\newcommand*{\eea}{\end{eqnarray}}
\newcommand*{\zb}{\bar{z}}

\newcommand*{\yb}{\bar{y}}
\newcommand*{\mb}{\overline{m}}
\newcommand*{\bp}{\bar{\partial}}
\newcommand*{\pb}{\bar{\partial}}

\newcommand*{\sfnd}{S_{\textrm{4ND}}}
\newcommand*{\efnd}{E_{\textrm{4ND}}}
\newcommand*{\send}{S_{\textrm{8ND}}}
\newcommand*{\eend}{E_{\textrm{8ND}}}
\newcommand*{\ssol}{\left . S \right |_{\textrm{sol}}}
\newcommand*{\esol}{\left . E \right |_{\textrm{sol}}}
\newcommand*{\zsol}{\left . Z \right |_{\textrm{sol}}}

\newcommand*{\C}[2][]{
\ifthenelse{\isempty{#1}}{
{C_{#2}}
}{
	{C_{#2,\,#1}}
}
}
\newcommand*{\CF}[2][]{
\ifthenelse{\isempty{#1}}{
{F_{#2}}
}{
	{F_{#2,\,#1}}
}
}

\title{Supersymmetric Holomorphic Masses in AdS/CFT with Flavour}
\author[1]{Pietro Capuozzo,}
\author[2]{Jack Holden,}
\author[3]{Andy O'Bannon,}
\author[1]{James Ratcliffe,}
\author[4]{Ronnie Rodgers,}
\author[1]{Benjamin Suzzoni}
\affiliation[1]{STAG Research Centre, Physics and Astronomy, University of Southampton,\\
    Highfield, Southampton SO17 1BJ, United Kingdom}
\affiliation[2]{Yau Mathematical Sciences Center, Tsinghua University, Beijing, 100084, China}
\affiliation[3]{Department of Chemistry and Physics, SUNY Old Westbury, Old Westbury, NY, 11568 USA}
\affiliation[4]{Nordita, Stockholm University and KTH Royal Institute of Technology,
Hannes Alfv\'{e}ns v\"{a}g 12, SE-106 91 Stockholm, Sweden}
\emailAdd{P.Capuozzo@soton.ac.uk}
\emailAdd{jholden@tsinghua.edu.cn}
\emailAdd{obannona@oldwestbury.edu}
\emailAdd{J.Ratcliffe@soton.ac.uk}
\emailAdd{ronnie.rodgers@su.se}
\emailAdd{b.suzzoni@benterre.com}
\preprint{NORDITA 2025-131}
\abstract{In type IIB supergravity (SUGRA), in the extremal background of a number $N_c$ of D3-branes we consider a number $N_f$ of probe D7-branes extended along all four D3-brane directions, $(x_0,x_1,x_2,x_3)$. Using $(x_2,x_3)$ to define complex coordinates $(z,\zb)$ and defining the D7-branes' worldvolume scalars as $(y,\yb)$, we prove that any holomorphic function $y(z)$ or antiholomorphic function $y(\zb)$ is a BPS solution of the D7-branes' equations of motion preserving $\N=(4,0)$ or $(0,4)$ supersymmetry (SUSY) along $(x_0,x_1)$, respectively. In the near-horizon geometry, five-dimensional Anti-de Sitter (AdS) spacetime times a five-sphere, the AdS/Conformal Field Theory correspondence states that type IIB SUGRA is holographically dual to four-dimensional $\mathcal{N}=4$ supersymmetric $SU(N_c)$ Yang-Mills theory at large $N_c$ and large 't Hooft coupling, and the $N_f$ D7-branes are dual to $N_f$ $\mathcal{N}=2$ hypermultiplets in the fundamental representation, i.e. flavour fields. Our D7-brane solutions are dual to a position-dependent hypermultiplet mass $m$ that is holomorphic, $m(z)$, or antiholomorphic, $m(\zb)$. We provide field theory proofs, valid for any $N_c$, $N_f$, and 't Hooft coupling, that such $m$ preserve $\N=(4,0)$ or $(0,4)$ SUSY along $(x_0,x_1)$. We also perform holographic calculations with probe D7-branes to show that the theory with such $m$ has zero energy and zero expectation value of the mass operator, and that a zero of $m$ is dual to a superconformal defect described by chiral fermions along $(x_0,x_1)$.}

\begin{document}
%%%%%%%%%%%%%%%%%%%%%%%%%%%%%%%%%%%%%%%%%%%%%%%%%%
%%%%%%%%%%%%%%%%%%%%%%%%%%%%%%%%%%%%%%%%%%%%%%%%%%

\maketitle

%%%%%%%%%%%%%%%%%%%%%%%%%%%%%%%%%%%%%%%%%%%%%%%%%%
%%%%%%%%%%%%%%%%%%%%%%%%%%%%%%%%%%%%%%%%%%%%%%%%%%
\section{Introduction and summary}
\label{sec:intro}
%%%%%%%%%%%%%%%%%%%%%%%%%%%%%%%%%%%%%%%%%%%%%%%%%%
%%%%%%%%%%%%%%%%%%%%%%%%%%%%%%%%%%%%%%%%%%%%%%%%%%

Many of the most important open questions in physics involve strongly coupled degrees of freedom without translational and/or rotational symmetry. Examples range from condensed matter physics, such as the Kondo lattice or Hubbard model, to particle physics, such as lattice Quantum Chromodynamics (QCD) or high density QCD, whose ground state may break these symmetries (colour superconductor LOFF phase, quarkyonic crystals, etc.). For some such systems, reliable approximation methods exist, such as Monte Carlo importance sampling. However, for many such systems no reliable approximation method currently exists, including in particular systems at non-zero density, where the sign problem renders Monte Carlo importance sampling practically useless. Moreover, \textit{exact} solutions for such systems are exceedingly rare. In this paper we present exact solutions for one such system by using a special symmetry, namely supersymmetry (SUSY).

We start in type IIB string theory in Minkowski spacetime with spacetime dimension $d=10$, including time coordinate $x_0$ and spatial coordinates $x_1$, $x_2$, $\ldots$, $x_9$. We introduce a number $N_c$ of coincident D3-branes along $(x_0,x_1,x_2,x_3)$ and a number $N_f$ of coincident D7-branes along $(x_0,\ldots,x_7)$, as shown in table~\ref{tab:4nd}. We focus exclusively on the low-energy theory on the D3-branes' worldvolume, which is $\N=4$ supersymmetric Yang-Mills (SYM) theory with gauge group $SU(N_c)$, arising from open strings with both ends on the D3-branes, plus a number $N_f$ of $\N=2$ hypermultiplet ``flavour'' fields in the fundamental representation of $SU(N_c)$, arising from open strings with one end on the D3-branes and the other end on the D7-branes. In each hypermultiplet the on-shell fields are a Dirac fermion ``quark'' plus its superpartners, namely two complex scalar ``squarks''.

\begin{table}
\begin{center}
        \begin{tabular}{|c|cccccccccc|}\hline
                & $x_0$ & $x_1$ & $x_2$ & $x_3$ & $x_4$ & $x_5$ & $x_6$ & $x_7$ & $x_8$ & $x_9$\\ \hline
$N_c$ D3&$\bullet$&$\bullet$&$\bullet$&$\bullet$&--&--&--&--&--&--\\\hline
$N_f$ D7&$\bullet$&$\bullet$&$\bullet$&$\bullet$&$\bullet$&$\bullet$&$\bullet$&$\bullet$&--&--\\\hline
        \end{tabular}
        \caption{Our starting point: in type IIB string theory in Minkowski spacetime, with time coordinate $x_0$ and spatial coordinates $(x_1,\ldots,x_9)$, the intersection of $N_c$ coincident D3-branes along $(x_0,x_1,x_2,x_3)$ and $N_f$ coincident D7-branes along $(x_0,\ldots,x_7)$. This intersection has 4 Neumann-Dirichlet (ND) directions, namely $(x_4,x_5,x_6,x_7)$.}
        \label{tab:4nd}
\end{center}
\end{table}

The $\N=4$ SYM theory has R-symmetry $SO(6)_R$, corresponding to the rotational symmetry of the directions transverse to the D3-branes, $(x_4,\ldots,x_9)$. When the D3- and D7-branes are coincident in the $(x_8,x_9)$ plane the D7-branes break $SO(6)_R$ to $SO(4) \times SO(2)$, where the $SO(4)$ corresponds to the rotational symmetry of the directions transverse to the D3-branes but along the D7-branes, $(x_4,x_5,x_6,x_7)$, while the $SO(2)$ corresponds to the rotational symmetry of the directions transverse to both stacks of branes, $(x_8,x_9)$. The $SO(4) \simeq SU(2)_R \times SU(2)$ and the $SO(2) \simeq U(1)_R$, where the $SU(2)_R \times U(1)_R \simeq U(2)_R$ is the $\N=2$ R-symmetry~\cite{Kruczenski:2003be,Erdmenger:2007cm}. The $U(1)_R$ acts on the quarks as a $U(1)$ axial symmetry, and has no effect on the squarks. 

The main character in our story will be the complex-valued hypermultiplet mass, $m$, which is determined by the length of the open strings between the D3- and D7-branes. When the D3-branes and D7-branes are coincident in the $(x_8,x_9)$ plane and so preserve $SO(2)$ rotational symmetry there, the open strings between them have zero length and hence zero energy. Translating to the SYM theory, we find $m=0$ and the axial symmetry $SO(2) \simeq U(1)_R$ is preserved. Separating the D3- and D7-branes in $(x_8,x_9)$ by a distance $\ell$ breaks the $SO(2)$ rotational symmetry there, and gives the open strings between them a non-zero length and hence non-zero energy, namely $\ell$ times their tension, $1/(2 \pi \alpha')$. Translating to the SYM theory, we find $|m| = \ell/(2 \pi \alpha')$, and the choice of $m$'s phase breaks $U(1)_R$. In all cases the D7-branes/hypermultiplets preserve $SO(4) \simeq SU(2)_R \times SU(2)$.

In this paper we replace the D3-branes with their closed string description, namely an extremal spacetime with near-horizon geometry $AdS_5 \times S^5$, where $AdS_5$ is $d=5$ Anti-de Sitter (AdS) spacetime and $S^5$ is a five-sphere, and asymptotic $d=10$ Minkowski spacetime, with $N_c$ units of Ramond-Ramond (RR) five-form flux on $S^5$ and all other fields trivial~\cite{Horowitz:1991cd}. The $AdS_5$ and $S^5$ each have radius of curvature $L$, given by $L^4 \equiv 4 \pi g_s N_c$, with closed string coupling $g_s$. We work exclusively in the supergravity (SUGRA) limit, namely $N_c \to \infty$ and $g_s \to 0$ such that $L^4/\alpha'^2$ is fixed, followed by $L^4 \gg \alpha'^2$. We also take the probe limit, $N_f \ll N_c$, wherein the D7-branes do not deform the SUGRA fields. In the near-horizon region the D7-branes asymptotically approach $AdS_5 \times S^3$.

The Minkowski vacuum of type IIB string theory or SUGRA is maximally SUSY, and in particular is invariant under SUSY transformations generated by 32 real supercharges. The D3-branes break those to 16 real supercharges, corresponding to the 16 Poincar\'e supercharges of $\N=4$ SYM. In fact $\N=4$ SYM is a Conformal Field Theory (CFT), which has an additional 16 real superconformal charges, for a total of 32. These ``extra'' superconformal charges emerge only near the extremal horizon: $AdS_5 \times S^5$ preserves 32 real supercharges~\cite{Maldacena:1997re}. In general the D7-branes break the Poincar\'e supercharges down to 8, and break all of the superconformal charges. However, in the probe limit and when $m=0$ the D7-branes also preserve 8 superconformal charges.

The AdS/CFT correspondence, also known as gauge-gravity duality or holography, is the statement that type IIB string theory in the near-horizon $AdS_5 \times S^5$ region is \textit{equivalent} to $\N=4$ SYM~\cite{Maldacena:1997re,Gubser:1998bc,Witten:1998qj}. The $N_f$ D7-branes along $AdS_5 \times S^3$ are dual to the $N_f$ hypermultiplets~\cite{Karch:2002sh}, where the worldvolume scalars $x_8$ and $x_9$ together are dual to the hypermultiplet's complex-valued mass operator, $\mathcal{O}$, and the worldvolume $U(N_f)$ gauge fields are dual to the conserved currents of the global $U(N_f)$ flavour symmetry. In what follows we will retain the full D3-brane geometry in sections~\ref{sec:cartesian} and~\ref{sec:kappa_symmetry}, and restrict to the near-horizon $AdS_5 \times S^5$ region, and invoke the AdS/CFT correspondence, only in sec.~\ref{sec:holodual}.

We will obtain solutions in which $m$ preserves some SUSY but can depend on position, and hence breaks some  translational and rotational symmetry. We do so as follows. First, from the spatial directions shared by both stacks of D-branes, $(x_1,x_2,x_3)$, we choose two directions at random, say $(x_2,x_3)$, and use them to define a complex coordinate,
\beq
\label{eq:zdef}
z \equiv \frac{1}{\sqrt{2}} \left(x_2 + i \, x_3\right).
\eeq
Next, we define a complex coordinate $y$ in the directions transverse to both stacks of branes,
\beq
\label{eq:ydef}
y \equiv \frac{1}{\sqrt{2}} \left(x_8 + i \, x_9\right).
\eeq
The probe D7-branes' worldvolume scalars are then $y$ and $\yb$, which are valued in the adjoint of the complexification of $U(N_f)$'s Lie algebra, that is, they are $N_f \times N_f$ complex matrices. Their action includes a non-Abelian Dirac-Born-Infeld (DBI) term plus Wess-Zumino (WZ) terms, describing their coupling to the worldvolume $U(N_f)$ gauge fields and to the SUGRA fields. We then make the following ansatz for $y$ and $\yb$. We take the D7-branes' worldvolume gauge fields to vanish, which implies the WZ terms vanish. Furthermore, we consider only coincident D7-branes, so that $y$ and $\yb$ reduce from matrices to scalars, and moreover $y$ and $\yb$ become complex conjugates of one another, so in what follows we will refer only to $y$. In terms of the $d=4$ SYM theory, $m = \sqrt{2} \, y/(2 \pi \alpha')$. The action for $y$ thus reduces to an Abelian DBI term. We also demand that $y$ preserve $d=2$ Poincar\'e symmetry along $(x_0,x_1)$ and the rotational symmetry along $(x_4,x_5,x_6,x_7)$. With these assumptions, the most general ansatz for $y$ is then $y(z,\zb,\rho)$, with $\rho$ the radial coordinate along $(x_4,x_5,x_6,x_7)$.

Our main result, obtained in secs.~\ref{sec:cartesian} and~\ref{sec:kappa_symmetry}, is: to preserve SUSY and solve the equation of motion, $y(z,\zb,\rho)$ must be independent of $\rho$ and holomorphic in $z$. In particular, in sec.~\ref{sec:cartesian} we show that any holomorphic function $y(z)$ solves the Abelian DBI equation of motion, and in fact saturates a BPS bound on the action. In sec.~\ref{sec:kappa_symmetry} we then show that a generic $y(z)$ preserves 4 real supercharges, which in terms of the $d=2$ SUSY in the directions $(x_0,x_1)$ all have the same chirality, which we take to be left-handed, i.e. $d=2$ $\N=(4,0)$ SUSY. In short, a generic $y(z)$ breaks $d=4$ $\N=2$ super-Poincar\'e symmetry along $(x_0,x_1,x_2,x_3)$ down to the subgroup with $d=2$ $\N=(4,0)$ super-Poincar\'e symmetry along $(x_0,x_1)$. An \textit{anti}-holomorphic function, $y(\zb)$, also solves the equation of motion and saturates a BPS bound, but preserves $d=2$ SUSY of opposite chirality, meaning $\N=(0,4)$ SUSY. For clarity, in the rest of the introduction we will refer mostly to holomorphic solutions, $y(z)$, although analogous statements apply to anti-holomorphic functions, $y(\zb)$. Also, in sec.~\ref{sec:poles} we argue that poles in $y$ require careful treatment, and in particular must be excised from the $z$ plane and replaced with appropriate boundary conditions, following refs.~\cite{Callan:1997kz,Constable:2001ag,Constable:2002xt,Arean:2007nh,Ammon:2012mu}. We will henceforth assume that replacement has been done, and in an abuse of language we will refer to $y$ as holomorphic, not meromorphic, even if $y$ has one or more poles.

In secs.~\ref{sec:cartesian} and~\ref{sec:kappa_symmetry} we also argue that as we approach a zero or pole of $y$ the 4ND D7-branes approach the 8ND D7-branes of tab.~\ref{tab:8nd}, which intersect the D3-branes only along $(x_0,x_1)$. Specifically, a zero or pole of degree $D$ describes a number $N_f D$ of these 8ND D7-branes. Refs.~\cite{Harvey:2007ab,Buchbinder:2007ar,Harvey:2008zz} studied these 8ND D7-branes in detail, and showed that they preserve $d=2$ $\N=(8,0)$ or $(0,8)$ SUSY along $(x_0,x_1)$, depending on their orientation. In sec.~\ref{sec:cartesian} we argue that the 4ND D7-branes approach the 8ND D7-branes at a zero or pole of $y$, and in sec.~\ref{sec:kappa_symmetry} we show that holomorphic or antiholomorphic $y$ exhibits SUSY enhancement from $d=2$ $\N=(4,0)$ to $(8,0)$ or $\N=(0,4)$ to $(0,8)$, respectively, at a zero of $y$. In short, a zero or pole of $y$ describes D7-branes that between $|z| \to \infty$ and the location of a zero or pole, interpolate between the 4ND D7-branes in tab.~\ref{tab:4nd} and the 8ND D7-branes in tab.~\ref{tab:8nd}.
\begin{table}
\begin{center}
        \begin{tabular}{|c|cccccccccc|}\hline
                & $x_0$ & $x_1$ & $x_2$ & $x_3$ & $x_4$ & $x_5$ & $x_6$ & $x_7$ & $x_8$ & $x_9$\\ \hline
$N_c$ D3&$\bullet$&$\bullet$&$\bullet$&$\bullet$&--&--&--&--&--&--\\\hline
$N_f D$ D7&$\bullet$&$\bullet$&--&--&$\bullet$&$\bullet$&$\bullet$&$\bullet$&$\bullet$&$\bullet$\\\hline
        \end{tabular}
        \caption{As we approach a pole or zero of $y$ of degree $D$, the $N_f$ D7-branes of tab.~\ref{tab:4nd} describe $N_f D$ D7-branes which intersect the $N_c$ coincident D3-branes only along $(x_0,x_1)$, as shown above. This intersection has 8 ND directions, namely $(x_2,\ldots,x_9)$.}
        \label{tab:8nd}
\end{center}
\end{table}

For strings, D-branes, and M-branes, holomorphic solutions for worldvolume scalars are in fact familiar in many contexts: see for example refs.~\cite{Wen:1985jz,Callan:1997kz,Constable:2002xt} and references therein. In particular, for M2-branes in $d=11$ Minkowski spacetime ref.~\cite{Gauntlett:1997ss} showed that holomorphic worldvolume scalars are BPS solutions of the Abelian DBI equation of motion. Ref.~\cite{Gauntlett:1997ss}'s arguments are essentially a Bogolmonyi trick, which easily generalises to strings, D-branes, and M-branes in Minkowski spacetime of any $d$. Our results are a straightforward generalisation of those of ref.~\cite{Gauntlett:1997ss} to a novel case, namely D7-branes in the closed string background produced by D3-branes, rather than $d=10$ Minkowski spacetime.

A major benefit of that generalisation is that we can take the near horizon limit, thus obtaining $AdS_5 \times S^5$, where we can invoke the AdS/CFT correspondence. All of the statements above apply in that limit: holomorphic $y(z)$ are BPS solutions of the Abelian DBI equation of motion, and preserve $d=2$ $\N=(4,0)$ SUSY, with analogous statements for antiholomorphic $y(\zb)$.\footnote{We suspect such solutions have been overlooked because most research of D3/D7 intersections assumed Poincar\'e symmetry in all the SYM directions $(x_0,x_1,x_2,x_3)$, including the holomorphic solutions of ref.~\cite{Ouyang:2003df}, which are (therefore) distinct from our solutions.} In sec.~\ref{sec:nearhorizon} we study the holographic duals of holomorphic $y(z)$. The $\N=4$ SYM 't Hooft coupling $\lambda = L^4/\alpha'^2 \equiv 4 \pi g_s N_c$, so the type IIB SUGRA limit corresponds to the 't Hooft limit, $N_c \to \infty$ with $\lambda$ fixed, followed by $\lambda \to \infty$, meaning the SYM theory is strongly coupled. In AdS/CFT the holographically dual theory is $d=4$ $\N=4$ $SU(N_c)$ SYM with large $N_c$ and large $\lambda$ coupled to $N_f \ll N_c$ $d=4$ $\N=2$ hypermultiplets in the fundamental representation of $SU(N_c)$, i.e. flavour fields. Holomorphic $y(z)$ is dual to a holomorphic hypermultiplet mass,
\beq
\label{eq:mofz}
m(z) = \frac{\sqrt{2}}{2\pi\alpha'} \, y(z),
\eeq
which preserves $d=2$ $\N=(4,0)$ SUSY in the directions $(x_0,x_1)$. Our BPS holomorphic solutions are thus holographically dual to a position-dependent coupling that preserves SUSY, in a fashion similar to the field theories of refs.~\cite{Festuccia:2011ws,Dumitrescu:2012ha,Closset:2013vra,Closset:2014uda}.

In sec.~\ref{sec:nearhorizon} we start on the SYM side of the correspondence: in sec.~\ref{sec:fieldtheory} we prove in the SYM theory (with no reference to D-branes or SUGRA) that holomorphic or antiholomorphic $m$ preserve $d=2$ $\N=(4,0)$ or $(0,4)$ SUSY, respectively. We also show that such $m$ describe non-trivial holonomies of a background, non-dynamical $U(1)_R$ gauge field (in our case equivalent to an axial gauge field) around any zeroes and poles of $m$.

In sec.~\ref{sec:holodual} we turn to the AdS side of the correspndence, and use holographic renormalisation for the probe D7-branes~\cite{deHaro:2000vlm,Bianchi:2001kw,Skenderis:2002wp,Karch:2005ms,Karch:2006bv,Hoyos:2011us} to show that, in the presence of $m(z)$, the renormalised energy vanishes and the vacuum expectation value (VEV) of the complex operator dual to $y(z)$, which we denote $\cO$, vanishes, i.e. $\langle \mathcal{O}\rangle=0$. When $y(z)$ is constant, $d=4$ $\N=2$ SUSY requires both of these properties, because $\mathcal{O}$ is the SUSY variation of another operator: if $Q$ and $\tilde{Q}$ are the chiral superfields whose lowest components are the squarks, then $\mathcal{O}$ is the F-term of $\bar{Q}Q$~\cite{Babington:2003vm,Erdmenger:2007cm}. For generic $y(z)$, our results suggest that $d=2$ $\N=(4,0)$ SUSY preserves the supercharge under which $\mathcal{O}$ is the SUSY variation of another operator, so that both the energy and $\langle \mathcal{O}\rangle$ must still vanish.

A generic $y(z)$ breaks $U(1)_R$ but preserves $SO(4)\simeq SU(2)_R \times SU(2)$ rotational symmetry in $(x_4,x_5,x_6,x_7)$. The $SU(2)_R$ now acts as the (non-maximal) R-symmetry for the remaining $d=2$ $\N=(4,0)$ SUSY. However, special solutions may preserve more symmetry. For example, as mentioned above, in the probe limit $m \propto y = 0$ preserves $d=4$ $\N=2$ superconformal symmetry, and constant $m \propto y \neq 0$ preserves $d=4$ $\N=2$ SUSY.

In sec.~\ref{sec:holodual} we study what happens to the 4ND D7-branes at a zero or pole of $y$, where the 4ND D7-branes become 8ND D7-branes and SUSY enhances from $d=2$ $\N=(4,0)$ to $(8,0)$ or from $\N=(0,4)$ to $(0,8)$ for holomorphic or antiholomorphic $y$, respectively. The 8ND D7-branes intersect the D3-branes only along $(x_0,x_1)$ and hence describe flavour fields supported only on a $d=2$ defect inside $d=4$ $\N=4$ SYM. Refs.~\cite{Harvey:2007ab,Buchbinder:2007ar,Harvey:2008zz} showed that in the probe flavour sector the only on-shell degrees of freedom are chiral fermions neutral under both SUSY and the R-symmetry, and coupled only to the $\N=4$ SYM gauge fields. Indeed, both anomaly inflow arguments~\cite{Callan:1984sa} and index theorems for the Dirac operator on a background with scalar vortices~\cite{Jackiw:1981ee,Weinberg:1981eu} require the existence of a net number of such $d=2$ chiral fermions. The flavour sector also trivially preserves $SO(6)_R$, as visible in tab.~\ref{tab:8nd}, where the D7-branes clearly preserve the rotational symmetry in $(x_4,\ldots,x_9)$, albeit now a $d=2$ chiral half of the original $SO(6)_R$~\cite{Harvey:2007ab,Buchbinder:2007ar,Harvey:2008zz}. In sec.~\ref{sec:holodual} we argue that a zero of $y$ describes these 8ND D7-branes, which  are extended along $AdS_3 \times S^5$ inside $AdS_5 \times S^5$. The isometries of the $AdS_3$ are dual to $d=2$ defect conformal symmetry, meaning the subgroup of the $d=4$ conformal group that leaves invariant the $d=2$ surface at $m$'s zero, while the isometries of the $S^5$ are dual to the $SO(6)_R$ R-symmetry. The holographically dual statement is that a zero of $m$ describes the $d=2$ $\N=(8,0)$ or $(0,8)$ chiral fermions coupled to $d=4$ $\N=4$ SYM as a conformal defect or ``quantum wire'' along $(x_0,x_1)$.

A textbook theorem of complex analysis states that a non-constant holomorphic function $y(z)$ can have only \textit{isolated} zeroes~\cite{ahlfors}. Said differently, in the complex $z$-plane $y(z)$ can vanish only on a surface of codimension two, also known as a point, and cannot vanish along a surface of codimension one, also known as a curve. In general, we expect any zero to produce a defect of the kind above, namely a $d=2$ $\N=(8,0)$ superconformal defect. For example, the solution $y(z) = \frac{1}{k} \sin(kz)$, with wave number $k$, describes a ``lattice'' of such superconformal defects, namely one defect at each zero, $z=n \pi /k$ with $n \in \mathbb{Z}$.

Another textbook theorem of complex analysis states that a holomorphic function $y(z)$ can have only isolated poles~\cite{ahlfors}. In terms of D-branes, a pole corresponds to the D-branes forming an infinitely-long ``spike'' in a direction orthogonal to its original worldvolume~\cite{Callan:1997kz}. Obviously, a pole, such as $y(z) \propto z^{-D}$ with integer degree $D$, implies $m(z) \propto z^{-D}$, and hence represents a hypermultiplet mass that diverges at $z=0$, decreases linearly as $|z|$ increases, and $\to 0$ as $|z| \to \infty$. A single pole thus describes an infinitely heavy, immobile $d=2$ defect off of which the quarks and squarks could scatter, i.e. a ``scattering centre.'' Solutions with periodically-spaced poles, such as $y(z) = 1/( k \sin(kz))$, could describe a fixed lattice of such scattering centres.

In the complex plane the most general holomorphic function without an essential singularity at infinity is a rational function, $y(z) = \frac{(z-a_1)(z-a_2) \ldots}{(z-b_1) (z-b_2) \ldots}$ with complex-valued constants $(a_1,a_2,\ldots)$ and $(b_1,b_2,\ldots)$. The freedom to choose any holomorphic $y(z)$ thus allows infinite possibilities for holographic descriptions of strongly coupled $d=4$ fields interacting with $d=2$ fields, whether massless (quantum wires of chiral fermions) or infinitely massive (scattering centres), arranged however we like in the $z$-plane (through the choice of $(a_1,a_2,\ldots)$ and $(b_1,b_2,\ldots)$).

We thus hope that this class of exact solutions may be useful for studying problems involving strongly-coupled degrees of freedom without translational and/or rotational symmetry, as mentioned above. For instance, for many such systems, such as the Kondo lattice, Hubbard model, or finite-density QCD, some of the most important questions involve charge transport, and specifically electrical conductivity. Translational symmetry implies momentum conservation, which in turn implies that electrical conductivity must either vanish or diverge. Introducing impurities, arranged either in a lattice or randomly, i.e. random disorder, break translational symmetry, thus producing momentum dissipation and a finite conductivity. How does that conductivity depend on the shape of the lattice, or the concentration of impurities? Holography has inspired conjectures for various bounds on transport coefficients in the presence of momentum dissipation or random disorder, including on conductivity~\cite{Grozdanov:2015qia}, the charge diffusion constant~\cite{Hartman:2017hhp}, and others~\cite{Hartnoll:2016apf}. What systems obey these bounds, and why? Do real strongly-coupled systems obey these bounds? As mentioned above, few reliable techniques exist to answer these questions, and exact solutions describing translational symmetry breaking are exceedingly rare. Indeed, in AdS/CFT to our knowledge the only such solutions are so-called Q-lattice models~\cite{Donos:2013eha} and linear axion models~\cite{Andrade:2013gsa,Baggioli:2021xuv}, in which an exactly marginal operator's source depends on a spatial coordinate either as a complex exponential or linear function, respectively. Our solutions are vastly more flexible: our source is for a relevant operator, namely a mass operator, and our source can be \textit{any holomorphic function}, allowing for an infinite number of ways of breaking translational symmetry, including impurities and/or scattering centres, distributed either in a lattice or randomly. Our solutions thus fill a major gap in the literature, and could be the first step towards answering many crucial questions about conductivity in strongly-interacting systems, among many other things.

This paper is organised as follows. In sec.~\ref{sec:cartesian}, for the full, asymptotically flat D3-brane background we show that holomorphic or antiholomorphic $y$ solve the Abelian DBI equation of motion and saturate a BPS bound. We also show that the 4ND D7-branes approach the 8ND D7-branes at a zero or pole of $y$. In sec.~\ref{sec:kappa_symmetry} we show that a generic holomorphic or antiholomorphic $y$ preserves $d=2$ $\N=(4,0)$ or $\N=(0,4)$ SUSY along $(x_0,x_1)$, respectively. We also show that at a zero or pole of $y$ SUSY enhances from $d=2$ $\N=(4,0)$ to $(8,0)$ or from $\N=(0,4)$ to $(0,8)$. In sec.~\ref{sec:nearhorizon} we focus on the near-horizon $AdS_5 \times S^5$ where we can invoke holographic duality. In the SYM theory we present two independent proofs that holomorphic or antiholomorphic $m$ preserves $d=2$ $\N=(4,0)$ or $\N=(0,4)$ SUSY along $(x_0,x_1)$, respectively. We then show how in $AdS_5 \times S^5$ holomorphic or antiholomorphic $y$ are dual to states with vanishing energy and $\langle \mathcal{O} \rangle=0$, and how the D7-branes' wodlvolume geometry encodes 8ND D7-branes at a zero of $y$. In sec.~\ref{sec:summary} we conclude with a summary and suggestions for future research. In the appendix we provide details of the holographic renormalisation of holomorphic or antiholomorphic $y$.

Our methods are very general, and indeed the companion paper ref.~\cite{Ratcliffe:2026exv} shows that holomorphic or antiholomorphic worldvolume scalar are BPS and SUSY solutions for many other probe D$q$-branes in extremal D$p$-brane backgrounds.

%%%%%%%%%%%%%%%%%%%%%%%%%%%%%%%%%%%%%%%%%%%%%%%%%%
%%%%%%%%%%%%%%%%%%%%%%%%%%%%%%%%%%%%%%%%%%%%%%%%%%
\section{Holomorphic probe brane solutions}
\label{sec:cartesian}
%%%%%%%%%%%%%%%%%%%%%%%%%%%%%%%%%%%%%%%%%%%%%%%%%%
%%%%%%%%%%%%%%%%%%%%%%%%%%%%%%%%%%%%%%%%%%%%%%%%%%

%%%%%%%%%%%%%%%%%%%%%%%%%%%%%%%%%%%%%%%%%%%%%%%%%%
\subsection{Review: background D3-brane solution}
\label{sec:background}
%%%%%%%%%%%%%%%%%%%%%%%%%%%%%%%%%%%%%%%%%%%%%%%%%%

As mentioned in sec.~\ref{sec:intro}, in type IIB string theory we start with $N_c$ D3-branes along $(x_0,x_1,x_2,x_3)$. We replace these D3-branes with their closed string description, namely an extremal solution with all fields vanishing except for a constant dilaton, describing the constant closed string coupling $g_s$, a RR four-form $C_4$ whose form we will not need, and an Einstein-frame metric \(ds^2 = G_{\mu\nu} dx^{\mu} dx^{\nu}\) with $\mu,\nu=0,\ldots, 9$ given by~\cite{Horowitz:1991cd} 
\begin{subequations}
\label{eq:metric}
\beq
    ds^2= f(r) \left( - dx_0^2 + dx_1^2 + dx_2^2 + dx_3^2\right) + f(r)^{-1}  \left( dx_4^2 + dx_5^2 + dx_6^2 + dx_7^2 + dx_8^2 + dx_9^2\right),
    \eeq
    \beq
    \label{eq:frdef}
f(r) \equiv \left(1 + \frac{L^4}{r^4}\right)^{-1/2},
\eeq
\end{subequations}
with $L^4 \equiv 4 \pi g_s N_c \a'^2$ and $r$ the radial coordinate transverse to the D3-branes, defined via
\beq
\label{eq:rdef0}
r^2 \equiv x_4^2 + x_5^2 + x_6^2 + x_7^2 + x_8^2 + x_9^2.
\eeq
In all that follows we will work in SUGRA limit, $N_c \to \infty$ and $g_s \to 0$ with $g_s N_c$ fixed, followed by $g_s N_c \to \infty$, so that $L^4 \gg \a'^2$.

The limit $r \gg L$ implies $f(r) \approx 1$, so that the metric in eq.~\eqref{eq:metric} becomes that of $d=10$ Minkowski spacetime. The opposite, near-horizon limit $r \ll L$ implies \(f(r) \approx r^2/L^2\), so that the metric in eq.~\eqref{eq:metric} becomes, in terms of a round, unit-radius $S^5$ metric $ds^2_{S^5}$,
\begin{subequations}
\label{eq:ads5s5metric}
\begin{eqnarray}
ds^2 & = & \frac{r^2}{L^2} \left( - dx_0^2 + dx_1^2 + dx_2^2 + dx_3^2\right) + \frac{L^2}{r^2}  \left( dr^2 + r^2 \, ds^2_{S^5}\right) \\ & = & \frac{r^2}{L^2} \left( - dx_0^2 + dx_1^2 + dx_2^2 + dx_3^2\right) + \frac{L^2}{r^2} dr^2  + L^2 \, ds^2_{S^5},
\end{eqnarray}
\end{subequations}
which is the metric of $AdS_5 \times S^5$ with each factor having radius of curvature $L$. In these coordinates the $AdS_5$ boundary is at $r \to \infty$ and the Poincar\'e horizon is at $r \to 0$.

In the SUGRA limit the degrees of freedom in the asymptotic $d=10$ Minkowski region decouple, and the degrees of freedom in the near-horizon $AdS_5 \times S^5$ region are holographically dual to \(\N=4\) SYM with gauge group \(SU(N_c)\) and 't Hooft coupling \(\l  = L^4/\a'^2 \equiv 4 \pi g_s N_c\), in the 't Hooft limit \(N_c \gg 1\), followed by \(\l \gg 1\)~\cite{Maldacena:1997re,Gubser:1998bc,Witten:1998qj}. The holographic coordinate $r$ is dual to the SYM energy scale~\cite{Susskind:1998dq,Peet:1998wn}: the region near the $AdS_5$ boundary, $r\to\infty$, is dual to the ultraviolet (UV) of the SYM theory, while the region near the Poincar\'e horizon, $r \to 0$, is dual to the infrared (IR).

In what follows we will work in the SUGRA limit but we will retain the full function $f(r)$ except in sec.~\ref{sec:nearhorizon} and in the appendix, where we restrict to the near-horizon $AdS_5 \times S^5$ region. We will retain the full $f(r)$ to demonstrate that our results are more general than AdS/CFT (i.e. are valid for all $r$), and in particular to connect our results to those of ref.~\cite{Gauntlett:1997ss} for branes in Minkowski spacetimes, as we will discuss further below.

%%%%%%%%%%%%%%%%%%%%%%%%%%%%%%%%%%%%%%%%%%%%%%%%%%
%%%%%%%%%%%%%%%%%%%%%%%%%%%%%%%%%%%%%%%%%%%%%%%%%%
\subsection{D7-brane (anti-)holomorphic solutions}
\label{sec:holosols}
%%%%%%%%%%%%%%%%%%%%%%%%%%%%%%%%%%%%%%%%%%%%%%%%%%
%%%%%%%%%%%%%%%%%%%%%%%%%%%%%%%%%%%%%%%%%%%%%%%%%%

In this section we will introduce the $N_f$ D7-branes of tab.~\ref{tab:4nd}, which are 4ND with respect to the $N_c$ D3-branes. We will work exclusively in the probe limit, $N_f \ll N_c$, and prove that holomorphic or antiholomorphic $y$ solve the probe D7-branes' equation of motion.

In this subsection we take the ``direct approach:'' we will derive the probe D7-brane equation of motion and show explicitly that holomorphic or antiholomorphic $y$ are solutions. In the next subsection we will show \textit{why} such $y$ solve the equation of motion, by showing that they saturate a BPS bound on the D7-branes' action and energy and hence, in a sector with fixed BPS charge, they extremise the action.

Returning to the $N_c$ D3-branes along $(x_0,x_1,x_2,x_3)$, we add $N_f$ D7-branes along $(x_0,\ldots,x_7)$, as summarised in tab.~\ref{tab:4nd}. In the $\N=4$ SYM theory on the D3-branes' worldvolume, the open strings between the D3- and D7-branes give rise to a number $N_f$ of $d=4$ $\N=2$ hypermultiplets in the fundamental representation of $SU(N_c)$. When we take $N_c \to \infty$ and replace the D3-branes with their closed string description, namely the solution in eq.~\eqref{eq:metric}, we will keep $N_f$ fixed. In other words, we will work in the probe limit, $N_f \ll N_c$, in which the D7-branes' back-reaction on the SUGRA fields is suppressed.

As explained in sec.~\ref{sec:intro}, using two spatial coordinates along both sets of D-branes, $(x_2,x_3)$, we define the complex coordinates $(z,\zb)$ in eq.~\eqref{eq:zdef}, and using the two coordinates transverse to both sets of D-branes, $(x_8,x_9)$, we define the complex coordinates $(y,\yb)$ in eq.~\eqref{eq:ydef}. In the directions transverse to the D3-branes but along the D7-branes, $(x_4,x_5,x_6,x_7)$, we will use spherical coordinates,
\beq
dx_4^2 + dx_5^2 + dx_6^2 + dx_7^2 = d\rho^2 + \rho^2 \, ds^2_{S^3},
\eeq
with the radial coordinate $\rho$ defined via
\beq
\label{eq:rhodef}
\rho^2 \equiv x_4^2 + x_5^2 + x_6^2 + x_7^2,
\eeq
and angular coordinates $(\chi_1,\chi_2,\chi_3)$ of a round, unit-radius $S^3$ with metric
\beq
ds^2_{S^3} = d\chi_1^2 + \sin^2 \chi_1 \left(d\chi_2^2 + \sin^2 \chi_2 \, d\chi_3^2 \right).
\eeq
Using all of these coordinates, the metric in eq.~\eqref{eq:metric} becomes
\beq
\label{eq:metric_complex_coords}
    ds^2 = f(r) (-dx_0^2 + dx_1^2 + 2 \, dz \, d\zb) + f(r)^{-1} (d\r^2 + \r^2 ds_{S^3}^2  + 2 \, dy \, d\yb),
\eeq
where now
\beq
\label{eq:rdef}
r^2 \equiv \rho^2 + 2 \, y \, \yb.
\eeq
We can reach $r \to \infty$ by taking $\rho \to \infty$ with $|y|^2$ fixed, or $\rho$ fixed with $|y|^2 \to \infty$, or both $\rho \to \infty$ and $|y|^2 \to \infty$ simultaneously. We can reach $r \to 0$ only by taking $\rho \to 0$ and $|y|^2\to0$ simultaneously. Tab.~\ref{tab:embedding} summarises the D3/D7 intersection in these coordinates.
\begin{table}
    \begin{center}\begin{tabular}{|c || c c c c | c  | c  c c | c c|}
    \hline
         & \multicolumn{4}{c |}{\(\mathbb{R}^{1,3}\)} & & \multicolumn{3}{c |}{\(S^3\)} &  & 
        \\
        & \(x_0\)\! & \(x_1\) \! & \(z\) & \(\zb\) & \(\r\) & \(\chi_1\) & \(\chi_2\) & \(\chi_3\) & \(y\) & \(\yb\)
        \\
        \hline
        $N_c$ D3 & \textbullet & \textbullet & \textbullet & \textbullet & -- & -- & -- & -- & -- & --
        \\ \hline
        $N_f$ D7 & \textbullet & \textbullet & \textbullet & \textbullet & \textbullet & \textbullet & \textbullet & \textbullet & -- & --\\ \hline
    \end{tabular}\end{center}
    \caption{The 4ND D3/D7 intersection of tab.~\ref{tab:4nd}, in the coordinates of eq.~\eqref{eq:metric_complex_coords}.}
    \label{tab:embedding}
\end{table}

For the D7-branes' worldvolume coordinates, $\xi_a$ with $a=0,\ldots,7$, we choose
\beq
\label{eq:d7coords}
(\xi_0,\xi_1,\xi_2,\xi_3,\xi_4,\xi_5,\xi_6,\xi_7) = (x_0,x_1,z,\zb,\r,\chi_1,\chi_2,\chi_3),
\eeq
which includes static gauge, $\xi_0 = x_0$. The D7-branes' worldvolume theory is then the $d=8$ maximally SUSY $U(N_f)$ YM theory with worldvolume field strength $F$ and scalars $(y,\yb)$, all valued in the complexification of $U(N_f)$. The worldvolume action is a non-Abelian DBI action plus WZ terms. As explained in sec.~\ref{sec:intro}, we consider only Abelian configurations of worldvolume fields, so that the action reduces to the Abelian DBI action plus WZ terms. As mentioned above, in the D3-brane background the only non-trivial RR field is $C_4$. As a result, the D7-branes' only WZ term is proportional to $\int F \wedge F \wedge P[C_4]$, where $P[C_4]$ is the pullback of $C_4$ to the D7-branes' worldvolume. We choose an ansatz in which $F=0$, so that the WZ term vanishes, and contributes nothing to the D7-brane equations of motion. We will thus ignore the WZ term from now on, and discuss only the Abelian DBI action,
\beq
\label{eq:D7-action}
    S = - T_{D7} \, N_f\int d^8 \xi \sqrt{|\det g_{ab}|},
\eeq
with $T_{D7} = (2\pi)^{-7} {\a'}^{-4} g_s^{-1}$ the D7-brane tension and $g_{ab}$ the pullback of the bulk metric to the D7-branes' worldvolume.

As explained in sec.~\ref{sec:intro}, we choose to preserve the $d=2$ Poincar\'e symmetry along $(x_0,x_1)$ and the $SO(4)$ rotational symmetry along $(x_4,x_5,x_6,x_7)$. The most general ansatz for $y$ and $\yb$ is then $y(z,\zb,\rho)$ and $\yb(z,\zb,\rho)$. We denote
\beq
\label{eq:derivnotation}
\partial y \equiv \frac{\partial y}{\partial z}, \qquad \pb y \equiv \frac{\partial y}{\partial \zb}, \qquad y' \equiv \frac{\partial y}{\partial \rho},
\eeq
and similarly for $\yb$. We can actually eliminate $y$ and $\yb$'s dependence on $\rho$, as follows. For the ansatz $y(z,\zb,\rho)$ and $\yb(z,\zb,\rho)$ a straightforward calculation shows that $\delta S/\delta y'$ is a sum of two terms, one term $\propto y'$ and the other term $\propto \yb'$, and the same is true of $\delta S/\delta\yb'$. We may thus self-consistently set $y'=0$ and $\yb'=0$, so that $y$ and $\yb$ no longer depend on $\rho$. Crucially, in sec.~\ref{sec:kappa_symmetry} we will also show that $d=2$ SUSY along $(x_0,x_1)$ \textit{requires} $y'=0$ and $\yb'=0$. In short, $(y,\yb)$ that depend on $(z,\zb)$ but do not depend on $\rho$ provide a \textit{self-consistent} ansatz preserving $d=2$ Poincar\'e symmetry along $(x_0,x_1)$, and provide the \textit{most general} ansatz preserving $d=2$ SUSY along $(x_0,x_1)$.

Our ansatz thus becomes $y(z,\zb)$ and $\yb(z,\zb)$, in which case
\beq
\label{eq:metric_determinant}
    |\det g_{ab}| = \r^6\,\sin^4 \chi_1 \, \sin^2 \chi_2\le[
        \le(1 + f(r)^{-2}\left(|\p y|^2 + |\pb y|^2\right) \ri)^2 - 4 f(r)^{-4} |\p y|^2 |\pb y|^2\ri].
\eeq
Plugging eq.~\eqref{eq:metric_determinant} into the action $S$ in eq.~\eqref{eq:D7-action} and integrating over the $S^3$ directions, producing a factor of a unit-radius $S^3$ volume, $2 \pi^2$, then gives
\beq
\label{eq:D7-action-ansatz}
    S = - 2\pi^2 T_{D7} N_f  \int dx_0 d x_1 dz d \zb \, d\r \,\r^3\sqrt{
        \le[1 + f(r)^{-2} \left(|\p y|^2 + |\pb y|^2 \ri)\ri]^2 - 4 f(r)^{-4} |\p y|^2 |\pb y|^2}.
\eeq
Defining
\begin{subequations}
\begin{align}
 A &\equiv f(r)^2 + | \p y |^2 + | \pb y |^2,\\
    \cD[\bullet] &\equiv \pb y \, \pb \yb \, \p^2 \bullet + \p y \, \p \yb \, \pb^2 \bullet - A \, \p \pb \bullet,
\end{align}
\end{subequations}
allows us to write \(y\)'s equation of motion compactly as
\beq \label{eq:yeom}
    \p \yb \, \pb \yb \,\cD[y] - \frac{1}{2} A \, \cD[\yb]
    - \frac{1}{r}\,f \, \p_r f \, \p \yb \, \pb \yb\, \le(A y - 2 \yb  \, \p y \, \pb y \ri) = 0,
\eeq
where $\yb$'s equation of motion is the complex conjugate of eq.~\eqref{eq:yeom}, so for brevity we omit it. We now come to the main result of this subsection: if \(y\) is holomorphic, meaning $\pb y=0$ and $\p \yb=0$, or if $y$ is antiholomorphic, meaning $\p y=0$ and $\pb \yb = 0$, then each of the three terms on the left-hand side of eq.~\eqref{eq:yeom} vanishes independently. Holomorphic or antiholomorphic $y$ thus solve the D7-brane equation of motion, as advertised.

%%%%%%%%%%%%%%%%%%%%%%%%%%%%%%%%%%%%%%%%%%%%%%%%%%
%%%%%%%%%%%%%%%%%%%%%%%%%%%%%%%%%%%%%%%%%%%%%%%%%%
\subsection{Subtleties about poles}
\label{sec:poles}
%%%%%%%%%%%%%%%%%%%%%%%%%%%%%%%%%%%%%%%%%%%%%%%%%%
%%%%%%%%%%%%%%%%%%%%%%%%%%%%%%%%%%%%%%%%%%%%%%%%%%

Strictly speaking, only holomorphic or antiholomorphic $y$ solve the equation of motion eq.~\eqref{eq:yeom}, while meromorphic or antimeromorphic $y$ do not. Put more simply, poles in $y$ do not solve the equation of motion. To see why, consider for example a simple pole at the origin, $y(z) = 1/z$, for which $\pb y\neq 0$ at the pole, instead being a delta function:
\beq
\label{eq:polederiv}
\bp \left(\frac{1}{z}\right) = 2 \pi \,\delta^{(2)} (z,\zb).
\eeq
As a result, each term on the left-hand side of eq.~\eqref{eq:yeom} no longer vanishes, so the equation of motion is violated locally, precisely at the pole. Meromorphic or antimeromorphic $y$ still solve the equation of motion \textit{away} from any poles, however. Analogous statements apply to higher-degree poles, such as $y \propto 1/z^D$ with integer $D>1$, because they also have $\pb y \neq 0$: taking $D-1$ derivatives $\p$ of eq.~\eqref{eq:polederiv} gives
\beq
\pb \left(\frac{1}{z^D}\right) \propto (-1)^{D-1}\p^{D-1} \delta^{(2)} (z,\zb),
\eeq
which again does not solve the equation of motion locally, at the pole.

Refs.~\cite{Callan:1997kz,Constable:2001ag,Constable:2002xt,Arean:2007nh,Ammon:2012mu} encountered similar issues about poles, and suggested two methods to modify the equation of motion such that poles become solutions.

The first method to make poles in $y$ solve the equation of motion is to add source terms to the equation, such as delta function sources on the right-hand side of eq.~\eqref{eq:yeom} that can cancel any delta functions produced by poles on the left-hand side. A pole describes a ``spike'' of D7-brane that extends to spatial infinity, and delta function sources in general represent other string theory objects, such as strings or D-branes, ending on or dissolved into the D7-branes at the tip of the spike, at spatial infinity. In our case such objects are presumably other D7-branes: anything else would source worldvolume gauge field flux on our D7-branes, but in our solutions the gauge fields vanish.

The second method to make a pole in $y$ solve the equation of motion is to \textit{excise} all points in the $z$ plane where poles are located. In general we must then introduce boundary conditions around those excised points. In other words, we replace the pole with boundary conditions. A detailed discussion of this method appears for example in sec.~$4$ of ref.~\cite{Ammon:2012mu}.

In what follows we will use the second method to make poles in $y$ solve the equations of motion, that is, we will excise points in the $z$ plane where any poles are located. Furthermore, we will assume that we can impose boundary conditions around any excised points so as to preserve the BPS bound of sec.~\ref{sec:bpsbound} and to preserve the SUSY we find in sec.~\ref{sec:kappa_symmetry}. In an abuse of language, from now on we will always refer to $y$ as holomorphic or antiholomorphic, even when $y$ appears to have a pole---because in fact we will implicitly be excising any such poles from the $z$ plane.

%%%%%%%%%%%%%%%%%%%%%%%%%%%%%%%%%%%%%%%%%%%%%%%%%%
%%%%%%%%%%%%%%%%%%%%%%%%%%%%%%%%%%%%%%%%%%%%%%%%%%
\subsection{D7-brane BPS bound}
\label{sec:bpsbound}
%%%%%%%%%%%%%%%%%%%%%%%%%%%%%%%%%%%%%%%%%%%%%%%%%%
%%%%%%%%%%%%%%%%%%%%%%%%%%%%%%%%%%%%%%%%%%%%%%%%%%

In this section we will prove a BPS bound on the D7-branes' action and energy, and prove that holomorphic or antiholomorphic $y$ saturate this BPS bound. In a sector with fixed BPS charge, such $y$ thus extremise the action and solve the equation of motion. The BPS bound will thus explain why these solutions exist, and what they describe.

Our proof will closely follow that of ref.~\cite{Gauntlett:1997ss} for BPS bounds on actions and energies of various types of probe branes in $d=10$ and $d=11$ Minkowski spacetimes, in terms of central charges of SUSY algebras. To follow the proof of ref.~\cite{Gauntlett:1997ss} we must make a crucial distinction between two different SUSY algebras: the target space SUSY algebra (also known as the background or spacetime SUSY algebra) and the worldvolume SUSY algebra. We will briefly review some key properties of these SUSY algebras that we will need for our proof, leaving the full details to the references.

In string and M-theory, the \textit{target space} SUSY algebra's central charges correspond to stable branes~\cite{deAzcarraga:1989mza,Sato:1998yx,Sato:1998ax,Sato:1998yu,Callister:2007jy}. Indeed, branes are solitons of string and M-theories, and the central charges provide quantum numbers that protect such solitons from decay. We can define the number of $p$-branes, $N_p$, via an asymptotic integral of Ramond-Ramond and/or Neveu-Schwarz $(d-p-2)$-forms in the directions transverse to the $p$-branes. In the target space SUSY algebra the $p$-branes' central charge \textit{density} then appears as a $p$-form along the $p$-brane's spatial directions, with coefficient $\propto N_p$. Integrating that $p$-form over the $p$-branes' spatial directions then gives the central charge itself, $Z$. The central charge $Z$ is non-zero only if the $p$-branes wrap a non-trivial compact cycle or fill an infinite direction in the target space, since otherwise they would collapse due to their tension (unless something provides a stabilising counter-force, such as worldvolume gauge field flux). The central charge $Z$ thus essentially counts the number of $p$-branes wrapping the given cycle/direction, which is topological in the sense that any continuous deformation of the cycle/direction cannot change the number of wrapped $p$-branes. The target space SUSY algebra then gives a BPS condition equating $Z$ to the $p$-branes' total energy. If the $p$-branes are static, then their action is minus their energy, integrated over time. A bound on static $p$-branes' energy in terms of $Z$ can thus translate to a bound on their action in terms of $Z$.

On the worldvolume of $N_p$ coincident $p$-branes, the \textit{worldvolume} SUSY algebra's central charges correspond to stable solitons of the worldvolume theory, including other stable branes that the $p$-branes' worldvolume fields can describe. For example, worldvolume fields can describe lower-dimensional branes ``dissolved'' into the $p$-brane~\cite{Douglas:1995bn}, they can  ``polarise'' to describe higher-dimensional branes (the Myers effect~\cite{Myers:1999ps}), and so on. Of particular importance to us are solitons comprised exclusively of worldvolume scalars, with all other worldvolume fields set to zero. Since the worldvolume scalars' eigenvalues describe the position of the $p$-branes in transverse directions, purely scalar solitons can describe the $p$-branes bending in space, or equivalently $p$-branes in different locations ``merging'' into a single stack. As mentioned below eq.~\eqref{eq:d7coords}, we will exclusively consider configurations of worldvolume scalars that are Abelian, and more specifically that are diagonal.

The target space and worldvolume SUSY algebra central charges must be consistent with one another, since they can both describe the same objects from different perspectives. We should therefore be able to express a BPS bound on brane energy or action in terms of either the target space or worldvolume SUSY algebra. Ref.~\cite{Gauntlett:1997ss} proved such BPS bounds in terms of worldvolume SUSY central charges. In the following we will appeal to both types of SUSY algebras, using whichever is most convenient at any given time.

Crucially, ref.~\cite{Gauntlett:1997ss} considered only the Minkowski vacuum of $d=10$ or $d=11$ SUGRA, and proved various BPS bounds on $p$-brane energies and actions, for various $p$. Our proof will generalise ref.~\cite{Gauntlett:1997ss}'s proof for purely scalar solitons on M2-branes. The generalisation is actually trivial for a Minkowski background in any $d$, for a DBI action of any $p$-brane with at least two worldvolume spatial directions, $p\geq 2$, so that we can define worldvolume complex coordinates $(z,\zb)$, and at least two transverse directions, $d-(p+1) \geq 2$, so that we can define worldvolume complex scalars $(y,\yb)$. In particular, the proof of ref.~\cite{Gauntlett:1997ss} trivially generalises to D7-branes in $d=10$ Minkowski spacetime. In other words, when $f(r)=1$ we can trivially generalise the proof in ref.~\cite{Gauntlett:1997ss} to show that holomorphic or antiholomorphic $y$ solve the D7-brane equation of motion, and describe D7-branes extended along a curve in the $(y,\yb)$ plane or equivalently D7-branes at various locations in the $(y,\yb)$ plane merging into a single stack. Our main result will thus be to generalise ref.~\cite{Gauntlett:1997ss}'s proof from the $d=10$ Minkowski background, with $f(r)=1$, to a background produced by a different type of brane, namely the D3-brane background with $f(r)$ in eq.~\eqref{eq:metric}. From our point of view, we will extend ref.~\cite{Gauntlett:1997ss}'s proof from the asymptotic Minkowski region at $r \gg L$ to all values of $r$, including the near-horizon $AdS_5 \times S^5$ region at $r \ll L$ where we can invoke the AdS/CFT correspondence, which we will discuss in sec.~\ref{sec:nearhorizon}.

We define a Lagrangian density $\cL$ via
\beq
\label{eq:D7-action-lag-density_1}
S \equiv - 2 \pi^2 \, T_{D7} \, N_f  \int dx_0 \, dx_1 \, dz \, d\zb \, d\rho \, \rho^3 \, \cL,
\eeq
where explicitly
\beq
\label{eq:D7-action-lag-density_2}
\cL \equiv \sqrt{\le[1 + f(r)^{-2} \left(|\p y|^2 + |\pb y|^2 \ri)\ri]^2 - 4 f(r)^{-4} |\p y|^2 |\pb y|^2}.
\eeq
Our ansatz is static, so we can write $S$ in terms of the D7-branes' total energy, $E$, as $S \equiv - \int d x_0 \, E$. In terms of the Lagrangian $\cL$ defined in eqs.~\eqref{eq:D7-action-lag-density_1} and~\eqref{eq:D7-action-lag-density_2},
\beq
\label{eq:D7-energy}
E \equiv 2 \pi^2 \, T_{D7} \, N_f  \int dx_1 \, dz \, d\zb \, d\rho \, \rho^3 \, \cL.
\eeq
We will actually prove a bound on $\cL$, which will immediately imply bounds on $S$ and $E$.

Our bound will involve two types of BPS D7-branes, namely the trivial embedding of the 4ND D7-branes of tab.~\ref{tab:4nd} and the trivial embedding of the 8ND D7-branes of tab.~\ref{tab:8nd}. Notice that the distinction between 4ND and 8ND D7-branes does not exist in $d=10$ Minkowski spacetime, that is, this distinction exists only in the presence of the D3-branes. The action and energy of the trivial embedding of the 4ND D7-branes are given by eqs.~\eqref{eq:D7-action-lag-density_1} and~\eqref{eq:D7-energy} with $y=0$ and $\yb=0$, respectively,
\begin{subequations}
\label{eq:D7-action-energy-4ND}
\beq
\label{eq:D7-action-4ND}
S_{\textrm{4ND}} \equiv - 2 \pi^2 \, T_{D7}  \, N_f \int dx_0 \, dx_1 \, dz \, d\zb \, d\rho \, \rho^3,
\eeq
\beq
\label{eq:D7-energy-4ND}
E_{\textrm{4ND}} \equiv 2 \pi^2 \,  T_{D7}\, N_f  \int dx_1 \, dz \, d\zb \, d\rho \, \rho^3.
\eeq
\end{subequations}
In both $S_{\textrm{4ND}}$ and $E_{\textrm{4ND}}$ all of the integrals trivially diverge due to the infinite extent of the D7-branes. Regularising these using simple long-distance cutoffs will suffice for our purposes. We will henceforth assume such cutoffs are in place. (In the near-horizon $AdS_5 \times S^5$ region we can eliminate the large-$\rho$ divergences via holographic renormalisation, that is, by introducing a large-$\rho$ cutoff and then adding localised counterterms on that cutoff surface: see the appendix.) From the target space SUSY algebra perspective, $E_{\textrm{4ND}}$ in eq.~\eqref{eq:D7-energy-4ND} comes from integrating a $7$-form central charge density over the trivial 4ND D7-branes' spatial directions, $(x_1,z,\zb,\rho,\chi_1,\chi_2,\chi_3)$, to obtain a central charge that counts the number of trivial 4ND D7-branes, that is, a central charge $\propto N_f$.

In the coordinates of eq.~\eqref{eq:metric_complex_coords}, the 8ND D7-branes are along $(x_0,x_1,y,\yb,\rho,\chi_1,\chi_2,\chi_3)$. Tab.~\ref{tab:embedding8ND} summarises the trivial 8ND D3/D7 intersection in these coordinates.
\begin{table}
    \begin{center}\begin{tabular}{|c || c c c c | c  | c  c c | c c|}
    \hline
         & \multicolumn{4}{c |}{\(\mathbb{R}^{1,3}\)} & & \multicolumn{3}{c |}{\(S^3\)} &  & 
        \\
        & \(x_0\)\! & \(x_1\) \! & \(z\) & \(\zb\) & \(\r\) & \(\chi_1\) & \(\chi_2\) & \(\chi_3\) & \(y\) & \(\yb\)
        \\
        \hline
        $N_c$ D3 & \textbullet & \textbullet & \textbullet & \textbullet & -- & -- & -- & -- & -- & --
        \\ \hline
        $N_8$ D7 & \textbullet & \textbullet & -- & -- & \textbullet & \textbullet & \textbullet & \textbullet & \textbullet & \textbullet\\ \hline
    \end{tabular}\end{center}
    \caption{Our 8ND intersection of $N_c$ D3-branes and $N_8$ D7-branes, similar to that in tab.~\ref{tab:8nd} but now in the coordinates of eq.~\eqref{eq:metric_complex_coords}.}
    \label{tab:embedding8ND}
\end{table}
The action and energy of the trivial embedding of a number $N_8$ of the 8ND D7-branes are, respectively,
\begin{subequations}
\label{eq:D7-action-energy-8ND}
\beq
\label{eq:D7-action-8ND}
S_{\textrm{8ND}} \equiv - 2 \pi^2 \, T_{D7}  \, N_8 \int dx_0 \, dx_1 \, dy \, d\yb \, d\rho \, \frac{\rho^3}{f(r)^2},
\eeq
\beq
\label{eq:D7-energy-8ND}
E_{\textrm{8ND}} \equiv 2 \pi^2 \, T_{D7} \, N_8  \int dx_1 \, dy \, d\yb \, d\rho \, \frac{\rho^3}{f(r)^2}.
\eeq
\end{subequations}
Like $\sfnd$ and $\efnd$, in both $S_{\textrm{8ND}}$ and $E_{\textrm{8ND}}$ all integrals diverge due to the D7-branes'  infinite extent, so we assume long-distance cutoffs are in place. (In the near-horizon $AdS_5 \times S^5$ region we could perform holographic renormalisation to eliminate large-$\rho$ divergences, but we will not need to do so.) From the target space SUSY algebra perspective, $E_{\textrm{8ND}}$ in eq.~\eqref{eq:D7-energy-8ND} comes from integrating a $7$-form central charge density over the trivial 8ND D7-branes' spatial directions to obtain a central charge that counts the number of trivial 8ND D7-branes, that is, a central charge $\propto N_8$. Again, the distinction between 4ND and 8ND D7-branes exists only in the presence of the D3-branes: the integrands in eqs.~\eqref{eq:D7-action-energy-4ND} and~\eqref{eq:D7-action-energy-8ND} differ only by a factor of $f(r)^{-2}$, and so are practically identical in Minkowski spacetime, where $f(r)=1$, but differ in the presence of the D3-branes, where $f(r)\neq 1$.

We now have all the ingredients to prove our bound for the 4ND D7-brane with non-trivial $y(z,\zb)$ and $\yb(z,\zb)$, closely following the analogous proof in ref.~\cite{Gauntlett:1997ss}. As mentioned above, we will actually prove a bound on the Lagrangian $\cL$ in eq.~\eqref{eq:D7-action-lag-density_2}. We first re-write the quantity under the square root in two different ways, each of which is a sum of squares, so that each term is manifestly $ \geq 0$,
\begin{subequations}
\label{eq:lag_density_1}
\bea
    \cL
      &= \sqrt{\le[1 + f(r)^{-2} \le(|\p y|^2 -  |\pb y|^2 \ri)\ri]^2 +4  f(r)^{-2} | \pb y|^2}
      \\
    &= \sqrt{\le[1 - f(r)^{-2} \le(|\p y|^2 -  |\pb y|^2 \ri)\ri]^2 +4  f(r)^{-2} | \p y  |^2}.
\eea
\end{subequations}
Following ref.~\cite{Gauntlett:1997ss} we identify the following term under the square root as a number density of scalar solitons on the D7-branes' worldvolume,
\beq
\label{eq:qdef}
\mathcal{Z}\equiv f(r)^{-2}(|\p y|^2 -  |\pb y|^2).
\eeq
Indeed, ref.~\cite{Gauntlett:1997ss} emphasised that $\mathcal{Z}$ is a worldvolume two-form, $\mathcal{Z} \propto f(r)^{-2} dy \wedge d\yb$, whose integral over the $z$-plane, $\int_{(z,\zb)} \mathcal{Z}$, counts the number of solitons whose worldvolumes extend along the directions transverse to the $z$-plane, namely $(x_1,\rho,\chi_1,\chi_2,\chi_3)$, in the same way that in SUGRA the integral of a Ramond-Ramond and/or Neveu-Schwarz form gives a number of branes. In the worldvolume SUSY algebra, the corresponding central charge density is $\int_{(z,\zb)} \mathcal{Z}$ times the volume form along $(x_1,\rho,\chi_1,\chi_2,\chi_3)$. Integrating that central charge density over $(x_1,\rho,\chi_1,\chi_2,\chi_3)$ at fixed time $x_0$ gives the worldvolume SUSY central charge, $Z$. Explicitly, after performing the integral over the $S^3$ directions $(\chi_1,\chi_2,\chi_3)$,
\beq
\label{eq:qdef2}
Z \equiv 2 \pi ^2 \,  T_{D7} \, N_f  \int dx_1 \, dz \, d\zb \, d\rho \, \rho^3 \, \mathcal{Z}.
\eeq
Like $\sfnd$, $\efnd$, $\send$, and $\eend$, again in $Z$ all integrals diverge, so again we assume long-distance cutoffs are in place. From the target space SUSY algebra perspective, the factor \((|\p y|^2 - |\pb y|^2)\) in eq.~\eqref{eq:qdef} is simply the Jacobian for the coordinate transformation from \((z,\zb)\) to \((y,\yb)\), which suggests that $Z$ is related to the central charge of D7-branes along $(x_1,y,\yb,\rho,\chi_1,\chi_2,\chi_3)$, that is, the central charge of 8ND D7-branes. We will clarify the relation between $Z$ and the 8ND D7-branes below, in eqs.~\eqref{eq:qdefsol} to~\eqref{eq:n8def}.

Crucially for our bound, constant $y$ have $\mathcal{Z} =0$ and hence $Z=0$, while holomorphic $y$ have $\mathcal{Z}=f(r)^{-2}|\p y|^2 \geq0$ and hence $Z \geq 0$, and antiholomorphic $y$ have $\mathcal{Z}=-f(r)^{-2}|\pb y|^2\leq0$ and hence $Z \leq 0$. In particular, in these cases the Jacobian for a generic coordinate transformation from the $z$-plane to the $y$-plane, \((|\p y|^2 - |\pb y|^2)\), reduces to either the Jacobian for a holomorphic coordinate transformation, \(|\p y|^2 \), or minus the absolute value of the Jacobian of an antiholomorphic coordinate transformation, \(- |\pb y|^2\), where the minus sign indicates orientation reversal.

Still following ref.~\cite{Gauntlett:1997ss}, we next write the $\cL$ in eq.~\eqref{eq:lag_density_1} in terms of $\mathcal{Z}$, 
\begin{subequations}
\label{eq:energy_density_2}
\bea
    \cL 
      &= &  \sqrt{\le[1 + \mathcal{Z}\ri]^2 +4  f(r)^{-2} | \pb y|^2}
      \label{eq:energy_density_2_holo}
      \\
    &= & \sqrt{\le[1 - \mathcal{Z}\ri]^2 +4  f(r)^{-2} | \p y  |^2}.
     \label{eq:energy_density_2_antiholo}
\eea
\end{subequations}
Because the quantity under the square root is a sum of squares, and in particular because the second term under the square root is $\geq0$, we immediately obtain a bound
\beq
\label{eq:energy_density_bound_1}
    \cL \geq 1  \pm \mathcal{Z},
\eeq
where holomorphic $y$ saturate this bound with the plus sign, coming from eq.~\eqref{eq:energy_density_2_holo} with $\pb y=0$, while antiholomorphic $y$ saturate this bound with the minus sign, coming from eq.~\eqref{eq:energy_density_2_antiholo} with $\p y=0$. Since holomorphic $y$ have $\mathcal{Z} \geq 0$ and antiholomorphic $y$ have $\mathcal{Z} \leq 0$, we can re-write eq.~\eqref{eq:energy_density_bound_1} as
\beq
\label{eq:energy_density_bound_2}
    \cL \geq 1  +  | \mathcal{Z} |.
\eeq
Plugging eq.~\eqref{eq:energy_density_bound_2} into eqs.~\eqref{eq:D7-action-lag-density_2} and~\eqref{eq:D7-energy} then gives BPS bounds on the D7-branes' action and energy, respectively
\begin{subequations}
\label{eq:actionenergybounds1}
\begin{eqnarray}
S & \leq & - 2 \pi^2 \, T_{D7} \, N_f  \int dx_0 \, dx_1 \, dz \, d\zb \, d\rho \, \rho^3 \left(1 + |\mathcal{Z}|\right),\\
E & \geq & 2 \pi^2 \, T_{D7} \, N_f  \int dx_1 \, dz \, d\zb \, d\rho \, \rho^3 \, \left(1 +| \mathcal{Z} |\right).
\end{eqnarray}
\end{subequations}
Using eqs.~\eqref{eq:D7-action-energy-4ND} for $\sfnd$ and $\efnd$ and eq.~\eqref{eq:qdef2} for $Z$ we can write eq.~\eqref{eq:actionenergybounds1} as
\begin{subequations}
\label{eq:actionenergybounds2}
\begin{eqnarray}
S & \leq & \sfnd - \int dx_0 |Z|,\\
E & \geq & \efnd + |Z|.
\end{eqnarray}
\end{subequations}
The bounds in eqs.~\eqref{eq:actionenergybounds1} and~\eqref{eq:actionenergybounds2} are only meaningful with long-distance cutoffs in place: otherwise every quantity in eqs.~\eqref{eq:actionenergybounds1} and~\eqref{eq:actionenergybounds2} diverges, rendering the bounds meaningless (i.e. ``$\infty \leq \infty + \infty$''). With long-distance cutoffs in place, the BPS bounds in eq.~\eqref{eq:actionenergybounds2} explain why holomorphic or antiholomorphic $y$ solve the equation of motion: in a sector with fixed $|Z|$, such $y$ saturate the bounds in eq.~\eqref{eq:actionenergybounds2}, meaning they extremise the action in that sector and hence solve the equation of motion.

Liouville's theorem of complex analysis states that the only \textit{bounded} holomorphic functions are the constant functions, $y(z)=y_0$ with complex constant $y_0$~\cite{ahlfors}. Any non-constant holomorphic function diverges as $|z| \to \infty$, i.e. has a pole at the point at infinity. From that perspective, our main result is that \textit{any} holomorphic function $y(z)$ is a BPS solution, whether bounded or not. Analogous statements apply for antiholomorphic $y$.

As mentioned above, holomorphic or antiholomorphic $y$ have $\mathcal{Z}$ proportional to the Jacobian or minus the Jacobian of the map from $(z,\zb)$ to $(y,\yb)$. In each case, using this Jacobian in the definition of $Z$ in eq.~\eqref{eq:qdef2} we can re-write the integral over $(z,\zb)$ as an integral over $(y,\yb)$. For example for a holomorphic map, $\int dz d\zb  |\partial y|^2 = D \int dy d\yb$, where $D$ is the \textit{degree} of the holomorphic map: in the complex plane any holomorphic map without an essential singularity at infinity must be a rational function, $y(z) = P(z)/Q(z)$ with polynomials $P(z)$ and $Q(z)$ (away from zeroes of $Q(z)$), and $D$ is the maximum of the degrees of $P(z)$ and $Q(z)$. Intuitively, $D$ is the number of \(y\)-planes we must integrate over to cover the whole \(z\)-plane. Analogous statements apply for antiholomorphic $y$. If $\zsol$ denotes $Z$ evaluated on a holomorphic or antiholomorphic solution $y$, then re-writing the integrals over $(z,\zb)$ as integrals over $(y,\yb)$ gives
\beq
\label{eq:qdefsol}
\zsol  = \pm 2 \pi ^2 \,  T_{D7} \, N_f \, D \int dx_1 \, dy \, d\yb \, d\rho \, \frac{\rho^3}{f(r)^2},
\eeq
with the plus or minus sign for holomorphic or antiholomorphic $y$, respectively. Crucially, eq.~\eqref{eq:qdefsol} is precisely of the form of the trivial 8ND D7-brane energy, $\eend$ in eq.~\eqref{eq:D7-energy-8ND}:
\beq
\label{eq:qdefsol2}
\zsol = \pm \eend,
\eeq
where we identify the number of trivial 8ND D7-branes as
\beq
\label{eq:n8def}
N_8 = N_f \, D.
\eeq
 Eq.~\eqref{eq:qdefsol2} provides an interpretation of $\zsol$ from the perspective of the target space SUSY algebra, as the central charge arising from a number $N_8=N_f \, D$ of trivial 8ND D7-branes~\cite{Callister:2007jy}. Moreover, using eq.~\eqref{eq:qdefsol2} we can re-write the BPS bounds in eq.~\eqref{eq:actionenergybounds2}, at saturation, in terms of $\send$ and $\eend$: if $\ssol$ and $\esol$ denote the action and energy evaluated on a holomorphic or antiholomorphic solution $y$, respectively, then
\begin{subequations}
\label{eq:actionenergybounds3}
\begin{eqnarray}
\label{eq:actionenergybounds3action}
\ssol & = & \sfnd + |\send|,\\
\esol & = & \efnd + |\eend|,
\end{eqnarray}
\end{subequations}
which shows that holomorphic or antiholomorphic $y$ describe an exactly marginal bound state of a number $N_f$ of 4ND D7-branes with a number $N_8 = N_f D$ of 8ND D7-branes.

Broadly speaking, 8ND D7-branes play a role in the BPS bound simply because $y$ are maps from the $z$-plane, where the 4ND D7-branes are extended, to the $y$-plane, where the 8ND D7-branes are extended. We can be more specific, however: the 8ND D7-branes appear wherever $y$ has a zero or a pole.

When $y$ has a zero we must find 8ND D7-branes because the 4ND D7-branes and D3-branes have mutually transverse directions, $(y,\yb)$, where we can separate them, whereas the 8ND D7-branes fill all directions transverse to the D3-branes, so we cannot separate them. The 8ND D7-branes thus always touch the D3-branes, so the 4ND D7-branes can describe the 8ND D7-branes at points in the $z$-plane where the 4ND D7-branes touch the D3-branes, meaning points where the separation distance $r\to 0$ and hence both $\rho \to 0$ and $y\to0$ (see below eq.~\eqref{eq:rdef}), that is, at points where $y$ has a zero. We can see such behaviour in eq.~\eqref{eq:actionenergybounds3}. For example, in $\esol = \efnd + |\eend|$, the integrand of the $|\eend|$ term dominates, so that $|\eend| \gg \efnd$, when \(f(r) \to 0\), which requires \(r \to 0\) and hence both \(\r \to 0\) and \(y \to 0\). We can also see such behavior in the D7-branes' induced metric. For example, for holomorphic $y(z)$ the D7-branes' induced metric is $ds^2_{D7}\equiv g_{ab} d\xi^a d\xi^b$ is
\beq
\label{eq:holoyinducedg}
ds^2_{D7} = f(r) \left(-dx_0^2 + dx_1^2 + 2  \left(1 +f(r)^{-2} |\partial y|^2 \right) dz \, d\zb \right) + f(r)^{-1} \left(d\rho^2 + \rho^2 ds^2_{S^3}\right),
\eeq
with $f(r)$ in eq.~\eqref{eq:frdef} and $r^2 \equiv \rho^2 + 2 |y|^2$. As $r \to 0$ and hence $f(r) \to 0$, because in general $y$ must vanish as a positive power and so $y\to0$ faster than $\partial y \to 0$, we can approximate $1 +f(r)^{-2} |\partial y|^2 \approx f(r)^{-2} |\partial y|^2$. Plugging that into eq.~\eqref{eq:holoyinducedg} and identifying $|\partial y|^2$ as the Jacobian of the map from $(z,\zb)$ to $(y,\yb)$ then gives
\beq
\left(1 +f(r)^{-2} |\partial y|^2 \right) dz \, d\zb \approx f(r)^{-2}\, \partial y \,\pb \yb \, dz \, d\zb = f(r)^{-2} \, dy \, d\yb.
\eeq
As a result, when $r \to 0$ the D7-branes' induced metric behaves as
\begin{subequations}
\label{eq:8ndfrom4ndzero}
\bea
ds^2_{D7} & \underset{r \to 0}\approx & f(r) \left(-dx_0^2 + dx_1^2 + f(r)^{-2} \, 2\, dy \, d\yb \right)+ f(r)^{-1} \left(d\rho^2 + \rho^2 ds^2_{S^3}\right),\\
& = & f(r) \left(-dx_0^2 + dx_1^2 \right)+ f(r)^{-1} \left(d\rho^2 + \rho^2 ds^2_{S^3} + 2 \, dy \, d\yb\right),
\eea
\end{subequations}
which is the induced metric of the trivial 8ND D7-branes in tab.~\ref{tab:embedding8ND}.

Back in the original 4ND D3/D7 intersection in tab.~\ref{tab:embedding}, in the D3-branes' worldvolume theory the 4ND D7-branes give rise to $d=4$ $\N=2$ SUSY hypermultiplets, where the separation from the D3-branes, $y$, corresponds to the hypermultiplet mass, $m \propto y$. A holomorphic $y$ thus gives rise to a holomorphic mass $m$, that is, a mass that depends holomorphically on the spatial coordinate $z$. The 8ND D7-branes give rise to $d=2$ $\N=(8,0)$ SUSY chiral fermions, and the fact that we cannot separate the 8ND D7-branes from the D3-branes corresponds to the fact that for such chiral fermions the symmetries forbid any mass terms. As a result, the $d=4$ $\N=2$ SUSY hypermultiplets can only describe the $d=2$ $\N=(8,0)$ SUSY chiral fermions at points in the $z$-plane where their mass vanishes, $m\propto y \to 0$. Analogous statements of course apply for antiholomorphic $y$.

When $y$ has a pole we find 8ND D7-branes because the pole describes a ``spike'' of D7-brane extending to the asymptotically flat region, all the while turning in space from being extended along $(z,\zb)$ at small $r$ to being extended along $(y,\yb)$ at large $r$. The asymptotically flat region is infinitely far from the D3-branes that produced the background geometry, which muddies the distinction between 4ND and 8ND D7-branes, similar to our observations below eq.~\eqref{eq:D7-action-energy-8ND}. We can see such behaviour in eq.~\eqref{eq:actionenergybounds3}. For example, if $y$ has a pole at the origin, $y \propto 1/z^D$, then as $z\to0$ we find $r \to \infty$ and $f(r) \to 1$, meaning we approach the asymptotically flat region, and $\efnd$ and $\eend$ become practically identical (they evaluate to the same number), but $\efnd$ has an integral over the $z$-plane while $\eend$ has an integral over the $y$-plane (times $D$). We can also see such behavior in the D7-branes' induced metric, eq.~\eqref{eq:holoyinducedg}. A pole in $y$ gives $|\partial y|^2 \to \infty$, $r \to \infty$ and $f(r) \to 1$, so that
\beq
\left(1 +f(r)^{-2} |\partial y|^2 \right) dz \, d\zb \approx \partial y \,\pb \yb \, dz \, d\zb =  dy \, d\yb.
\eeq
As a result, the D7-branes' induced metric behaves as
\beq
\label{eq:8ndfrom4ndpole}
ds^2_{D7}  \underset{r \to \infty}\approx  -dx_0^2 + dx_1^2 + \, 2 \, dy \, d\yb + d\rho^2 + \rho^2 ds^2_{S^3},
\eeq
which is the induced metric of the trivial 8ND D7-branes in $d=10$ Minkowski spacetime.

In summary, for the ansatz $y(z,\zb)$ and $\yb(z,\zb)$, following ref.~\cite{Gauntlett:1997ss} we have proven the BPS bound on the action $S$ and energy $E$ of the 4ND D7-branes in eq.~\eqref{eq:actionenergybounds2}, in terms of the worldvolume SUSY central charge $Z$ in eq.~\eqref{eq:qdef2}. The BPS bound is saturated for holomorphic or antiholomorphic $y$, which thus solve the D7-branes' equation of motion in the sector with that $Z$. Such a $y$ describes an exactly marginal bound state of $N_f$ trivial 4ND D7-branes with $N_f D$ trivial 8ND D7-branes, with $D$ the degree of $y$.

%%%%%%%%%%%%%%%%%%%%%%%%%%%%%%%%%%%%%%%%%%%%%%%%%%
%%%%%%%%%%%%%%%%%%%%%%%%%%%%%%%%%%%%%%%%%%%%%%%%%%
\subsection{8ND D7-brane perspective}
\label{sec:8ndpov}
%%%%%%%%%%%%%%%%%%%%%%%%%%%%%%%%%%%%%%%%%%%%%%%%%%
%%%%%%%%%%%%%%%%%%%%%%%%%%%%%%%%%%%%%%%%%%%%%%%%%%

Eq.~\eqref{eq:actionenergybounds3} is invariant under the exchange of the 4ND and 8ND D7-branes, suggesting that if we start with 8ND D7-branes along $(y,\yb)$ and write an ansatz $z(y,\yb)$ and $\zb(y,\yb)$, then we will find that holomorphic or antiholomorphic $z$ are BPS solutions describing exactly marginal bound states with 4ND D7-branes. Indeed, that is straightforward to show by suitably adapting the proof above, although we will not present the details.

Such BPS solutions for 8ND D7-branes are not necessarily identical to our BPS solutions for 4ND D7-branes. In particular, a holomorphic solution for the 4ND D7-branes, $y(z)$, provides a holomorphic solution for the 8ND D7-branes, $z(y)$, only when $y(z)$ has a holomorphic inverse, $y^{-1}=z(y)$. An inverse exists only in domains where $y$ is one-to-one (i.e. injective), which is equivalent to a non-vanishing first derivative, $\partial y \neq 0$, which in turn is equivalent to a non-zero Jacobian, $|\partial y|^2\neq0$. The inverse, when defined, is automatically holomorphic: $y \circ y^{-1} = 1$ implies $\frac{\partial}{\partial \yb} \left(y\circ y^{-1}\right)=0$, so that the chain rule gives $\partial y \left( \frac{\partial}{\partial \yb}y^{-1}\right)+\pb y \left(\frac{\partial}{\partial \yb}y^{-1}\right)=0$, and then imposing that $y$ is holomorphic gives $\partial y \left( \frac{\partial}{\partial \yb}y^{-1}\right)=0$, and then imposing that $y$ is injective gives $\frac{\partial}{\partial \yb}y^{-1}=0$. Analogous statements of course apply for antiholomorphic $y$ and $y^{-1}$.

An illustrative example is a power law solution, $y(z)=a \, z^D$ with complex constant $a\neq0$ and integer $D=0,1,2,3,\ldots$. All such functions are holomorphic in $z$, and hence are BPS solutions for the 4ND D7-branes. However, such functions are not always one-to-one, and hence do not always have a holomorphic inverse that provides a BPS solution for the 8ND D7-branes. The case $D=0$ gives the constant function $y(z)=a$, which maps all points in the $z$-plane to one point in the $y$-plane, which is clearly not one-to-one. Correspondingly, $\partial y=0$ for all $z$. As a result, $y(z)=a$ has no inverse, and hence provides no BPS solution for the 8ND D7-branes. In contrast, $D=1$ gives a linear function, $y(z)=az$, which is one-to-one for all $z$, and correspondingly $\partial y =a \neq 0$ for all $z$. As a result, $y(z)=az$ has a holomorphic inverse, $z(y)=y/a$, which does provide a BPS solution for the 8ND D7-branes. Indeed, a linear function is not only holomorphic but entire, and in fact is the only entire function with an entire inverse. In contrast to both of those cases, $D>1$ gives $y(z)=az^D$, which is one-to-one everywhere except the origin, $z=0$, which is a branch point. Correspondingly, $\partial y(z)= a\, D\, z^{D-1}$ vanishes at the origin. As a result, $y(z)=az^D$ has a holomorphic inverse, $z(y) = (y/a)^{1/D}$, only away from $z=0$, and so provides a BPS solution for the 8ND D7-branes only in that domain of the $z$-plane. A non-trivial source could potentially make the branch cut in $z(y)\propto y^{1/D}$ a solution.

When a BPS solution for the 4ND D7-branes has an inverse, the corresponding solution for the 8ND D7-branes can provide a useful alternative perspective. For example, what does $N_8 = N_f \, D$ mean? Why are the numbers of 4ND and 8ND D7-branes different? Why do the 8ND D7-branes describe $D$ copies of the $y$-plane wrapping the $z$-plane? To answer these questions suppose for simplicity that $N_f=1$ and hence $N_8=N_fD=D$. From the 8ND D7-brane perspective, the $D$ copies of the $y$-plane are simply $D$ different 8ND D7-branes. These have a $U(N_8)$ worldvolume gauge group, with the worldvolume scalar $z$ is valued in the adjoint of $U(N_8)$'s complexified Lie algebra, which can describe $N_f < N_8$ units of 4ND D7-brane charge. An example is $z(y) = \left(y/a\right)^{1/D}$, which has $D$ sheets, where each sheet is a single 8ND D7-brane and the solution describes their bound state. Further examples of such non-Abelian solutions appear for instance in ref.~\cite{Li:1998ce}, which showed that the degree $D$ of a pole determines the number of branes in a transverse direction. 

In general, however, the 4ND D7-brane perspective is more intuitive than that of the 8ND D7-branes, largely because the former describe flavour fields propagating in all $d=4$ directions of the D3-brane worldvolume theory while the latter describe flavour fields propagating on a $d=2$ defect. A BPS solution for the 4ND D7-branes, $y$, describes a spatially-dependent mass $m = y/(2 \pi \alpha')$, which is relatively intuitive. In contrast, a BPS solution for the 8ND D7-branes, $z$, describes the position of the defect changing as we move in the $y$-plane, which is less intuitive. How do we interpret such a solution? At the moment we cannot answer this question completely, but we can provide two useful observations. First, in the D3-branes' worldvolume theory $y$ and $\yb$ are adjoint scalars, so a solution like $z(y)$ or $z(\yb)$ describes a coupling to an adjoint scalar that, in effect, changes the defect's position. How can a defect's position change? That brings up our second observation: like any non-topological defect, this defect's operator spectrum includes the displacement operator, which describes local changes in the defect's position. A solution $z(y)$ or $z(\yb)$ must somehow describe a coupling between the adjoint scalar $y$ or $\yb$ and the displacement operator such that SUSY is preserved, $SO(6)_R$ breaks to $SO(4)$, and so on. We leave a complete analysis of the BPS solutions of 8ND D7-branes for future research.

%%%%%%%%%%%%%%%%%%%%%%%%%%%%%%%%%%%%%%%%%%%%%%%%%%
%%%%%%%%%%%%%%%%%%%%%%%%%%%%%%%%%%%%%%%%%%%%%%%%%%
\section{Supersymmetry analysis}
\label{sec:kappa_symmetry}
%%%%%%%%%%%%%%%%%%%%%%%%%%%%%%%%%%%%%%%%%%%%%%%%%%
%%%%%%%%%%%%%%%%%%%%%%%%%%%%%%%%%%%%%%%%%%%%%%%%%%

For the 4ND probe D7-branes in the D3-brane background of eq.~\eqref{eq:metric}, constant $y$ solutions of the D7-branes' equation of motion preserve $d=4$ $\N=2$ SUSY in the directions $(x_0,x_1,x_2,x_3)$ or equivalently $(x_0,x_1,z,\zb)$~\cite{Karch:2002sh}. In the previous section we showed that $y$ holomorphic or antiholomorphic in $z$ are BPS solutions. We thus expect them to preserve some SUSY. In this section, we will show that a generic holomorphic or antiholomorphic $y$ preserves $d=2$ $\N=(4,0)$ or $\N=(0,4)$ SUSY along $(x_0,x_1)$, respectively.

Let $\ve$ denote a constant doublet of spinors in $d=10$, which has $64$ real components. Type IIB SUGRA's Minkowski vacuum has $32$ real supercharges, corresponding to the $32$ components of $\ve$ that survive the $d=10$ Majorana-Weyl projection condition
\beq
\label{eq:ads5_chirality_condition}
 \le(\id_2 \otimes \G_{\sharp} \ri)  \ve  = (\id_2 \otimes \id) \ve,
\eeq
with  \(2\times2\) identity matrix \(\id_2\) acting on the doublet index and $d=10$ Minkowski Dirac matrices $\Gamma_A$ with $A = 0,1,\ldots,9$, which obey \(\{\G_A, \G_B\} = 2 \h_{AB} \id\) with mostly-plus Minkowski metric $\eta_{AB}$ and $32 \times 32$ identity matrix $\id$. Our $d=10$ chirality matrix is \(\G_\sharp \equiv \G_{01\cdots9}\), where we normalise our totally antisymmetric products, for example  \(\G_{AB} \equiv \frac{1}{2} (\G_A \G_B - \G_B \G_A)\).

The D3-branes impose an additional projection condition, from their $\kappa$ symmetry~\cite{Bergshoeff:1997kr}
\beq
\label{eq:ads5_susy_condition}
    \le(i \s_2 \otimes \G_{0123} \ri) \ve = \ve,
\eeq
with second Pauli matrix $\s_2$. The matrices on the left-hand sides of the two projection conditions eqs.~\eqref{eq:ads5_chirality_condition} and~\eqref{eq:ads5_susy_condition} are linearly independent, so that eq.~\eqref{eq:ads5_susy_condition} sets to zero another half of $\ve$'s components. The D3-brane background thus preserves $16$ real supercharges. The corresponding Killing spinors are~\cite{Kehagias:1998gn,Grana:2000jj}
\beq
\label{eq:d3killingspinors}
\hat{\varepsilon} \equiv f(r)^{1/4} \ve.
\eeq
The \(AdS_5 \times S^5\) solution, which emerges in the near-horizon region of the D3-brane background, has an additional $16$ supercharges, for a total of $32$. The initial $16$ supercharges in eq.~\eqref{eq:d3killingspinors} are dual to $\N=4$ SYM's Poincar\'e supercharges, while the $16$ additional supercharges are dual to $\N=4$ SYM's superconformal supercharges.

%%%%%%%%%%%%%%%%%%%%%%%%%%%%%%%%%%%%%%%%%%%%%%%%%%
%%%%%%%%%%%%%%%%%%%%%%%%%%%%%%%%%%%%%%%%%%%%%%%%%%
\subsection{(Anti-)holomorphic solutions preserve $4$ supercharges}
\label{sec:number_of_susy}
%%%%%%%%%%%%%%%%%%%%%%%%%%%%%%%%%%%%%%%%%%%%%%%%%%
%%%%%%%%%%%%%%%%%%%%%%%%%%%%%%%%%%%%%%%%%%%%%%%%%%

The D7-branes impose their own $\kappa$ symmetry projection condition on the Killing spinor,
\beq
\label{eq:kappa_symmetry_condition}
    \Gamma \,\hat{\ve} =  \hat{\ve},
\eeq
with \(\k\) symmetry projector \(\G\) determined by the D7-branes' worldvolume fields. To write $\Gamma$ explicitly, we need the $d=10$ vielbeins, \(e_{\m}^{A}\), which we use to define the $d=10$ curved-space Dirac matrices, $e_{\m}^{A} \, \G_{A}$, whose pullbacks to the D7-branes' worldvolume are
\beq
    \g_a \equiv (\p_a x^{\m}) \, e_{\m}^{A} \, \G_{A}.
\eeq
For D7-branes with zero worldvolume gauge field the $\kappa$ symmetry projector is then (recall $g_{ab}$ denotes the pullback of the metric to the D7-branes' worldvolume)
\beq
\label{eq:kappa_projector}
    \G \equiv \frac{1}{\sqrt{|\det g_{ab}|}} \, \s_2 \otimes \g_{01234567}.
\eeq
For our general ansatz $y(z,\zb,\rho)$, and in the worldvolume coordinates of eq.~\eqref{eq:d7coords}, we choose the $d=10$ vielbeins $e_{\m}^{A}$ such that (recall our notation for derivatives of $y$ in eq.~\eqref{eq:derivnotation})
\begin{subequations}
\label{eq:gamma_matrices_pullback}
\begin{align}
    \g_0 &= \sqrt{f(r)} \,\G_{0},
    \\
    \g_1 &= \sqrt{f(r)} \,\G_{1},
    \\
    \g_2 &= \sqrt{\frac{f(r)}{2}} \le(\G_{2} - i \G_{3} \ri) + \frac{1}{\sqrt{2f(r)}} \le[
        \p y \le( \G_8 - i \G_9 \ri)
        + \p \yb \le( \G_8 + i \G_9 \ri)
    \ri],
    \\
    \g_3 &= \sqrt{\frac{f(r)}{2}} \le(\G_{2} + i \G_{3} \ri) + \frac{1}{\sqrt{2f(r)}} \le[
        \pb y \le( \G_8 - i \G_9 \ri)
        + \pb \yb \le( \G_8 + i \G_9 \ri)
    \ri],
    \\
    \g_4 &= \frac{1}{\sqrt{f(r)}}\, \G_4 + \frac{1}{\sqrt{2 f(r)}} \le[ y' (\G_8 - i \G_9) + \yb' (\G_8 + i \G_9) \ri],
    \\
    \g_5 &= \frac{\r}{\sqrt{f(r)}} \,\G_5,
    \\
    \g_6 &= \frac{\r}{\sqrt{f(r)}} \sin \chi_1 \,\G_6,
    \\
    \g_7 &= \frac{\r}{\sqrt{f(r)}} \sin \chi_1 \sin \chi_2 \, \G_7.
\end{align}
\end{subequations}

Plugging the constant solution, $y = y_0$ with complex constant $y_0$, into eqs.~\eqref{eq:kappa_projector} and~\eqref{eq:gamma_matrices_pullback} gives $\Gamma = i \s_2 \otimes \G_{01234567}$. In that case, the matrices on the left-hand sides of all three projection conditions, eqs.~\eqref{eq:ads5_chirality_condition},~\eqref{eq:ads5_susy_condition}, and~\eqref{eq:kappa_symmetry_condition} are linearly independent. As a result, $8$ of $\ve$'s components survive all three projection conditions, corresponding to the $8$ Poincar\'e supercharges of $d=4$ $\N=2$ SUSY. From this perspective, we will prove that our non-constant BPS solutions preserve $4$ of these $8$ supercharges.

In the $AdS_5 \times S^5$ solution, out of the $16$ supercharges dual to $\N=4$ SYM's Poincar\'e supercharges and the $16$ dual to superconformal supercharges, the trivial solution $y=0$ preserves $8$ of each, for a total of $16$. A non-zero constant solution $y = y_0\neq0$ preserves only the $8$ Poincar\'e supercharges. The holographically dual statements are that in the probe limit flavours with \textit{zero mass} preserve $d = 4$ $\N=2$ superconformal symmetry while flavours with \textit{non-zero constant mass} preserve only $d=4$ $\N=2$ Poincar\'e SUSY, respectively. In what follows, we will consider only $y\neq0$, so we will not consider the superconformal supercharges any further, i.e. our arguments will apply only to the Poincar\'e supercharges. 

Returning to our general ansatz $y(z,\zb,\rho)$, in eq.~\eqref{eq:kappa_projector} for $\Gamma$ we can reduce $\g_{01234567}$  to a sum of terms, each of which has a product of just two $\G_A$, as follows. A straightforward calculation shows that the only non-zero anticommutators between the $\g_a$ are
\begin{subequations} 
\begin{align}
    \{ \g_2 , \g_3 \} &= 2 g_{23} = 2 \le( f(r) + \frac{\p y \, \pb \yb + \pb y \, \p \yb}{f(r)} \ri),
    \\
    \{ \g_2 , \g_4 \} &= 2 g_{24} = \frac{y' \p \yb + \yb' \p y}{f(r)},
    \\
    \{ \g_3 , \g_4 \} &= 2 g_{34} = \frac{y' \pb \yb + \yb' \pb y}{f(r)}.
\end{align}
\end{subequations}
In eq.~\eqref{eq:kappa_projector} for $\Gamma$ we may thus write
\begin{align}
\label{eq:gamma_product}
    \g_{01234567} &= - \g_{01567} \g_{234},
\end{align}
where, in terms of the central charge density $\mathcal{Z}$ in eq~\eqref{eq:qdef},
\begin{subequations}
\label{eq:gamma_sub_products}
\begin{align}
\label{eq:gamma_01567}
\g_{01567} &= \frac{\r^3}{\sqrt{f(r)}} \sin^2 \chi_1 \sin \chi_2\, \G_{01567},
\\
\label{eq:gamma_234}
    \frac{\g_{234}}{\sqrt{f(r)}} &=  i  \G_{234} + i \mathcal{Z} \,\G_{489}
    + \frac{i}{\sqrt{2}} \le[y' (\G_{238} - i \G_{239})  + \yb' (\G_{238} + i \G_{239})\ri]
    \nonumber \\ &\phantom{=}
    +\frac{1}{2 f(r)}(\p y - \pb \yb) (\G_{248} + \G_{349}) 
    +\frac{1}{2 f(r)} (\p \yb - \pb y) (\G_{248} - \G_{349})
    \nonumber \\ &\phantom{=}
    -\frac{i}{2 f(r)} (\p y + \pb \yb) (\G_{249} - \G_{348})
    +\frac{i}{2 f(r)} (\p \yb + \pb y) (\G_{249} + \G_{348})
    \nonumber \\ &\phantom{=}
    + i\frac{y' (\p - \pb) \yb - \yb' (\p - \pb)y}{\sqrt{2} \, f(r)}   \G_{289}
    - \frac{y' (\p + \pb) \yb - \yb' (\p + \pb)y }{\sqrt{2}\, f(r)}  \G_{389}.
\end{align}
\end{subequations}
In the \(\k\) symmetry condition eq.~\eqref{eq:kappa_symmetry_condition} the product \(\g_{01234567} \e\) is thus proportional to a linear combination of terms of the form \((i\s_2 \otimes\G_{01567} \G_{A_1 A_2 A_3}) \e_0\) with $\{A_1, A_2, A_3\} \in \{2,3,4,8,9\}$. We can reduce such products to a product of only two $\G_A$ using eq.~\eqref{eq:ads5_chirality_condition} combined with the identity \(\G_{01567} \G_{A_1 A_2 A_3} = - \frac{1}{2}\e_{01567 A_1 A_2A_3 B_1 B_2} \G_{B_1 B_2} \G_\sharp\), with summation over the indices $\{B_1,B_2\}$ between the $d=10$ totally antisymmetric symbol and $\G_{B_1 B_2}$. For example,
\beq
\label{eq:gamma_product_simplification}
    (i \s_2 \otimes \G_{01567} \G_{234}) \ve = (i \s_2 \otimes \G_{89} \G_\sharp )\ve = ( i \s_2 \otimes \G_{89} )\ve.
\eeq
By plugging eqs.~\eqref{eq:kappa_projector},~\eqref{eq:gamma_product} and~\eqref{eq:gamma_sub_products} into the $\kappa$ symmetry condition eq.~\eqref{eq:kappa_symmetry_condition} and performing simplifications such as those in eq.~\eqref{eq:gamma_product_simplification}, we can reduce the factor $\g_{01234567}$ in eq.~\eqref{eq:kappa_projector} to a sum of terms each with a product of just two $\G_A$, as advertised:
\beq
\begin{aligned}
\label{eq:kappa_condition_total}
    \sqrt{\Sigma} \, \ve&= -i\sigma_2\otimes\biggl\{
       \G_{89} +\mathcal{Z}\,\G_{23}
    - \frac{1}{\sqrt{2}} \G_4\le[ y' (\G_{9} + i \G_{8})  + \yb' (\G_{9} -i \G_{8})\ri]
    \\ &\phantom{=}
    -\frac{i}{2 f(r)}(\p y - \pb \yb) ( \G_{28} + \G_{39}) 
    +\frac{i}{2 f(r)} (\p \yb - \pb y) (\G_{28} - \G_{39} )
    \\ &\phantom{=}
    -\frac{1}{2 f(r)} (\p y + \pb \yb) (\G_{29} - \G_{38} )
    -\frac{1}{2 f(r)} (\p \yb + \pb y) ( \G_{29} + \G_{38})
    \\ &\phantom{=}
    + \frac{y' (\p - \pb) \yb - \yb' (\p - \pb)y }{\sqrt{2} \, f(r)}   \G_{34}
    - i\frac{y' (\p + \pb) \yb - \yb' (\p + \pb)y}{\sqrt{2} \, f(r)}  \G_{24}
    \biggr\} \ve,
\end{aligned}
\eeq
where for brevity on the left-hand side we defined
\begin{subequations}
\label{eq:sigmadef}
\begin{align}
    \Sigma & \equiv  \frac{|\det g_{ab}| }{\r^6 \sin^4\a_1 \sin^2 \a_2}\\
    &= (1 + \mathcal{Z})^2 + \frac{4}{f(r)^2} |\pb y|^2 + 2 |y'|^2 + \frac{2}{f(r)^2} |\yb' \p y - y' \p \yb |^2
    \\
    &= (1 - \mathcal{Z})^2 + \frac{4}{f(r)^2} |\p y|^2 + 2 |y'|^2 + \frac{2}{f(r)^2} |\yb' \p y - y' \p \yb |^2.
\end{align}
\end{subequations}

Our goal now is to find how many of $\ve$'s $16$ components survive the D7-branes' $\kappa$ symmetry projection condition in eq.~\eqref{eq:kappa_condition_total}. We will start by proving that if $y$ has any $\rho$ dependence then the projection condition sets all of $\ve$'s components to zero, and hence the D7-branes break all SUSY. In other words, we will prove that the D7-branes can preserve SUSY only if \(y\) is independent of \(\r\), that is if $y' = 0$, as mentioned below eq.~\eqref{eq:derivnotation}.

As a warm-up, first suppose that \(y\) depends on \(\r\) but is independent of \(z\) and \(\zb\), that is suppose $y(\rho)$. In that case eq.~\eqref{eq:kappa_condition_total} reduces to
\beq
\label{eq:kappa_condition_rho_only}
 -i \s_2 \otimes \le\{ \G_{89} - \frac{1}{\sqrt{2}} \G_4 \le[y' (\G_{9} + i \G_{8}) + \yb' (\G_{9} - i \G_{8}) \ri] \ri\} \ve = \sqrt{1 + 2 |y'|^2} \, \ve.
\eeq
Suppose $y'\neq 0$ depends on $\rho$, which is generically the case. Each term in eq.~\eqref{eq:kappa_condition_rho_only} then has different \(\r\) dependence, hence eq.~\eqref{eq:kappa_condition_rho_only} can be satisfied only if each term vanishes independently. In particular the right-hand side must vanish, which occurs only if \(\ve=0\), meaning all SUSY is broken. The only exception is the special case of a linear function, \(y(\r) = y_0 + y_1 \r\) with complex constants \(y_0\) and \(y_1\neq0\), so that \(y'=y_1\neq0\) is independent of $\rho$. In that special case all terms in eq.~\eqref{eq:kappa_condition_rho_only} are independent of \(\r\), and as a result $8$ of $\ve$'s components survive the projection condition in eq.~\eqref{eq:kappa_condition_rho_only}, i.e. half of the D3-brane background's SUSY. However, we can exclude a linear function for at least two reasons. First, a linear function diverges as $\rho \to \infty$, which is an inadmissable non-normalisable boundary condition. For example, for linear $y(\rho)$ as $\rho \to \infty$ the D7-brane's Lagrangian is $\propto L^8 T_{D7} N_f\, \rho^3 \sqrt{1+y'^2} \to L^8 T_{D7} N_f \,\rho^3 \sqrt{1+y_1^2}$, in effect re-scaling the dimensionless parameter in front of the D7-branes' action, $L^8 T_{D7} N_f$. We can interpret that as rescaling either the D7-branes tension, $T_{D7}$, or the number of D7-branes, $N_f$. Either change of boundary condition changes the definition of our system, and hence must be excluded as an inadmissable non-normalisable boundary condition.\footnote{In $AdS_5 \times S^5$ the analogous statements are as follows. With the $AdS_5$ boundary at $\rho \to \infty$, $y$'s normalisable mode is $\propto \rho^{-2}$, whose coefficient encodes $\langle \mathcal{O} \rangle$, while $y$'s non-normalisable mode is $\propto y^0$, whose coefficient encodes $m$, as we review in the appendix. A linear mode $\propto y_1 \, \rho$ diverges as $\rho \to \infty$, and in effect rescales the dimensionless parameter in front of the D7-branes' action, $L^8 T_{D7} N_f$. Roughly speaking, that dimensionless parameter is dual to the number of flavour fields in the SYM theory: using $\lambda = L^4 \, \alpha'^{-2}=4\pi g_sN_c$ from below eq.~\eqref{eq:ads5s5metric} and $T_{D7}\propto \alpha'^{-4} g_s^{-1}$ from below eq.~\eqref{eq:D7-action}, we find $L^8 T_{D7} N_f\propto \lambda N_c N_f$. Changing the $\rho^{-2}$ or $\rho^0$ mode thus changes the state or the Hamiltonian of the same SYM theory, respectively, while changing the $\rho$ mode changes the SYM theory itself, by changing the number of fields. In other words, changing the $\rho$ mode changes the definition of our system. To remain within the same SYM theory we must therefore exclude linear asymptotic behaviour $y \propto \rho$.} Second, and more importantly, we have checked explicitly that \(y(\r) = y_0 + y_1 \r\) with \(y_1\neq0\) does not solve the D7-branes' equation of motion. We thus find that for an ansatz $y(\rho)$ only the case $y'=0$ solves the D7-branes' equation of motion and preserves SUSY---and indeed is simply the constant solution $y(\rho)=y_0$ that preserves $d=4$ $\N=2$ SUSY, as mentioned above.

Returning again to our general ansatz $y(z,\zb,\rho)$, we can use the lessons learned from the special ansatz $y(\rho)$ to prove that any $\rho$ dependence must break all SUSY, as follows. For generic \(y(z,\zb,\rho)\) each term in eq.~\eqref{eq:kappa_condition_total} has different dependence on \((z,\zb,\rho)\), hence eq.~\eqref{eq:kappa_condition_total} can be satisfied only if each term vanishes independently. In particular the right-hand side of eq.~\eqref{eq:kappa_condition_total} must vanish, which occurs only if \(\ve=0\), meaning all SUSY is broken. However, now two exceptions exist.

The first exception is for $y(z,\zb,\rho)$ that makes every term in eq.~\eqref{eq:kappa_condition_total} constant. For \(y'\neq0\) that means \(y' \, \s_2 \otimes \G_4(\G_{9} + i \G_{8}) \ve\) and \(\yb' \, \s_2 \otimes \G_4(\G_{9} - i \G_{8}) \ve\) are constants, which can potentially happen in two ways---both of which we can eliminate, as follows. The first way is when $y'$ is constant, meaning $y$ is linear in $\rho$, and in particular $y(z,\zb,\rho) = y_0(z,\zb) + y_1 \rho$ with constant $y_1$. We can exclude $y$ linear in $\rho$ for the same reasons as above, i.e. for having an inadmissable non-normalisable boundary condition and for not solving the equation of motion. The second way that \(y' \, \s_2 \otimes \G_4(\G_{9} + i \G_{8}) \ve\) and \(\yb' \, \s_2 \otimes \G_4(\G_{9} - i \G_{8}) \ve\) could be constant is if $\varepsilon$ could solve the two equations $\id_2 \otimes (\G_{9} \pm i \G_{8}) \ve = 0$ simultaneously. However, that is impossible: $\ve$ can solve either one of these equations (i.e. with a single sign) because the Clifford algebra implies \(\G_9 \pm i \G_8\) are nilpotent and therefore all their eigenvalues vanish, but summing the two equations gives \(2 (\id_2 \otimes \G_9) \ve = 0\), which $\ve$ cannot solve because \(\id_2 \otimes \G_9\) has eigenvalues \(\pm 1\), not zero. We have thus eliminated the possibility that $y(z,\zb,\rho)$ would allow for non-trivial solutions of eq.~\eqref{eq:kappa_condition_total} by making every term in eq.~\eqref{eq:kappa_condition_total} constant.

The second exception that could allow $\ve$ to solve eq.~\eqref{eq:kappa_condition_total} is if \(\sqrt{\Sigma}\) is a polynomial of derivatives of \(y\) and \(\yb\) (and could also depend on \(f(r)\)). In that case we have the chance to cancel terms between the left- and right-hand sides of eq.~\eqref{eq:kappa_condition_total}. That would require \(\ve\) to be a simultaneous eigenspinor of the matrices of any non-zero terms on the right-hand side of eq.~\eqref{eq:kappa_condition_total}, and in particular if \(y'\neq0\) then \(\ve\) would have to be a simultaneous eigenspinor of \(\s_2 \otimes \G_4(\G_{9} \pm i \G_{8}) \ve\). However, that is impossible, as we argued in the previous paragraph. We have thus eliminated the possibility that $y(z,\zb,\rho)$ could allow non-trivial solutions of eq.~\eqref{eq:kappa_condition_total} by allowing terms on the left- and right-hand sides to cancel.

In summary, for our general ansatz $y(z,\zb,\rho)$ we have shown that any $\rho$ dependence cannot preserve SUSY and solve the D7-branes' equation of motion. From now on we will therefore exclusively consider the ansatz that $y$ depends only on $(z,\zb)$, that is, $y(z,\zb)$.

Plugging the ansatz $y(z,\zb)$ into eq.~\eqref{eq:sigmadef} gives \(\sqrt{\Sigma} = \mathcal{L}\), the Lagrangian density in eq.~\eqref{eq:energy_density_2}. In the D7-branes' $\kappa$ symmetry projection condition eq.~\eqref{eq:kappa_condition_total}, plugging in $\sqrt{\Sigma}=\cL$ and $y'=0$ gives
\beq
\begin{aligned}
\label{eq:kappa_condition_final_1}
   \cL \, \ve &= -i \s_2 \otimes \le(\G_{89} + \mathcal{Z} \, \G_{23} \ri) \ve
    \\
        &\phantom{=} +\frac{1}{2 f(r)}\biggl[-\phantom{i}(\p y - \pb \yb) \s_2 \otimes (\G_{28} + \G_{39}) +\phantom{i}(\p \yb - \pb y ) \s_2 \otimes (\G_{28}-\G_{39})
    \\
        &\phantom{=\hspace{1.53cm}}+i (\p y + \pb \yb) \s_2 \otimes (\G_{29}-\G_{38}) +i ( \p \yb + \pb y ) \s_2 \otimes (\G_{29}+ \G_{38})
    \biggr] \ve.
\end{aligned}
\eeq
Our goal now is to find how many of $\ve$'s $16$ components survive the D7-branes' $\kappa$ symmetry projection condition, in the form in eq.~\eqref{eq:kappa_condition_final_1}. To begin, recall from eq.~\eqref{eq:rdef} that \(r^2 \equiv \r^2 + 2 |y|^2\), and observe that every term in the top row of eq.~\eqref{eq:kappa_condition_final_1} depends on $r$ and hence on $\rho$, while the second and third rows are independent of \(\r\). As a result, eq.~\eqref{eq:kappa_condition_final_1} can be satisfied for all \(\r\) and for non-zero \(\ve\) only if the top row cancels, that is if
\beq
\label{eq:kappa_condition_final_first_row}
\cL \, \ve =   -i \s_2 \otimes \le(\G_{89} + \mathcal{Z} \, \G_{23} \ri) \ve,
\eeq
which in turn requires that \(\ve\) is an eigenspinor of both matrices on the right-hand side,
\begin{subequations}
\label{eq:irreducible_kappa_condition_1}
\begin{eqnarray}
\label{eq:irreducible_kappa_condition_1a}
    -i \s_2\otimes \G_{89} \, \ve & = & \pm\ve, \\
    \label{eq:irreducible_kappa_condition_1b}
    -i \s_2 \otimes \G_{23}\, \ve & = & \pm \ve,
\end{eqnarray}
\end{subequations}
with the $\pm$ signs uncorrelated. Plugging eq.~\eqref{eq:irreducible_kappa_condition_1} into eq.~\eqref{eq:kappa_condition_final_first_row} gives
\beq
\label{eq:bps_bound_from_kappa_symmetry_1}
    \cL \ve = \left(\pm 1 \pm \mathcal{Z}\right)\ve,
\eeq
again with the $\pm$ signs uncorrelated. However, the BPS bound in eq.~\eqref{eq:energy_density_bound_1}, $\cL \geq 1 \pm \cZ$, means we can allow only the $+$ sign in eq.~\eqref{eq:irreducible_kappa_condition_1a} and hence in the $\pm 1$ in the first term on the right-hand side of eq.~\eqref{eq:bps_bound_from_kappa_symmetry_1}. With that choice, and dropping $\ve$, we find
\beq
\label{eq:bps_bound_from_kappa_symmetry_2}
    \cL =  1 \pm \mathcal{Z},
\eeq
which is precisely the saturation of the BPS bound. We have thus found that $\ve$ can have non-zero components, and hence the D7-branes can preserve SUSY, only if $y$ \textit{saturates} the BPS bound, as expected on general grounds for SUSY.

From now on, we will therefore exclusively consider $\ve$ obeying the projection conditions in eq.~\eqref{eq:irreducible_kappa_condition_1}, so that $y(z,\zb)$ saturates the BPS bound, i.e. satisfies eq.~\eqref{eq:bps_bound_from_kappa_symmetry_2}. In that case the top row of eq.~\eqref{eq:kappa_condition_final_1} cancels, leaving
\beq
\begin{aligned}
\label{eq:kappa_condition_final_2}
  0 &=  \biggl[-\phantom{i}(\p y - \pb \yb) \s_2 \otimes (\G_{28} + \G_{39}) +\phantom{i}(\p \yb - \pb y ) \s_2 \otimes (\G_{28}-\G_{39})
    \\
        &\hspace{0.66cm}+i (\p y + \pb \yb) \s_2 \otimes (\G_{29}-\G_{38}) +i( \p \yb + \pb y ) \s_2 \otimes (\G_{29}+ \G_{38})
    \biggr] \ve.
\end{aligned}
\eeq
Components of $\ve$ can survive the D7-branes' $\kappa$ symmetry projection condition in eq.~\eqref{eq:kappa_condition_final_2} for all $(z,\zb)$ only if all four terms in the square brackets vanish simultaneously:
\begin{subequations}
\label{eq:kappa_condition_2}
\begin{align}
    (\p y - \pb \yb) \s_2 \otimes (\G_{28} + \G_{39}) \ve &= 0,
     \label{eq:kappa_condition_2_a}
    \\
    (\p y + \pb \yb) \s_2 \otimes (\G_{29}-\G_{38})  \ve &= 0,
     \label{eq:kappa_condition_2_b}
   \\
    (\p \yb - \pb y )   \s_2 \otimes (\G_{28}-\G_{39}) \ve & = 0,
    \label{eq:kappa_condition_2_c}
    \\
    ( \p \yb + \pb y ) \s_2 \otimes (\G_{29}+ \G_{38}) \ve &=0.
    \label{eq:kappa_condition_2_d}
\end{align}
\end{subequations}
As we explained in sec.~\ref{sec:cartesian}, if we choose the plus or minus sign in eq.~\eqref{eq:bps_bound_from_kappa_symmetry_2} then \(y\) must be a holomorphic or antiholomorphic function of \(z\), respectively. For holomorphic \(y\) the prefactors in eqs.~\eqref{eq:kappa_condition_2_c} and~\eqref{eq:kappa_condition_2_d} vanish identically, $\p \yb \pm \pb y =0$, so only the projectors in eqs.~\eqref{eq:kappa_condition_2_a} and~\eqref{eq:kappa_condition_2_b} remain non-trivial. Conversely, for antiholomorphic $y$ the prefactors in eqs.~\eqref{eq:kappa_condition_2_a} and~\eqref{eq:kappa_condition_2_b} vanish identically, $\p y \pm \pb \yb =0$, so only the projectors in eqs.~\eqref{eq:kappa_condition_2_c} and~\eqref{eq:kappa_condition_2_d} remain non-trivial. We can summarise these non-trivial projection conditions for both cases succinctly as
\begin{subequations}
\label{eq:irreducible_kappa_condition_2}
\begin{align}
    \s_2 \otimes (\G_{28} \pm \G_{39}) \ve &= 0,
    \label{eq:irreducible_kappa_condition_2_a}
    \\\
    \s_2 \otimes (\G_{29} \mp \G_{38}) \ve &= 0,
    \label{eq:irreducible_kappa_condition_2_b}
\end{align}
\end{subequations}
with the upper or lower signs for holomorphic or antiholomorphic $y$, respectively. Multiplying eqs.~\eqref{eq:irreducible_kappa_condition_2_a} and~\eqref{eq:irreducible_kappa_condition_2_b} from the left by \(\id_2 \otimes \G_{29}\) and \(\id_2 \otimes \G_{28}\), respectively, we find that both are equivalent to a single projection condition, namely
\beq
\label{eq:kappa_condition_final_3}
    (\s_2 \otimes \G_{23})\ve = \pm (\s_2 \otimes \G_{89}) \ve,
\eeq
again with the upper or lower sign for holomorphic or antiholomorphic $y$, respectively. Ultimately, then, our goal is to find how many of $\ve$'s $16$ components survive the D7-branes' $\kappa$ symmetry projection condition, in the form in eq.~\eqref{eq:kappa_condition_final_3}

Crucially, however, plugging any solution of the projection conditions of eq.~\eqref{eq:irreducible_kappa_condition_1}, which follow from saturation of the BPS bound, into the simplified D7-brane $\kappa$ symmetry projection condition eq.~\eqref{eq:kappa_condition_final_3}, and recalling that in eq.~\eqref{eq:irreducible_kappa_condition_1a} the BPS bound requires us to choose the upper sign, we find $\pm \ve = \pm \ve$. In other words, once the BPS bound is saturated eq.~\eqref{eq:kappa_condition_final_3} places no additional constraint on \(\ve\). We therefore only need to find how many of $\ve$'s $16$ components survive the projection conditions in eq.~\eqref{eq:irreducible_kappa_condition_1}, again recalling that the BPS bound forces us to choose the upper sign in eq.~\eqref{eq:irreducible_kappa_condition_1a}. The two matrices on the left-hand side of eq.~\eqref{eq:irreducible_kappa_condition_1} are linearly independent. As a result, only one-quarter of $\ve$'s $16$ components survive the D7-branes' projection condition in eq.~\eqref{eq:irreducible_kappa_condition_1}, that is, $4$ of $\ve$'s components survive, corresponding to $4$ real supercharges.

In summary, then, we have shown that to preserve SUSY $y$ must saturate the BPS bound $\cL \geq 1 \pm \cZ$, and hence must be either holomorphic or antiholomorphic in $z$, in which case $y$ preserves  $4$ real supercharges i.e. one-quarter of the D3-brane background's SUSY.

%%%%%%%%%%%%%%%%%%%%%%%%%%%%%%%%%%%%%%%%%%%%%%%%%%
%%%%%%%%%%%%%%%%%%%%%%%%%%%%%%%%%%%%%%%%%%%%%%%%%%
\subsection{All preserved supercharges have the same $d=2$ chirality}
\label{sec:chirality_of_susy}
%%%%%%%%%%%%%%%%%%%%%%%%%%%%%%%%%%%%%%%%%%%%%%%%%%
%%%%%%%%%%%%%%%%%%%%%%%%%%%%%%%%%%%%%%%%%%%%%%%%%%

A generic holomorophic or antiholomorphic function $y$ will break translational symmetry in the $z$-plane, and since supercharges anti-commute into generators of translations we expect such $y$ to break all SUSY in the $z$-plane. We thus expect any preserved supercharges to be those that anti-commute to the generators of translations in the $(x_0,x_1)$ plane. In this section we answer a key question about the $4$ real supercharges preserved by holomorphic or antiholomorphic $y$: is their $d=2$ chirality $\N=(4,0)$, $(2,2)$, or $(0,4)$?

The chirality operator in the directions $(x_0,x_1)$ is $\id_2 \otimes \G_{01}$. We will define left-handed spinors by $\left(\id_2 \otimes \G_{01}\right) \ve = +\ve$ and right-handed spinors by $\left(\id_2 \otimes \G_{01}\right) \ve = - \ve$. To determine the chirality of our preserved supercharges we will use two projection conditions. The first is the D3-branes' $\kappa$ symmetry projection condition from eq.~\eqref{eq:ads5_susy_condition}, $i \s_2 \otimes \G_{0123} \ve = \ve$. The second is the D7-branes' $\kappa$ symmetry projection condition in eq.~\eqref{eq:irreducible_kappa_condition_1}, and specifically the projection condition in eq.~\eqref{eq:irreducible_kappa_condition_1b}, whose sign is sensitive to whether $y$ is holomorphic or antiholomorphic: $\le(i\s_2 \otimes \G_{23} \ri)\varepsilon=\pm\varepsilon$, with the upper or lower sign for holomorphic or antiholomorphic $y$, respectively. Using these projection conditions we find
\begin{align}
    (\id_2 \otimes \G_{01}) \ve &=  \pm(\id_2 \otimes \G_{01}) (i \s_2 \otimes \G_{23}) \ve
    \nonumber \\
    &= \pm(i \s_2 \otimes \G_{0123}) \ve
    \label{eq:2d_chirality_holomorphic}
    \\
    &= \pm \ve,
    \nonumber
\end{align}
hence holomorphic $y$ preserve $4$ left-handed supercharges, or $d=2$ $\N=(4,0)$ SUSY, while antiholomorphic $y$ preserve $4$ right-handed supercharges, or $d=2$ $\N=(0,4)$ SUSY.

As a simple check, for a constant solution $y = y_0\neq0$ our arguments apply unchanged up to eq.~\eqref{eq:kappa_condition_final_1}. If $y$ is constant then $\cZ=0$. Plugging that into eqs.~\eqref{eq:kappa_condition_final_1} and~\eqref{eq:kappa_condition_final_first_row} leads only to the projection condition in eq.~\eqref{eq:irreducible_kappa_condition_1a}, where again the BPS bound forces us to choose the upper sign, $-i \s_2\otimes \G_{89} \, \ve = +\ve$, and we do not impose the projection condition in eq.~\eqref{eq:irreducible_kappa_condition_1b}. As a result, both chiralities in eq.~\eqref{eq:2d_chirality_holomorphic} survive, so the constant $y$ solution preserves $d=2$ $\N=(4,4)$ SUSY, which is equivalent to $d=4$ $\N=2$ SUSY, as expected.

As mentioned below eq.~\eqref{eq:actionenergybounds3}, back in the original 4ND D3/D7 intersection of tab.~\ref{tab:embedding}, at points where the D7-branes touch the D3-branes, meaning $y \to 0$ and $\rho \to 0$ so that $r^2 \equiv \rho^2 + 2 |y|^2 \to 0$, we obtain 8ND D8-branes. These 8ND D8-branes preserve either $d=2$ $\N=(8,0)$ or $(0,8)$ SUSY, depending on their orientiation. How does the SUSY enhancement from $d=2$ $\N=(4,0)$ or $(0,4)$ occur? If $|\p y|^2 -  |\pb y|^2 \neq 0$ as $r \to 0$ then $f(r) \propto r^2 \to 0$ and \(\mathcal{Z}=f(r)^{-2}(|\p y|^2 -  |\pb y|^2) \to \infty\), in which case $\cL = 1 \pm \cZ \to \pm \infty$. In the first line of eq.~\eqref{eq:kappa_condition_final_1} we can then drop the term \(\propto \s_2 \otimes \G_{89}\) relative to the term \(\propto \cZ \left(\s_2 \otimes \G_{23}\right)\), so that we no longer need to impose the projection condition involving \(\s_2 \otimes \G_{89}\) in eq.~\eqref{eq:irreducible_kappa_condition_1a}. Dropping that projection condition doubles the number of preserved supercharges but does not change the chirality analysis in eq.~\eqref{eq:2d_chirality_holomorphic}, thus leading to the doubling of SUSY from $d=2$ $\N=(4,0)$ or $(0,4)$ to $\N=(8,0)$ or $(0,8)$, respectively. We emphasise that such SUSY enhancement emerges \textit{only} when $r \to 0$.

%%%%%%%%%%%%%%%%%%%%%%%%%%%%%%%%%%%%%%%%%%%%%%%%%%
%%%%%%%%%%%%%%%%%%%%%%%%%%%%%%%%%%%%%%%%%%%%%%%%%%
\section{Near-horizon limit and holographic duality}
\label{sec:nearhorizon}
%%%%%%%%%%%%%%%%%%%%%%%%%%%%%%%%%%%%%%%%%%%%%%%%%%
%%%%%%%%%%%%%%%%%%%%%%%%%%%%%%%%%%%%%%%%%%%%%%%%%%

For the general $f(r)$ in eq.~\eqref{eq:metric}, in sec.~\ref{sec:holosols} we showed that holomorphic $y(z)$ solve the equation of motion in eq.~\eqref{eq:yeom}, in sec.~\ref{sec:bpsbound} we showed that such $y$ saturate the BPS bounds in eq.~\eqref{eq:actionenergybounds2}, and in sec.~\ref{sec:kappa_symmetry} we showed that such $y$ preserve $\N=(4,0)$ SUSY in the $(x_0,x_1)$ directions. Analogous statements apply for antiholomorphic $y(\zb)$. Because the arguments of those sections were valid for the general $f(r)$ in eq.~\eqref{eq:metric}, they are valid in the near-horizon limit where $f(r) \approx r^2/L^2$ and the metric becomes that of $AdS_5 \times S^5$: for 4ND D7-branes in $AdS_5 \times S^5$, holomorphic $y(z)$ solve the equation of motion in eq.~\eqref{eq:yeom}, saturate the BPS bounds in eq.~\eqref{eq:actionenergybounds2}, and preserve $d=2$ $\N=(4,0)$ SUSY in $(x_0,x_1)$, and analogous statements apply for antiholomorphic $y(\zb)$.

However, a key feature of the near-horizon limit is that we can invoke the AdS/CFT correspondence, wherein type IIB SUGRA in $AdS_5 \times S^5$ is holographically dual to $d=4$ $SU(N_c)$ $\N=4$ SYM in the limits $N_c \to \infty$ and $\lambda \to \infty$, and the $N_f \ll N_c$ probe 4ND D7-branes are dual to probe $d=4$ $\N=2$ hypermultiplets in the fundamental representation of $SU(N_c)$, i.e. flavour fields. In the near-horizon limit, using the AdS/CFT dictionary we can translate our results to the dual SYM theory.

In this section we restrict to $f(r)=r^2/L^2$ such that the metric is that of $AdS_5 \times S^5$, and we study the holographic duals of holomorphic or antiholomorphic $y$. In sec.~\ref{sec:fieldtheory} we start on the SYM side of the correspondence, and prove that a holomorphic or antiholomorphic hypermultiplet mass $m$ preserves $d=2$ $\N=(4,0)$ or $(0,4)$ SUSY, respectively. Furthermore, we show that holomorphic or antiholomorphic $m$ describe holonomies of a non-dynamical, flat, background $U(1)_R$ gauge field around $m$'s zeroes and poles. In sec.~\ref{sec:holodual} we move to the AdS side of the correspondence, and use holographic renormalisation for the probe D7-branes~\cite{deHaro:2000vlm,Bianchi:2001kw,Skenderis:2002wp,Karch:2005ms,Karch:2006bv,Hoyos:2011us} to show that for holomorphic or antiholomorphic $y$ the probe D7-branes' contribution to the renormalised energy vanishes and the VEV of the operator dual to $y(z)$, which we denote $\cO$, vanishes, i.e. $\langle \mathcal{O}\rangle=0$. We also illustrate how the probe 4ND D7-branes' worldvolume geometry encodes holomorphic or antiholomorphic $m$, with emphasis on the D7-branes' behaviour near a zero of $m$ where we obtain the 8ND D7-branes and the SUSY enhances to $d=2$ $\N=(8,0)$ or $(0,8)$.

%%%%%%%%%%%%%%%%%%%%%%%%%%%%%%%%%%%%%%%%%%%%%%%%%%
%%%%%%%%%%%%%%%%%%%%%%%%%%%%%%%%%%%%%%%%%%%%%%%%%%
\subsection{SYM side: (anti-)holomorphic mass and SUSY}
\label{sec:fieldtheory}
%%%%%%%%%%%%%%%%%%%%%%%%%%%%%%%%%%%%%%%%%%%%%%%%%%
%%%%%%%%%%%%%%%%%%%%%%%%%%%%%%%%%%%%%%%%%%%%%%%%%%

In this subsection we will work exclusively in the SYM theory, and prove that a holomorphic or antiholomorphic hypermultiplet mass $m$ preserves $d=2$ $\N=(4,0)$ or $(0,4)$ SUSY along $(x_0,x_1)$, respectively. We will present two different field theory proofs. Our first proof, in sec.~\ref{sec:bgvec}, uses methods similar to those of refs.~\cite{Festuccia:2011ws,Dumitrescu:2012ha,Closset:2013vra,Closset:2014uda}, who showed that to preserve SUSY we can introduce classical background values of additional gauge or SUGRA multiplets. We will introduce a background $d=4$ $\N=2$ vector multiplet such that if the multiplet's scalar acquires a VEV then the hypermultiplets acquire a mass. We will show that a holomorphic or antiholomorphic VEV preserves $d=2$ $\N=(4,0)$ or $(0,4)$ SUSY along $(x_0,x_1)$, respectively. Our second proof, in sec.~\ref{sec:massivehypers} uses no background fields, and instead involves showing explicitly that the SYM theory's classical action is invariant under $d=2$ $(4,0)$ or $(0,4)$ SUSY along $(x_0,x_1)$ in the presence of holomorphic or antiholomorphic $m$. In sec.~\ref{sec:u1rholo} we also show that holomorphic or antiholomorphic $m$ describe holonomies of a non-dynamical, flat, background $U(1)_R$ gauge field around zeroes and poles of $m$.

In this subsection \(\g^M\) with $M = 0,1,2,3$ denote Dirac matrices of $d=4$ Minkowski spacetime. Using a mostly-plus $d=4$ Minkowski inverse metric $\eta^{MN}$, our convention for the Clifford algebra is
\begin{equation}
    \{\g^M , \g^N\} = 2 \,\h^{MN}.
\end{equation}
We take \(\g^0\) to be anti-Hermitian and the $\g^M$ with $M=1,2,3$ to be Hermitian. The $d=4$ chirality matrix, \(\g_5 \equiv - i \g^{0123}\), is then Hermitian, where multiple indices denote a normalised antisymmetric product, as in sec.~\ref{sec:kappa_symmetry}. The $d=4$ chirality projectors \(P_\pm\) are then
\begin{equation}
    P_\pm \equiv \frac{1}{2} \le(\id \pm \g_5\ri).
\end{equation}
A $d=4$ Majorana spinor \(\z\) obeys the Majorana condition \(\z^* = B \z\), where the matrix \(B\) satisfies \((\g^M)^* = B \g^M B^{-1}\), and consequently \((\g_5)^* = - B \g_5 B^{-1}\). Since we use the complex coordinates $(z,\zb)$ defined in eq.~\eqref{eq:zdef}, we also define
\begin{subequations}
\beq
    \g^z \equiv \frac{1}{\sqrt{2}} \le(\g^2 + i \g^3\ri),
    \eeq
    \beq
    \g^{\zb} \equiv \frac{1}{\sqrt{2}} \le(\g^2 - i \g^3\ri).
\end{equation}
\end{subequations}

Our SYM theory has $d=4$ $\N=2$ SUSY, with R-symmetry $U(2)_R \simeq SU(2)_R \times U(1)_R$. The SUSY parameters are an $SU(2)_R$ doublet, which we denote $\xi_I$ with $I=1,2$. These satisfy the condition
\begin{equation}
\label{eq:sort_of_majorana_condition}
    \xi_I^* = i \e_{IJ} B \g_5 \xi_J ,
\end{equation}
with \(\e_{IJ}\) the $d=2$ Levi-Civita symbol with $\e_{12}=+1$. One way to derive $d=4$ $\N=2$ SUSY is via dimensional reduction from $d=6$ \(\N=1\) SUSY, where eq.~\eqref{eq:sort_of_majorana_condition} follows from the symplectic Majorana condition on the $d=6$ SUSY parameter. Each $\xi_I$ has $4$ complex components, for a total of $8$ real components, but eq.~\eqref{eq:sort_of_majorana_condition} reduces that to $4$ real components. Counting $I=1,2$ then gives the $8$ real supercharges of $d=4$ $\N=2$ SUSY.

%%%%%%%%%%%%%%%%%%%%%%%%%%%%%%%%%%%%%%%%%%%%%%%%%%
%%%%%%%%%%%%%%%%%%%%%%%%%%%%%%%%%%%%%%%%%%%%%%%%%%
\subsubsection{Background $\N=2$ vector multiplet}
\label{sec:bgvec}
%%%%%%%%%%%%%%%%%%%%%%%%%%%%%%%%%%%%%%%%%%%%%%%%%%
%%%%%%%%%%%%%%%%%%%%%%%%%%%%%%%%%%%%%%%%%%%%%%%%%%

In this proof we start with \textit{massless} hypermultiplets, and introduce a background $\N=2$ vector multiplet in the Cartan of the $U(N_f)$ flavour symmetry. The field content of the $\N=2$ vector multiplet is a field strength, which we set to zero, an $SU(2)_R$ doublet of gauginos, $\Lambda_I$ with $I=1,2$, and a complex scalar, $\Phi$. We will give $\Phi$ a non-zero VEV, $\langle \Phi \rangle \neq 0$, which in turn will give the hypermultiplets a non-zero mass, $m \propto \langle \Phi \rangle$. We let $\Phi$ depend on $(z,\zb)$ so the hypermultiplet mass $m(z,\zb) \propto \langle \Phi(z,\zb) \rangle$ depends on $(z,\zb)$ as well.

Under a $d=4$ $\N=2$ SUSY transformation, the $\N=4$ SYM vector multiplet, the hypermultiplets, and the background $\N=2$ vector multiplet all have non-trivial variations. However, when $\Phi$ depends on position the action is invariant under the SUSY variations of the $\N=4$ SYM vector multiplet and the hypermultiplets, but not necessarily under the variations of the background $\N=2$ vector multiplet. In particular, to preserve SUSY we must find the conditions under which $\Lambda_I$ has vanishing SUSY variations~\cite{Seiberg:1993vc}. When the background $\N=2$ vector multiplet has zero field strength, when the background spacetime is Minkowski, and when $\Phi = |\Phi| e^{i \psi}$ depends on position, $\Lambda_I$'s SUSY variations are
\begin{subequations}
\label{eq:gauginovariations1}
\begin{align}
\delta \Lambda_I & = i \sqrt{2} \gamma^M \p_M \left(|\Phi | e^{-i\psi \gamma_5} \right)\xi_I\\
&= i \sqrt{2}\,\g^M \,\p_M\Phi \, P_-\xi_I + i \sqrt{2}\, \g^M\partial_M\bar{\Phi}P_+\xi_I.\label{eq:gauginovariations1b}
\end{align}
\end{subequations}
One way to derive eq.~\eqref{eq:gauginovariations1} is via dimensional reduction of the SUSY variations of the gauginos in a $d=6$ $\N=1$ vector multiplet. Plugging $\Phi(z,\zb)$ into eq.~\eqref{eq:gauginovariations1} and demanding that $\delta \Lambda_I=0$ then gives
\begin{subequations}
\label{eq:gauginovariations2}
\beq 
(\gamma^z\, \p \Phi + \gamma^{\zb} \, \pb \Phi ) \,P_-\xi_I= 0,
\eeq
\beq
(\gamma^{\zb} \, \pb \bar{\Phi} + \gamma^{z} \, \p \bar{\Phi} ) \,P_+\xi_I= 0.
\eeq
\end{subequations}
For generic $\Phi(z,\zb)$ the only solution of eq.~\eqref{eq:gauginovariations2} is to demand $P_-\xi_I = 0$ and $P_+ \xi_I=0$, meaning all SUSY is broken.

However, suppose we demand that $\Phi(z,\zb)$ is holomorphic in $z$, so that $\pb \Phi=0$ and $\p \bar{\Phi}=0$. In that case eq.~\eqref{eq:gauginovariations2} reduces to
\begin{subequations}
\label{eq:gauginovariations3}
\begin{align}
\label{eq:gauginovariations3a}
\g^z P_-\xi_I &= 0,\\
\g^{\zb} P_+ \xi_I&=0.
\label{eq:gauginovariations3b}
\end{align}
\end{subequations}
Actually, if eq.~\eqref{eq:gauginovariations3a} is satisfied then eq.~\eqref{eq:gauginovariations3b} is satisfied too, because eq.~\eqref{eq:sort_of_majorana_condition} gives
\beq
\label{eq:sort_of_majorana_condition_2}
   \g^{\zb} P_+ \xi_I = -i\, \e_{IJ} \le( \g^z P_- \xi_J \ri)^*.
\eeq
For holomorphic $\Phi(z)$ we thus find that $\delta \Lambda_I = 0$ reduces to the single projection condition in eq.~\eqref{eq:gauginovariations3a}. Half of the components of $\xi_I$ survive that projection condition, meaning $4$ real supercharges survive. We thus find that holomorphic $\Phi(z)$ preserves $4$ real supercharges.

Analogously, suppose that $\Phi(z,\zb)$ is antiholomorphic in $z$, so that $\p \Phi=0$ and $\pb \bar{\Phi}=0$. In that case eq.~\eqref{eq:gauginovariations2} reduces to
\begin{subequations}
\label{eq:gauginovariations4}
\begin{align}
\label{eq:gauginovariations4a}
\g^{\zb} P_-\xi_I &= 0,\\
\g^z P_+ \xi_I&=0,
\label{eq:gauginovariations4b}
\end{align}
\end{subequations}
where again eq.~\eqref{eq:sort_of_majorana_condition} guarantees that if eq.~\eqref{eq:gauginovariations4a} is satisfied then eq.~\eqref{eq:gauginovariations4b} is satisfied too. As a result, $\delta \Lambda_I=0$ again reduces to a single projection condition, namely that in eq.~\eqref{eq:gauginovariations4a}. Half of the components of $\xi_I$ survive that projection condition, meaning $4$ real supercharges. We thus find that antiholomorphic $\Phi(\zb)$ preserves $4$ real supercharges.

A generic holomorphic or antiholomorphic $\Phi$ breaks translations in the $z$-plane, so we expect the $4$ preserved supercharges to be those that anti-commute to generators of translations in the $(x_0,x_1)$ directions. What is the $d=2$ chirality of the preserved supercharges: $\N=(4,0)$, $(2,2)$, or $(0,4)$? In sec.~\ref{sec:chirality_of_susy} we showed that a holomorphic or antiholomorphic D7-brane solution $y$ preserves $d=2$ $\N=(4,0)$ or $\N=(0,4)$ SUSY along $(x_0,x_1)$, respectively. We can straightforwardly prove the analogous statement in the SYM theory, as follows. Multiplying eqs.~\eqref{eq:gauginovariations3b} and~\eqref{eq:gauginovariations4b} from the left by \(\g^2\) and rearranging gives us
\begin{equation}
\label{eq:combined_mass_projection_condition}
    i \g^{23} P_+ \xi_I = \pm P_+ \xi_I,
\end{equation}
with the plus or minus sign for holomorphic or antiholomorphic $\Phi$, respectively. Note the similarity between eq.~\eqref{eq:combined_mass_projection_condition} and eq.~\eqref{eq:irreducible_kappa_condition_1b} for the SUSY preserved by a holomorphic or antiholomorphic D7-brane solution $y$. Multiplying eq.~\eqref{eq:combined_mass_projection_condition} from the left by \(\g_5\) produces
\beq
\label{eq:field_theory_2d_chirality}
    \g_{01} P_+\xi_I = \pm P_+\xi_I ,
\eeq
again with the plus or minus sign for holomorphic or antiholomorphic $\Phi$, respectively. Since \(\g_{01}\) is the $d=2$ chirality operator for the directions \((x_0,x_1)\), eq.~\eqref{eq:field_theory_2d_chirality} shows that a holomorphic $\Phi$ preserves left-handed supercharges, meaning $d=2$ $\N=(4,0)$ SUSY, while an antiholomorphic $\Phi$ preserves right-handed supercharges, meaning $d=2$ $\N=(0,4)$ SUSY, just as occurred in our SUGRA analysis of SUSY in sec.~\ref{sec:kappa_symmetry}.

We have thus proven that holomorphic or antiholomorphic $\Phi$ preserves $d=2$ $\N=(4,0)$ or $(0,4)$ SUSY in $(x_0,x_1)$, respectively. Since the hypermultiplet mass $m \propto \langle \Phi \rangle$, the same results apply for holomorphic or antiholomorphic $m$, namely these preserve $d=2$ $\N=(4,0)$ or $(0,4)$ SUSY in $(x_0,x_1)$, respectively.

This method of proof is extremely general, relying only on the existence of a background $d=4$ $\N=2$ vector multiplet and its coupling to the hypermultiplet, such that $m \propto \langle \Phi \rangle$. The proof is therefore valid for any values of $N_c$, $\lambda$, and $N_f$. The proof also clearly agrees with the SUGRA proof in sec.~\ref{sec:kappa_symmetry} where the two proofs overlap, namely in the near-horizon limit of the SUGRA proof where we can invoke holographic duality with SYM in the 't Hooft large $N_c$ limit, the strong coupling limit $\lambda \to \infty$, and the probe limit $N_f \ll N_c$.

Of course holography also provides results that are difficult to derive directly in the SYM theory, even with SUSY. A prime example is from sec.~\ref{sec:bpsbound}, where we argued that a zero of $m$ should exhibit SUSY enhancement from $d=2$ $\N=(4,0)$ or $(0,4)$ to $(8,0)$ or $(0,8)$, respectively. Proving that statement directly in the SYM theory is non-trivial, requiring a renormalisation group (RG) flow that integrates out everything except the massless degrees of freedom. Proving such SUSY enhancement may be within reach given the relatively high amount of SUSY present, but doing so is beyond the scope of this paper.

Being very general, the method of proof in this subsection can be straightforwardly adapted to practically any system in any $d$ that admits a background vector multiplet, although the interpretation will depend on how the vector multiplet scalar $\Phi$ couples to other fields. Ref.~\cite{Ratcliffe:2026exv} provides further discussion of more general cases. In the next subsection we present our second proof, which is less general, relying crucially on the form of the action of massive $d=4$ $\N=2$ hypermultiplets.

%%%%%%%%%%%%%%%%%%%%%%%%%%%%%%%%%%%%%%%%%%%%%%%%%%
%%%%%%%%%%%%%%%%%%%%%%%%%%%%%%%%%%%%%%%%%%%%%%%%%%
\subsubsection{SUSY variations of massive hypermultiplets}
\label{sec:massivehypers}
%%%%%%%%%%%%%%%%%%%%%%%%%%%%%%%%%%%%%%%%%%%%%%%%%%
%%%%%%%%%%%%%%%%%%%%%%%%%%%%%%%%%%%%%%%%%%%%%%%%%%

In our second proof we will start with a \textit{free} $d=4$ $\N=2$ hypermultiplet with \textit{constant} mass $m$, whose action is invariant under SUSY variations of $8$ real supercharges, and show that a holomorphic or antiholomorphic $m$ preserves $4$ real supercharges all of the same chirality in $(x_0,x_1)$. We will then argue that coupling one or more such hypermultiplets to $d=4$ \(\N=4\) SYM does not change our arguments, i.e. holomorphic or antiholomorphic $m$ still preserves $d=2$ $\N=(4,0)$ or $(0,4)$ SUSY along $(x_0,x_1)$, respectively.

The matter content of an on-shell \(\mathcal{N}=2\) hypermultiplet consists of a Dirac spinor $\chi$ and a pair of complex scalars \(q\) and \(\tilde{q}\). The theory has R-symmetry $U(2)_R \simeq U(1)_R \times SU(2)_R$. The $U(1)_R$ acts as an axial symmetry on the Dirac fermions $\chi$ and leaves invariant the two scalars \(q\) and \(\tilde{q}\). Under $SU(2)_R$ the Dirac fermion is invariant while the two scalars form a doublet,  \((q,\tilde{q}^\dag)\), which we denote \(q_I\) where $I=1,2$ with \(q_1 \equiv q\) and \(q_2 \equiv \tilde{q}^\dag\). The classical action for a free \(\mathcal{N}=2\) hypermultiplet with complex mass \(m = |m| e^{i\y}\) is then
\beq
\label{eq:hypermultiplet_action}
    S_{\mathcal{N} = 2} = \int d^4 x \le(-  \partial^M q_I^\dag \partial_M q_I - |m|^2 q_I^\dag q_I + i \bar{\chi}\slashed{\p}\chi - i |m| \bar{\chi} e^{i\y\g_5}\chi \ri),
\eeq
where \(\slashed{\p} \equiv \g^M \p_M\) and \(\bar{\chi} \equiv \chi^\dag \g^0\), and repeated R-symmetry doublet indices are summed.

When \(m\) is constant, the free hypermultiplet action in eq.~\eqref{eq:hypermultiplet_action} is invariant under the $d=4$ \(\mathcal{N}=2\) SUSY transformations:
\begin{subequations}
\label{eq:N=2_susy_transformation}
\begin{align}
    \d q_I &= \sqrt{2} \, i \, \bar{\xi}_I \chi ,
    \\
    \d \chi &= - \sqrt{2} \, \g^M \xi_I \, \p_M q_I - \sqrt{2} \, |m| e^{-i\y\g_5} \xi_I q_I,
\end{align}
\end{subequations}
where \(\xi_I\) are again the SUSY parameters, which obey eq.~\eqref{eq:sort_of_majorana_condition}. One way to derive eq.~\eqref{eq:N=2_susy_transformation} is via dimensional reduction of the SUSY variations of a $d=6$ $\N=1$ hypermultiplet.

When \(m\) depends on position, the action in eq.~\eqref{eq:hypermultiplet_action} is no longer invariant under the entire $d=4$ $\N=2$ SUSY transformation in eq.~\eqref{eq:N=2_susy_transformation} due to spacetime derivatives acting on \(|m|\) and \(\y\). Specifically, when $m$ depends on position eq.~\eqref{eq:hypermultiplet_action}'s SUSY variation is
\begin{equation}\begin{aligned} \label{eq:hypermultiplet_action_transformation}
    \d S_{\mathcal{N}=2} =\sqrt{2} \, i \int d^4 x \biggl[&
    \le( \bar{\chi} \g^M P_- \xi_I \, q_I + q_I^\dag \, \bar{\xi}_I  P_- \g^M \chi \ri) \p_M m
    \\ &
    \hspace{-0.4cm}+ \le( \bar{\chi} \g^M P_+ \xi_I \, q_I + q_I^\dag \, \bar{\xi}_I P_+ \g^M  \chi \ri) \p_M \mb
    \biggr],
\end{aligned}\end{equation}
where \(\mb = |m|e^{-i\y}\). In eq.~\eqref{eq:hypermultiplet_action_transformation} the factors of \(P_\pm\) arise because \(|m|e^{i\psi \g_5} = m P_+ + \mb P_-\). If we let $m$ depend on \((z,\zb)\) but not on \((x_0,x_1)\), then eq.~\eqref{eq:hypermultiplet_action_transformation} becomes
\begin{equation}\begin{aligned} \label{eq:hypermultiplet_action_transformation_complex}
    \d S_{\mathcal{N}=2} =\sqrt{2} \, i \int d^4 & x \biggl[
    \le( \bar{\chi} \g^{\zb} P_- \xi_I \, q_I + q_I^\dag \, \bar{\xi}_I  P_- \g^{\zb} \chi \ri) \pb m
    \\ &
   + \le( \bar{\chi} \g^z P_+ \xi_I \, q_I + q_I^\dag \, \bar{\xi}_I P_+ \g^z  \chi \ri) \p \mb
    \\ &
    +\le( \bar{\chi} \g^z P_- \xi_I \, q_I + q_I^\dag \, \bar{\xi}_I  P_- \g^z \chi \ri) \p m
    \\ &
    + \le( \bar{\chi} \g^{\zb} P_+ \xi_I \, q_I + q_I^\dag \, \bar{\xi}_I P_+ \g^{\zb}  \chi \ri) \pb \mb
    \biggr].
\end{aligned}\end{equation}
We want to find the conditions under which $\d S_{\mathcal{N}=2}=0$ in eq.~\eqref{eq:hypermultiplet_action_transformation_complex}, indicating that the position-dependent $m$ preserves some SUSY.

Suppose that $m$ is holomorphic in $z$, so that $\pb m =0$ and $\p \mb=0$. In that case, the first two lines of eq.~\eqref{eq:hypermultiplet_action_transformation_complex} vanish. To make the third and fourth lines in eq.~\eqref{eq:hypermultiplet_action_transformation_complex} vanish we must impose the following projection conditions on the $\xi_I$, respectively
\begin{subequations}
\label{eq:holomorphic_mass_projection_condition}
\begin{align}
\label{eq:holomorphic_mass_projection_condition_a}
\g^z P_-\xi_I &= 0,\\
\g^{\zb} P_+ \xi_I&=0.
\label{eq:holomorphic_mass_projection_condition_b}
\end{align}
\end{subequations}
Eq.~\eqref{eq:holomorphic_mass_projection_condition} is in fact identical to eq.~\eqref{eq:gauginovariations3}, so the rest of our proof is formally identical to that of sec.~\ref{sec:bgvec}. If eq.~\eqref{eq:holomorphic_mass_projection_condition_a} is satisfied then eq.~\eqref{eq:holomorphic_mass_projection_condition_b} is satisfied too, because of eq.~\eqref{eq:sort_of_majorana_condition_2}. We thus find that if $m$ is holomorphic, $m(z)$, and we impose the projection condition in eq.~\eqref{eq:holomorphic_mass_projection_condition_a}, then $\d S_{\mathcal{N}=2}=0$ in eq.~\eqref{eq:hypermultiplet_action_transformation_complex}, so that some SUSY is preserved. Only half of $\xi_I$'s components survive the projection condition in eq.~\eqref{eq:holomorphic_mass_projection_condition}, meaning $4$ real supercharges. We thus find that a holomorphic $m(z)$ preserves $4$ real supercharges.

Analogously, suppose that $m$ is antiholomorphic, so that $\p m =0$ and $\pb \mb=0$. In that case, the last two lines of eq.~\eqref{eq:hypermultiplet_action_transformation_complex} vanish, and to make the first two lines vanish we must impose the projection conditions in eq.~\eqref{eq:gauginovariations4}. Half of $\xi_I$'s components survive those projection conditions, meaning antiholomorphic $m(\zb)$ preserves $4$ real supercharges.

What is the $d=2$ chirality of the preserved supercharges? The chirality analysis is formally identical to that in eqs.~\eqref{eq:combined_mass_projection_condition} and~\eqref{eq:field_theory_2d_chirality}, so the result is the same: holomorphic or antiholomorphic $m$ preserve $d=2$ $\N=(4,0)$ or $(0,4)$ SUSY along $(x_0,x_1)$, respectively

We have thus completed our proof that holomorphic or antiholomorphic $m$ preserve SUSY for a \textit{free} $d=4$ $\N=2$ hypermultiplet. We can argue that the same results apply for any number $N_f$ of hypermultiplets coupled to $\N=4$ SYM, in the fundamental representation of $SU(N_c)$, as follows. For that theory, the classical action has the form $S_{\textrm{SYM}} \equiv N_f \, S_{\N=2} + S_{\N=4} + S_\mathrm{int}$, where \(S_{\N=4}\) is the action of \(\N=4\) SYM and \(S_\mathrm{int}\) represents the interaction terms between the fields of \(\N=4\) SYM and the hypermultiplets. An explicit expression for $S_{\textrm{SYM}}$ in terms of $\N=1$ superfields appears for example in ref.~\cite{Erdmenger:2007cm}, and an expression in terms of on-shell component fields appears for example in ref.~\cite{Chesler:2006gr}. Compared to the free hypermultiplet, the SUSY variation of the hypermultiplet fermions, $\delta \chi$, now includes an additional contribution proportional to one of the adjoint scalar fields of \(\N=4\) SYM. Crucially, however, neither \(S_\mathrm{int}\) nor this ``new'' term in the SUSY variations depends on derivatives of $\chi$, nor on the mass \(m\). As a result, if we allow $m$ to depend on position then in the SUSY variation of $S_{\textrm{SYM}}$ the only non-zero terms that can arise are those of the free hypermultiplets. In other words, with position-dependent \(m\) we have the SUSY variation \(\d S_{\textrm{SYM}} = N_f \, \d S_{\N=2}\) with the \(\d S_{\N=2}\) in eq.~\eqref{eq:hypermultiplet_action_transformation_complex}. Consequently, a holomorphic or antiholomorphic $m$ preserves $1/2$ of the supercharges of \(\N=4\) SYM coupled to \(N_f\) \(\N=2\) hypermultiplets, again meaning $4$ real supercharges. By extension, the $d=2$ chirality of the preserved supercharges is also unchanged. We thus conclude that for any number $N_f$ of hypermultiplets coupled to $\N=4$ SYM, a holomorphic or antiholomorphic $m$ preserves $d=2$ $\N=(4,0)$ or $(0,4)$ SUSY, respectively.

This second proof is less general than our first proof in sec.~\ref{sec:bgvec}, but nevertheless still applies for any values of $N_c$, $\lambda$, and $N_f$, and agrees with the SUGRA analysis in sec.~\ref{sec:kappa_symmetry} in the near-horizon limit, dual to the SYM theory in the 't Hooft large $N_c$ limit, the strong coupling limit $\lambda \to \infty$, and the probe limit $N_f \ll N_c$. As we did in sec.~\ref{sec:bgvec}, here again we will leave for future research a proof that zeroes of $m$ exhibit SUSY enhancement from $d=2$ $\N=(4,0)$ or $(0,4)$ to $(8,0)$ or $(0,8)$, respectively.

%%%%%%%%%%%%%%%%%%%%%%%%%%%%%%%%%%%%%%%%%%%%%%%%%%
%%%%%%%%%%%%%%%%%%%%%%%%%%%%%%%%%%%%%%%%%%%%%%%%%%
\subsubsection{Background $U(1)_R$ holonomies}
\label{sec:u1rholo}
%%%%%%%%%%%%%%%%%%%%%%%%%%%%%%%%%%%%%%%%%%%%%%%%%%
%%%%%%%%%%%%%%%%%%%%%%%%%%%%%%%%%%%%%%%%%%%%%%%%%%

Typically, for position-dependent sources to preserve SUSY requires holonomies of flat background R-symmetry gauge fields~\cite{Festuccia:2011ws,Dumitrescu:2012ha,Closset:2013vra,Closset:2014uda}. We will end this subsection by showing that  holomorphic $m$ indeed describe a non-trivial background gauge field of the $U(1)_R$, which acts on $\chi$ as an axial $U(1)$. An analogous argument applies for antiholomorphic $m$. 

In the hypermultiplet action eq.~\eqref{eq:hypermultiplet_action} we can remove the mass term's factor of \(e^{i \psi \g^5}\) by a field redefinition, \(\chi \equiv e^{-i \psi \g^5/2} \chi'\) with new Dirac fermion \(\chi'\). If \(\psi\) depends on position then the derivative in $\chi$'s kinetic term acts on $\psi$, generating a new term in the action:
\beq
\label{eq:axial_coupling}
    i \bar{\chi} \g^M \p_M \chi - |m| \bar{\chi} e^{i \psi \g^5}\chi = i \bar{\chi}' \g^M \p_M \chi' - |m| \bar{\chi}' \chi' - \frac{1}{2} (\p_M\psi) \bar{\chi}' \g^M \g^5 \chi'.
\eeq
Since \(\frac{1}{2} \bar{\chi}' \g^M \g^5 \chi'\) is the axial current for the Dirac fermion \(\chi'\), the right-most term in eq.~\eqref{eq:axial_coupling} corresponds to a coupling of \(\chi'\) to a background axial vector field, $\cA_M \equiv \partial_M \psi$, or in form notation \(\cA \equiv d \psi\). Using
\beq
\psi = \frac{1}{2i} \log\left(\frac{m}{\mb}\right),
\eeq
we can write $\cA$ as
\beq
\label{eq:axialgaugefield}
    \cA = \frac{1}{2i} \left(\frac{\p m}{m} dz - \frac{\pb \mb}{\mb} d\zb\right).
\eeq
Since $\cA=d\psi$ is locally exact, and hence has vanishing field strength, $\cA$ can at most have non-trivial holonomies around zeroes or poles of $m$. Indeed, the integral of \(\cA\) along a closed contour \(\cC\) is the same as that in Cauchy's argument principle, so if we define $n$ as the number of zeroes of $m(z)$ enclosed by $\cC$, weighted by their multiplicity, and $p$ as the number of poles of $m(z)$ enclosed by $\cC$, weighted by their order, then
\beq
\label{eq:axial_vector_integral}
    \oint_{\cC} \cA = 2\pi (n - p).
\eeq
In short, holomorphic $m$ automatically describes the background $U(1)_R$/axial gauge field $\cA$ in eq.~\eqref{eq:axialgaugefield} whose only non-trivial content is the integer holonomy in the $z$-plane in eq.~\eqref{eq:axial_vector_integral}. Analogous statements apply for antiholomorphic $m$: in $\cA$'s definition eq.~\eqref{eq:axialgaugefield} simply replace $\p m/m \to - \p \mb/\mb$ and $\pb \mb /\mb \to - \pb m/m$ and in the holonomy equation eq.~\eqref{eq:axial_vector_integral} multiply the right-hand side by $-1$.

%%%%%%%%%%%%%%%%%%%%%%%%%%%%%%%%%%%%%%%%%%%%%%%%%%
%%%%%%%%%%%%%%%%%%%%%%%%%%%%%%%%%%%%%%%%%%%%%%%%%%
\subsection{Gravity side: worldvolume geometry of (anti-)holomorphic solutions}
\label{sec:holodual}
%%%%%%%%%%%%%%%%%%%%%%%%%%%%%%%%%%%%%%%%%%%%%%%%%%
%%%%%%%%%%%%%%%%%%%%%%%%%%%%%%%%%%%%%%%%%%%%%%%%%%

In this subsection we work exclusively with probe D7-branes in $AdS_5 \times S^5$. In sec.~\ref{sec:constantmass} we will review properties of the constant solution, $y = y_0$ with complex constant $y_0$, including in particular how the D7-branes' worldvolume geometry holographically encodes the hypermultiplet mass $m \propto y_0$. In sec.~\ref{sec:holomass} we will use holographic renormalisation to show that holomorphic or antiholomorphic solutions $y$ are dual to states in which the hypermultiplet mass operator, $\mathcal{O}$, has vanishing VEV, $\langle \mathcal{O}\rangle=0$. We will also discuss in detail the example of a linear function, $y\propto z$, including in particular how the D7-branes' worldvolume geometry encodes massless $d=2$ $\N=(8,0)$ fields at $z=0$.

%%%%%%%%%%%%%%%%%%%%%%%%%%%%%%%%%%%%%%%%%%%%%%%%%%
%%%%%%%%%%%%%%%%%%%%%%%%%%%%%%%%%%%%%%%%%%%%%%%%%%
\subsubsection{Review: holographic dual of constant mass}
\label{sec:constantmass}
%%%%%%%%%%%%%%%%%%%%%%%%%%%%%%%%%%%%%%%%%%%%%%%%%%
%%%%%%%%%%%%%%%%%%%%%%%%%%%%%%%%%%%%%%%%%%%%%%%%%%

In $AdS_5 \times S^5$ and when $y=0$ the $N_f$ 4ND D7-branes are extended along $AdS_5 \times S^3$, where the $S^3$ is equatorial on the $S^5$. These D7-branes are holographically dual to a number $N_f \ll N_c$ of \textit{massless} hypermultiplets in the fundamental representation of $SU(N_c)$~\cite{Karch:2002sh}. The fact that the D7-branes preserve the isometries of $AdS_5$ is dual to the fact that massless hypermultiplets in the probe limit preserve $d=4$ conformal symmetry, or more precisely $d=4$ $\N=2$ superconformal symmetry.  The fact that the D7-branes break the $SO(6)$ isometry of the $S^5$ down to the $SO(4)$ isometry of the $S^3$ is dual to the fact that massless hypermultiplets break the $SO(6)_R$ R-symmetry down to $SO(4) \times SO(2)_R$, where $SO(4) \simeq SU(2)_R \times SU(2)$, thus producing the $d=4$ $\N=2$ R-symmetry $SU(2)_R \times SO(2)_R$. The $SO(2)_R \simeq U(1)_R$ acts as an axial symmetry on the hypermultiplet fermion, i.e. the ``quark'' $\chi$, but does not affect the hypermultiplet scalars, i.e. the ``squarks'' $q_I$.

The D7-branes' complex-valued worldvolume scalar $y$ is holographically dual to the hypermultiplets' complex-valued mass operator, $\mathcal{O}$. We will write $y=|y|e^{i\psi}$ and denote $|y|$'s dual operator as $\Om$ and $\psi$'s dual operator as $\Opsi$. The D7-branes' $U(N_f)$ worldvolume gauge fields are dual to the $U(N_f)$ flavour symmetry currents. Explicit expressions for these operators in terms of SYM fields appear for example in refs.~\cite{Myers:2007we,Hoyos:2011us}.

In AdS/CFT we compute SYM correlators from variations of the on-shell action in AdS. The fact that SYM correlators generically have UV divergences that we renormalise using local counterterms is dual to the fact that generically the on-shell action in AdS has divergences at large $r$ that we remove using local counterterms on a large-$r$ cutoff surface---a procedure called ``holographic renormalisation''~\cite{deHaro:2000vlm,Bianchi:2001kw,Skenderis:2002wp}. The counterterms we need to renormalise our probe D7-branes' on-shell action appear in refs.~\cite{Karch:2005ms,Karch:2006bv,Hoyos:2011us}.

To our knowledge, the only non-trivial holomorphic solutions studied to date are the constant solutions, $y(z) = y_0$ with complex constant \(y_0\)~\cite{Karch:2002sh,Kruczenski:2003be}. These solutions are holographically dual to probe flavours with constant mass $m = \sqrt{2} \, y_0/(2 \pi \alpha')$. A non-zero constant mass preserves $d=4$ $\N=2$ SUSY, including translational and rotational symmetry in the SYM directions $(x_0,x_1,x_2,x_3)$, but breaks conformal symmetry explicitly. A non-zero constant mass also breaks axial symmetry explicitly, here meaning that $SO(4) \times SO(2)_R$ breaks down to $SO(4)$.

How do the D7-branes along $AdS_5 \times S^3$ encode the properties of massive hypermultiplets? When $m\propto y_0\neq 0$, at the $AdS_5$ boundary the D7-branes still wrap an equatorial $S^3 \subset S^5$, but now as the D7-branes extend into $AdS_5$ the $S^3$ ``slips'' on the $S^5$ (being a trivial cycle) and eventually shrinks to zero size at a pole of the $S^5$. In our coordinates the $S^3$ collapses to zero size at $\rho=0$, which via eq.~\eqref{eq:rdef} corresponds to a finite value of $r = \sqrt{\rho^2 + 2 |y|^2}$, which we denote $r_0 \equiv \sqrt{2} |y_0|$. The D7-branes in $AdS_5$ thus appear to ``end'' at $r_0$. Since $r$ is holographically dual to the SYM theory's energy scale, $r_0$ is the holographic dual of the flavour fields' mass gap: the fact that the D7-branes are present for $r>r_0$ and absent for $r<r_0$ is holographically dual to the fact that the flavour fields are present at energies above their mass gap and absent at energies below their mass gap, respectively. When $m\propto y_0\neq 0$ the D7-branes' worldvolume therefore only \textit{asymptotically} approaches $AdS_5 \times S^3$ near the $AdS_5$ boundary and is deformed as we move into the $AdS_5$ bulk. The holographically dual statement is that the SYM theory with massive hypermultiplets is \textit{asymptotically} superconformal in the UV but has a non-trivial RG flow such that in the IR the massive hypermultiplets decouple.

In the SYM theory with constant $m$, the $d=4$ $\N=2$ SUSY requires the total energy to vanish, including the probe flavour contribution, and also requires $\Omv=0$ and $\Opsiv=0$ because the mass operator is the SUSY variation of another operator~\cite{Babington:2003vm,Erdmenger:2007cm}. SUSY does \textit{not} require $\Om$ and $\Opsi$'s higher-point correlators to vanish, however. Holographic renormalisation reproduces these properties~\cite{Karch:2005ms,Karch:2006bv,Karch:2008uy,Hoyos:2011us}: for the constant solution $y=y_0$ the probe D7-branes' holographically renormalised on-shell action vanishes, dual to the fact that the probe flavour contribution to the energy vanishes, and the first variation of the on-shell action with respect to $y$'s value at the $AdS_5$ boundary vanishes, dual to the fact that $\Omv=0$ and $\Opsiv=0$. Higher-order variations generically do not vanish, however. For example $\Om$ and $\Opsi$'s two-point functions are non-zero, and when $m \neq 0$ they exhibit poles corresponding to massive mesons~\cite{Kruczenski:2003be}.

%%%%%%%%%%%%%%%%%%%%%%%%%%%%%%%%%%%%%%%%%%%%%%%%%%
%%%%%%%%%%%%%%%%%%%%%%%%%%%%%%%%%%%%%%%%%%%%%%%%%%
\subsubsection{Holographic dual of (anti-)holomorphic mass}
\label{sec:holomass}
%%%%%%%%%%%%%%%%%%%%%%%%%%%%%%%%%%%%%%%%%%%%%%%%%%
%%%%%%%%%%%%%%%%%%%%%%%%%%%%%%%%%%%%%%%%%%%%%%%%%%

We have shown than any holomorphic function $y(z)$ is a BPS solution of the 4ND D7-branes' equation of motion. Such a solution is dual to a position-dependent mass,
\begin{equation}
\label{eq:holomass}
    m(z) = \frac{\sqrt{2}}{2 \pi \a'} y(z).
\end{equation}
Similar statements apply for antiholomorphic $y(\zb)$.

In the appendix we perform holographic renormalisation for a generic holomorphic solution $y(z)$. In particular, we compute the probe flavour contribution to the energy, and we compute $\Omv$ and $\Opsiv$. We find that the probe flavour contribution to the energy is zero. That is no surprise, since probe branes dual to SUSY-preserving defects typically have vanishing energy~\cite{Karch:2005ms,Karch:2008uy}. Indeed, we could have guessed that the on-shell action and hence the energy would vanish: eq.~\eqref{eq:actionenergybounds3action} shows that our on-shell action is simply the sum of the trivial 4ND and 8ND D7-branes' actions, each of which vanishes independently after holographic renormalisation, due to SUSY. In the appendix we also show that our BPS solutions give $\Omv=0$ and $\Opsiv=0$. In other words, not just the constant solution $y=y_0$ but in fact \textit{any} holomorphic $y(z)$ gives $\Omv=0$ and $\Opsiv=0$. We suspect the reason is SUSY, and specifically we suspect that the mass operator is a $d=2$ $\N=(4,0)$ SUSY variation of another operator, though we have not proven so. Analogous statements apply for generic antiholomorphic $y(\zb)$, which preserve $d=2$ $\N=(0,4)$ SUSY.

How do the D7-branes along $AdS_5 \times S^3$ encode a holomorphic mass like $m(z)$ in eq.~\eqref{eq:holomass}? Broadly speaking, a non-trivial $y(z)$ describes D7-branes whose $S^3$ ``slips off'' the $S^5$, i.e. collapses to a pole of the $S^5$, at different values of $r = \sqrt{\rho^2 + 2 |y|^2}$ as we move in the $z$-plane. Said differently, the D7-branes's endpoint $r_0$ varies with $z$, dual to the fact that the flavour mass gap varies with $z$. The details of the D7-branes' worldvolume geometry clearly depend on the details of $y(z)$, and a comprehensive analysis of all possible worldvolume geometries is beyond the scope of this paper.

However, we will address a crucial question: what happens at zeroes or poles of $y(z)$? For the full, asymptotically flat geometry with $f(r)$ in eq.~\eqref{eq:frdef}, we argued in sec.~\ref{sec:bpsbound} that at a zero or pole of $y(z)$ the 4ND D7-brane describe trivial 8ND D7-branes. For example, below eq.~\ref{eq:actionenergybounds3} we showed that the 4ND D7-branes' worldvolume approaches that of trivial 8ND D7-branes both at a zero of $y(z)$ in eq.~\eqref{eq:8ndfrom4ndzero} and at a pole of $y(z)$ in eq.~\eqref{eq:8ndfrom4ndpole}.

How do those arguments change when $f(r) = r^2/L^2$ and the geometry is that of $AdS_5 \times S^5$? The arguments for zeroes of $y(z)$ are largely the same, because both $f(r)$ in eq.~\eqref{eq:frdef} and $f(r) = r^2/L^2$ vanish as $r^2$ as $r \to 0$. In contrast, the arguments for poles differ substantially because $f(r)$ in eq.~\eqref{eq:frdef} approaches $1$ as $r \to \infty$, describing the asymptotically flat region, whereas $f(r) = r^2/L^2$ diverges as $r^2$ as $r \to \infty$, describing the region near the $AdS_5$ boundary. That difference has a trivial origin: by definition, in the near-horizon limit we discard the asymptotically flat region. Given these differences, in what follows we will discuss zeros of $y(z)$ first and poles of $y(z)$ second.

When the geometry is $AdS_5 \times S^5$, at a zero of $y(z)$ the arguments in eqs.~\eqref{eq:holoyinducedg} to~\eqref{eq:8ndfrom4ndzero} still apply, but now with $f(r) \to r^2/L^2$. In particular, when $r \to 0$, which is only possible if $\rho \to 0$ and $y(z) \to 0$, i.e. at a zero of $y(z)$, we expect the 4ND D7-branes to describe trivial 8ND D7-branes. In $AdS_5 \times S^5$ the trivial 8ND D7-branes are extended along $AdS_3 \times S^5$~\cite{Harvey:2007ab,Buchbinder:2007ar,Harvey:2008zz}, and are holographically dual to a $d=2$ $\N=(8,0)$ SUSY multiplet, whose only on-shell fields are chiral fermions coupled as a defect to $d=4$ $\N=4$ SYM. Both anomaly inflow arguments~\cite{Callan:1984sa} and index theorems~\cite{Jackiw:1981ee,Weinberg:1981eu} require a net number of such chiral fermions. The fact that trivial 8ND D7-branes preserve the $AdS_3$ isometries is holographically dual to the fact that probe $d=2$ $\N=(8,0)$ chiral fermions preserve $d=2$ defect conformal symmetry. The fact that trivial 8ND D7-branes preserve the full $SO(6)$ isometry of the $S^5$ is holographically dual to the fact that $d=2$ $\N=(8,0)$ chiral fermions preserve an $SO(6)_R$ R-symmetry, namely a $d=2$ chiral half of the original $d=4$ $SO(6)_R$~\cite{Harvey:2007ab,Buchbinder:2007ar,Harvey:2008zz}.

Plugging $f(r) = r^2/L^2$ into eq.~\eqref{eq:8ndfrom4ndzero} indeed produces an $AdS_3 \times S^5$ worldvolume metric at $r \to 0$, as expected. Since $r$ is holographically dual to the SYM theory's energy scale, as explained in sec.~\eqref{sec:background}, approaching $r \to 0$ is holographically dual to approaching the IR of the SYM theory. The emergence of the trivial 8ND D7-branes at $r \to 0$ is thus holographically dual to the emergence of massless fields in the IR, namely the $d=2$ $\N=(8,0)$ chiral fermions. In other words, the $d=2$ $\N=(8,0)$ chiral fermions appear only when we perform an RG flow to integrate out everything except the zero of $m(z)$.

However, a curious ancillary question arises at a zero of $y(z)$, as follows. The 4ND D7-branes asymptotically approach $AdS_5 \times S^3$, where $S^3$ is obviously a trivial/collapsible cycle of $S^5$. The trivial 8ND D7-branes at a zero of $y(z)$ are extended along $AdS_3 \times S^5$, where the $S^5$ is obviously the unique non-trivial/non-collapsible cycle of the $S^5$. A topology change thus occurs at a zero of $y(z)$, from the trivial cycle $S^3$ to the non-trivial cycle $S^5$. How does that topology change happen in the worldvolume geometry?

We will answer that question using a simple example, namely a linear solution, $y(z) = c \, z$ with dimensionless constant $c$. We choose $c$ to be real and positive, without loss of generality: by a rotation in the $z$-plane we can map negative or complex $c$ to real and positive $c$. The function $y(z) = c\,z$ obviously has a zero at $z=0$, and so at that point should exhibit the topology change mentioned above. We choose a linear function as perhaps the simplest non-trivial holomorphic function with a zero, being not only holomorphic but entire, and being the only entire function with an entire inverse, as mentioned in sec.~\ref{sec:8ndpov}. A linear solution $y(z) = c\,z$ is holographically dual to probe hypermultiplets with a linearly increasing mass, $m(z) \propto c\,z$, with massless modes only at $z=0$.

The figure below is a cartoon depiction of the linear solution $y = c\,z$, which we will explain in detail in the following.

\begin{figure}[t!]
\centering
\begin{tikzpicture}[scale=0.75]
    \node (c1m2) at (-4,0) {};
    \node (c1m1) at (-2,0) {};
    \node (c10) at (0,0) {};
    \node (c11) at (2,0) {};
    \node (c12) at (4,0) {};
    \node (c2m1) at (-2,-2) {};
    \node (c20) at (0,-2) {};
    \node (c21) at (2,-2) {};
    \node (c30) at (0,-4) {};

    \coordinate (p1) at (7,-0.5);
    \coordinate (p2) at (7,-2.5);
    \coordinate (p3) at (7,-4.5);

    %----- Brane boundaries -----
    \draw[thick, pattern=crosshatch dots] (-4,0) -- (0,-4) -- (4,0) -- cycle;

    %----- Top Line -----
    \draw[white, fill=white] (c1m2) circle (0.65);
    \draw (c1m2) circle (0.5);
    \draw[thick, orange, fill=orange] (c1m2)+(0,-0.5) circle (0.05);

    \draw[white, fill=white] (c1m1) circle (0.65);
    \draw (c1m1) circle (0.5);
    \draw[thick, orange] (c1m1)+(0,-0.25) ellipse (0.4 and 0.4/2);

    \draw[white, fill=white] (c10) circle (0.65);
    \draw (c10) circle (0.5);
    \draw[thick, orange] (c10) ellipse (0.5 and 0.5/2);

    \draw[white, fill=white] (c11) circle (0.65);
    \draw (c11) circle (0.5);
    \draw[thick, orange] (c11)+(0,0.25) ellipse (0.4 and 0.4/2);

    \draw[white, fill=white] (c12) circle (0.65);
    \draw (c12) circle (0.5);
    \draw[thick, orange, fill=orange] (c12)+(0,0.5) circle (0.05);

    %----- Second Line -----
    \draw[white, fill=white] (c2m1) circle (0.65);
    \draw (c2m1) circle (0.5);
    \draw[thick, orange, fill=orange] (c2m1)+(0,-0.5) circle (0.05);

    \draw[white, fill=white] (c20) circle (0.65);
    \draw (c20) circle (0.5);
    \draw[thick, orange] (c20) ellipse (0.5 and 0.5/2);

    \draw[white, fill=white] (c21) circle (0.65);
    \draw (c21) circle (0.5);
    \draw[thick, orange, fill=orange] (c21)+(0,0.5) circle (0.05);

    %----- Third Line -----
    \draw[white, fill=white] (c30) circle (0.65);
    \begin{scope}
        %\clip ($(c30)-(1,0)$) rectangle ($(c30)+(1,1)$);
        \draw[orange, fill=orange!50!white] (c30) circle (0.5);
    \end{scope}
    \draw (c30) circle (0.5);

    %----- Axes -----
    \draw[thick, ->] (-4,1) -- (4,1) node[right]{$Re(z)$};
    \draw (0,.9) -- (0,1.1) node[above]{$0$};

    \draw[thick, ->] (-5,-4) -- (-5,0) node[above]{$r$};
    \draw (-5.1,-4) -- (-4.9,-4) node[right]{$0$};

    %----- Plots -----
    %----- Top Plot -----
    \draw[orange, thick] ($(p1)+(0,0)$) .. controls ($(p1)+(2,0.25)$) and ($(p1)+(3.5,0)$) .. ($(p1)+(4,1)$);
    \draw[->] (p1) -- ($(p1)+(5,0)$) node[right]{$|z|$};
    \draw[-|] (p1) node[left]{\footnotesize{$0$}} -- ($(p1)+(0,1)$) node[above]{$\theta$} node[left]{\footnotesize{$\pi/2$}};

    %----- Second Plot -----
    \draw[orange, thick] ($(p2)+(0,0)$) .. controls ($(p2)+(1,0.5)$) and ($(p2)+(1.8,0.5)$) .. ($(p2)+(2,1)$);
    \draw[->] (p2) -- ($(p2)+(5,0)$) node[right]{$|z|$};
    \draw[-|] (p2) node[left]{\footnotesize{$0$}} -- ($(p2)+(0,1)$) node[above]{$\theta$} node[left]{\footnotesize{$\pi/2$}};

    %----- Third Plot -----
    \draw[->] (p3) -- ($(p3)+(5,0)$) node[right]{$|z|$};
    \draw[-|] (p3) node[left]{\footnotesize{$0$}} -- ($(p3)+(0,1)$) node[above]{$\theta$} node[left]{\footnotesize{$\pi/2$}};
    \draw[orange, thick] (p3) -- ($(p3)+(0,1)$);
\end{tikzpicture}
\caption*{\textbf{Figure:} Cartoon of the probe D7-branes' worldvolume for the linear solution, $y(z)=c\,z$ with dimensionless real constant $c >0$. The vertical axis is $r = \sqrt{\rho^2 + 2 |y|^2}$ in arbitrary units, where the Poincar\'e horizon is at $r=0$ and the boundary is at $r \to \infty$. The horizontal axis is $\textrm{Re}(z)$ in arbitrary units. The shaded triangle depicts the D7-branes' extent inside $AdS_5$. The black circles represent the $S^5$, while the orange circles represent the D7-branes' worldvolume along an $S^3 \subset S^5$, the orange dots represent an $S^3\subset S^5$ that has collapsed to a pole of the $S^5$, and the orange disk represents D7-branes wrapping the entire $S^5$. On the right we plot the angle of latitude that parametrises the $S^3$ fibration, $\theta \in [0,\pi/2]$, versus $|z|$, for three representative values of $r$. When $\theta$ is zero, the $S^3$ is maximal, whereas when $\theta$ reaches $\pi/2$ the $S^3$ collapses to a pole of the $S^5$, as shown on the left and right edges of the triangle. If we move down along the left edge of the triangle, through $r=0$, and then up the right edge, then at $r=0$ the D7-branes must ``jump'' from the south pole to the north pole of the $S^5$ in a single point. At that point they thus extend along the entire range of $\theta \in [0,\pi/2]$, as we depict on the bottom right, so at that point they wrap the entire $S^5$.
}
\end{figure}
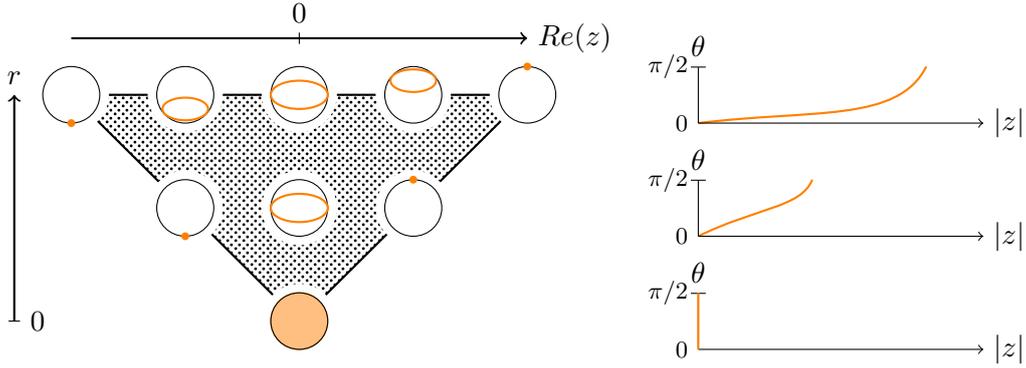

We expect to see a topology change from the trivial cycle $S^3 \subset S^5$ to the non-trivial cycle $S^5$ exactly at the Poincar\'e horizon, $r=0$, where $z=0$. To illustrate how that topology change occurs in the worldvolume geometry, we start with the D7-branes' induced worldvolume metric in eq.~\eqref{eq:holoyinducedg}, which we repeat here for convenience:
\beq
\label{eq:holoyinducedg2}
ds^2_{D7} = f(r) \left(-dx_0^2 + dx_1^2 + 2  \left(1 +f(r)^{-2} |\partial y|^2 \right) dz \, d\zb \right) + f(r)^{-1} \left(d\rho^2 + \rho^2 ds^2_{S^3}\right).
\eeq
Plugging in $f(r) = r^2/L^2$ and, for the linear solution $y(z)=c\,z$ also $|\partial y|^2=c^2$, gives
\beq
\label{eq:linear_solution_near_horizon}
    d s_{D7}^2=\frac{r^2}{L^2}\left[-d x_0^2+d x_1^2+2\left(1+\frac{L^4}{r^4}c^2\right)d z~d\bar{z}\right]+\frac{L^2}{r^2}(d \r^2 + \r^2 d  s_{S^3}^2)~.
\eeq
We will now show that eq.~\eqref{eq:linear_solution_near_horizon} smoothly interpolates between $AdS_5\times S^3$ near the $AdS_5$ boundary and $AdS_3\times S^5$ at the Poincar\'e horizon at $z=0$, respectively.

The $AdS_5$ boundary is at $r\to\infty$. Since $r^2 = \rho^2 + 2 |y|^2 = \rho^2 + 2 c^2 |z|^2$, we can approach $r \to \infty$ in two ways. First is $\rho\to\infty$ with fixed, finite $z$. In that limit eq.~\eqref{eq:linear_solution_near_horizon} approaches
\begin{align}
\label{eq:linearuv}
    d s_{D7}^2\approx\frac{\rho^2}{L^2}\left(-d x_0^2+d x_1^2+2~d z~d\bar{z}\right)+\frac{L^2}{\rho^2}\left(d\rho^2+\rho^2d s_{S^3}^2\right)~,
\end{align}
which is the metric of $AdS_5 \times S^3$, both with radius of curvature $L$. More specifically, eq.~\eqref{eq:linearuv} is the worldvolume metric of trivial 4ND D7-branes, holographically dual to $d=4$ hypermultiplets with $m=0$. Since $r \to \infty$ is holographically dual to the UV of the SYM theory, eq.~\eqref{eq:linearuv} shows that at any fixed, finite $z$ the extreme UV is always described by $d=4$ hypermultiplets with $m=0$.

The second way to approach $r \to \infty$ is to fix $\rho$ and send $|z|\to\infty$. In that case, in eq.~\eqref{eq:linear_solution_near_horizon} the factor in front of the $S^3$ metric $ds_{S^3}^2$ vanishes, so that the $S^3$ collapses already at the boundary. Translating to the SYM theory, if $|z|\to\infty$ then $|m|\to\infty$, so we are in a region of space where the hypermultiplets are infinitely massive and decouple no matter how far we go into the UV. We depict these two limits in the top row of the figure, which shows the $S^3$ approaching the equator of the $S^5$ as $r \to \infty$ with fixed $|z|$, at a rate that depends on $|z|$, and such that at $|z|\to\infty$ the $S^3$ collapses at the $AdS_5$ boundary.

As we move from $r\to \infty$ down into the bulk of $AdS_5$, towards finite $r$, the figure shows that the D7-branes' endpoint, where the $S^3$ collapses, varies with $z$, and in particular we see that the $S^3$ collapses more quickly at large $|z|$. The figure also shows that when $\textrm{Re}(z)<0$ the $S^3$ collapses to the south pole of the $S^5$, whereas when $\textrm{Re}(z)>0$ the $S^3$ collapses to the north pole of the $S^5$. Precisely at $|z|=0$, the $S^3$ is equatorial for all $r$.

The Poincar\'e horizon of $AdS_5$ is at $r\to0$, which we can reach only by sending both $\rho \to0$ and $|z| \to 0$ simultaneously. In that limit eq.~\eqref{eq:linear_solution_near_horizon} approaches
\begin{subequations}
\label{eq:linear_solution_poincare}
\begin{align}
    d s_{D7}^2&\approx\frac{r^2}{L^2}\left(-d x_0^2+d x_1^2\right)+\frac{L^2}{r^2}\left(2c^2\,d z~d\bar{z}\right)+\frac{L^2}{r^2} \left(d\rho^2+\rho^2d s_{S^3}^2\right)\\
    & = \frac{r^2}{L^2}\left(-d x_0^2+d x_1^2\right) +\frac{L^2}{r^2}\left(2c^2\,d z~d\bar{z}+d\rho^2+\rho^2d s_{S^3}^2\right).
\end{align}
\end{subequations}
Since $y=c\,z$ we can re-write eq.~\eqref{eq:linear_solution_poincare} as
\beq
\label{eq:linear_solution_poincare_2}
d s_{D7}^2 = \frac{r^2}{L^2}\left(-d x_0^2+d x_1^2\right) +\frac{L^2}{r^2}\left(2\,d y~d\bar{y}+d\rho^2+\rho^2d s_{S^3}^2\right),
\eeq
which is the same as plugging $f(r)=r^2/L^2$ into eq.~\eqref{eq:8ndfrom4ndzero}, as expected. Using $r^2 = \rho^2 + 2 |y|^2$, we can subsequently re-write eq.~\eqref{eq:linear_solution_poincare_2} as
\begin{subequations}
\label{eq:linear_solution_poincare_3}
\begin{align}
d s_{D7}^2 & = \frac{r^2}{L^2}\left(-d x_0^2+d x_1^2\right) +\frac{L^2}{r^2}\left(dr^2 + r^2 ds^2_{S^5}\right) \\
& =  \frac{r^2}{L^2}\left(-d x_0^2+d x_1^2\right) +\frac{L^2}{r^2} \, dr^2 + L^2 \, ds^2_{S^5},
\end{align}
\end{subequations}
which is the metric of $AdS_3 \times S^5$, both with radius of curvature $L$. To illustrate the topology change, starting from eq.~\eqref{eq:linear_solution_poincare} a more useful coordinate change is
\begin{subequations}
\label{eq:linear_variables}
\begin{align}
\rho&\equiv r\,\cos(\theta),\\
z&\equiv \frac{1}{\sqrt{2}\,c}\,r\,\sin(\theta) \,e^{i\psi},
\end{align}
\end{subequations}
with $\psi \in [0,2\pi]$ the azimuthal angle and $\theta \in [0,\pi/2]$ the latitudinal angle of the $S^3 \subset S^5$: plugging eq.~\eqref{eq:linear_variables} into eq.~\eqref{eq:linear_solution_poincare} gives
\beq
\label{eq:s5toroidal}
ds^2_{D7} =\frac{r^2}{L^2}\left(-d x_0^2+d x_1^2\right)+\frac{L^2}{r^2}d r^2+L^2\left(d\theta^2+\sin^2(\theta)~d\psi^2+\cos^2(\theta)\,d s_{S^3}^2\right),
\eeq
which is eq.~\eqref{eq:linear_solution_poincare_3}, but now with the $S^5$ metric written explicitly in toroidal coordinates.

If we fix $r$ and $\psi$ and then vary $\theta$, then we move horizontally across the triangle, as follows. Suppose we fix $\psi \in [\pi/2,3\pi/2]$, so that $\textrm{Re}(z) \leq 0$, and then vary $\theta$ from $\pi/2$ to $0$. When $\theta =\pi/2$ eq.~\eqref{eq:linear_variables} gives $\rho=0$ and $z = - r/(\sqrt{2}\,c)$, so we sit at a point on the left edge of the triangle, where from eq.~\eqref{eq:s5toroidal} the $S^3$ has collapsed to the south pole of the $S^5$, as depicted in the figure. If we then decrease $\theta$ from $\pi/2$ to $0$, then we move horizontally to the right. When we reach $\theta=0$ eq.~\eqref{eq:linear_variables} gives $\rho = r$ and $z=0$, so we now sit on the centre line that connects the top of the triangle at $r \to \infty$ to the bottom corner at $r=0$, where from eq.~\eqref{eq:s5toroidal} the $S^3$ has maximal radius on the $S^5$, i.e. wraps the equator of the $S^5$, as depicted in the figure. In the figure we also depict, on the right, the range that $\theta$ traverses as we change $|z|$ at fixed $r$. If we stay at the same $r$ but move to $\psi \in [0,\pi/2]$ or $[3\pi/2,2\pi]$, so that $\textrm{Re}(z) \geq 0$, and now increase $\theta$ from $0$ to $\pi/2$, then we continue moving horizontally to the right. When $\theta=0$ we start on the centre line, where the $S^3$ wraps the equator of $S^5$, and if we increase $\theta$ to $\pi/2$ we move from the centre line to the right edge of the triangle. When we reach $\theta=\pi/2$ eq.~\eqref{eq:linear_variables} gives $\rho=0$ and $z = + r/(\sqrt{2}\,c)$, where from eq.~\eqref{eq:s5toroidal} the $S^3$ has collapsed to the north pole of the $S^5$, as depicted in the figure. Again $\theta$ traverses the same range as we change $|z|$ at fixed $r$, but now with $\textrm{Re}(z) \geq 0$.

The key point for the topology change is that as we move horizontally in the figure, from one side of the triangle to the other, the $S^3 \subset S^5$ traverses the entire $S^5$ from the south pole to the north pole. However, at $r=0$ the $S^3$ must traverse the entire $S^5$ at a single point--hence the D7-branes wrap the entire $S^5$ at that point.

As another perspective, imagine approaching the point $r=0$ by moving down along the left edge of the triangle, towards $r=0$, then moving through the point $r=0$, and then moving up along the right edge of the triangle, away from $r=0$. When we pass through $r=0$ the $S^3 \subset S^5$ must jump from the south pole to the north pole at a single point, and in particular must traverse the entire range of $\theta \in [0,\pi/2]$ all at the same point, $r=0$, as we depict in the bottom row of the figure, on the right. As a result, precisely at the point $r=0$ the D7-branes must wrap the entire $S^5$.

We have thus seen how zeroes of $y$ describe trivial 8ND D7-branes along $AdS_3 \times S^5$ at the Poincar\'e horizon $r=0$. What about poles of $y$? As in sec.~\ref{sec:bpsbound} we expect a pole to describe a ``spike'' of 4ND D7-branes that approaches the trivial 8ND D7 branes. Indeed, in eq.~\eqref{eq:8ndfrom4ndpole} we showed that, for the full asymptotically flat geometry in eq.~\eqref{eq:metric}, a pole in $y$ produces a 4ND D7-brane worldvolume metric that asymptotically approaches the trivial 8ND D7-branes' worldvolume metric in the asymptotic $d=10$ Minkowski region. However, as mentioned above, in the near-horizon limit by definition we discard the asymptotically flat region of the geometry in eq.~\eqref{eq:metric}, so in $AdS_5 \times S^5$ we will not find the worldvolume metric in eq.~\eqref{eq:8ndfrom4ndpole}. Instead, at a pole of $y$ we expect the $S^3$ to collapse at the $AdS_5$ boundary, dual to the statement that the hypermultiplets have infinite mass at the pole.

As an illustrative example consider a simple pole, $y = c/z$ where again we choose real and positive constant $c$ without loss of generality. The constant $c$ has dimensions $\left(\textrm{length}\right)^2$. The hypermultiplets then have a mass $m \propto c/z$, such that they have infinite mass at $z=0$ and finite mass away from $z=0$ that decreases to zero as $|z| \to \infty$. We thus expect the dual 4ND D7-branes' worldvolume to have a collapsing $S^3$ at $z=0$. Plugging $f(r)=r^2/L^2$ and $y=c/z$ into the D7-branes' worldvolume metric in eq.~\eqref{eq:holoyinducedg2} gives
\begin{subequations}
\label{eq:polemetric1}
\begin{align}
ds^2_{D7} &= \frac{r^2}{L^2} \left(-dx_0^2 + dx_1^2 + 2  \left(1 + \frac{L^4}{r^4} \frac{c^2}{|z|^4} \right) dz \, d\zb \right) + \frac{L^2}{r^2}\left(d\rho^2 + \rho^2 ds^2_{S^3}\right),\\
r^2 &= \rho^2 + 2 |y|^2 = \rho^2 + \frac{2 c^2}{|z|^2}.
\end{align}
\end{subequations}
We can approach the $AdS_5$ boundary $r \to \infty$ in two ways. The first way is to fix $z$ and send $\rho \to \infty$. In that case, $r^2 \to \rho^2$ and the worldvolume metric reduces to that of $AdS_5 \times S^3$ in eq.~\eqref{eq:linearuv}. As mentioned below eq.~\eqref{eq:linearuv}, the dual statement is that at any fixed $z$ the extreme UV is described by $d=4$ hypermultiplets with $m=0$. The second way to reach $r \to \infty$ is to fix $\rho$ and send $|z| \to 0$, thus approaching the pole. In that limit $r^2 \to 2c^2/|z|^2$ and the worldvolume metric of eq.~\eqref{eq:polemetric1} approaches
\beq
\label{eq:polemetric2}
ds^2_{D7} \underset{z \to 0}\approx \frac{2c^2}{L^2 }\frac{1}{|z|^2} \left(-dx_0^2 + dx_1^2 \right) +  \left(\frac{2c^2}{L^2}+\frac{L^2}{2}\right)\frac{2}{|z|^2} dz\,d\zb + \frac{L^2}{2 c^2}\, |z|^2\left(d\rho^2 + \rho^2 ds^2_{S^3}\right).
\eeq
If we write $z = |z| e^{i \psi}$ then we can re-write eq.~\eqref{eq:polemetric2} as
\beq
\label{eq:polemetric3}
ds^2_{D7} \underset{z \to 0}\approx \frac{2c^2}{L^2} \frac{1}{ |z|^2}\left(-dx_0^2 + dx_1^2 + \left(1+\frac{L^4}{4 c^2}\right)d|z|^2\right) +  \left(\frac{2c^2}{L^2}+\frac{L^2}{2}\right) d\psi^2 + \frac{L^2}{2 c^2}\, |z|^2\left(d\rho^2 + \rho^2 ds^2_{S^3}\right).
\eeq
which is the metric of $\mathbb{R}^4$ spanned by $\rho$ and the $S^3$ fibered over an $AdS_3$ spanned by $(x_0,x_1,|z|)$ in direct product with an $S^1$ spanned by $\psi$. We thus see that at the pole the $S^3$ indeed collapses at the $AdS_5$ boundary, as expected. However, the remainder of the geometry is difficult to interpret cleanly. For example, we find an $AdS_3$, which is suggestive of the trivial 8ND D7-branes, but we also find that the $\rho$ direction collapses along with the $S^3$, and the $\psi$ direction appears to decouple from the other directions. More intuitive is the behaviour of the worldvolume metric near the Poincar\'e horizon: $r \to 0$ requires both $\rho \to 0$ and $|z|\to\infty$, so that the D7-branes reach the Poincar\'e horizon only infinitely far from the pole. In that limit, $r^2 \to \rho^2$ and the worldvolume metric again reduces to that of $AdS_5 \times S^3$ in eq.~\eqref{eq:linearuv}, representing trivial 4ND D7-branes. The dual statement is that infinitely far from the pole the extreme IR is always described by $d=4$ hypermultiplets with $m=0$. We leave a more thorough analysis of poles of $y$ for future research.

Finally, as mentioned in sec.~\ref{sec:intro}, we can think of zeroes of $m$ as conformal defects and poles of $m$ as scattering centres. A key question about such objects is: what results from their fusion? For example, suppose we have two zeroes or two poles of $m$. What happens if we move them closer to one another, until eventually they coincide? The answer is simple. As mentioned in sec.~\ref{sec:intro}, in the complex plane the most general holomorphic function without an essential singularity at infinity is a rational function, $y(z) = \frac{(z-a_1)(z-a_2) \ldots}{(z-b_1) (z-b_2) \ldots}$ with complex constants $(a_1,a_2,\ldots)$ and $(b_1,b_2,\ldots)$. Fusing two zeroes means making two of the $(a_1,a_2,\ldots)$ coincide, which converts two zeroes of degree one into a single zero of degree two. Similarly, fusing two poles means making two of the $(b_1,b_2,\ldots)$ coincide, which converts two poles of degree one into a single pole of degree two. Fusing a zero with a pole means making one of the $(a_1,a_2,\ldots)$ coincide with one of the $(b_1,b_2,\ldots)$ in which case they cancel between the numerator and denominator of $y(z)$. In short, the ``fusion algebra'' of zeroes and poles in $y$ is described simply by taking a signed sum of their degrees.

%%%%%%%%%%%%%%%%%%%%%%%%%%%%%%%%%%%%%%%%%%%%%%%%%%
%%%%%%%%%%%%%%%%%%%%%%%%%%%%%%%%%%%%%%%%%%%%%%%%%%
\section{Summary and outlook}
\label{sec:summary}
%%%%%%%%%%%%%%%%%%%%%%%%%%%%%%%%%%%%%%%%%%%%%%%%%%
%%%%%%%%%%%%%%%%%%%%%%%%%%%%%%%%%%%%%%%%%%%%%%%%%%

In the background geometry of eq.~\eqref{eq:metric}, produced by a large number $N_c$ of D3-branes along $(x_0,x_1,z,\zb)$, we introduced a probe number $N_f \ll N_c$ of 4ND D7-branes, and in sec.~\ref{sec:holosols} we proved that a worldvolume scalar $y$ holomorphic or antiholomorphic in $z$ solves the D7-branes' equations of motion. In sec.~\ref{sec:bpsbound} we showed that such solutions saturate a BPS bound and represent an exactly marginal bound state of 4ND and 8ND D7-branes, and in particular we argued that zeroes or poles of holomorphic or antiholomorphic $y$ describe 8ND D7-branes. Correspondingly, in sec.~\ref{sec:kappa_symmetry} we proved that holomorphic or antiholomorphic $y$ preserve $d=2$ $\N=(4,0)$ or $(0,4)$ SUSY along $(x_0,x_1)$, respectively, while zeroes of $y$ should exhibit SUSY enhancement to $d=2$ $\N=(8,0)$ or $(0,8)$, respectively. In sec.~\ref{sec:nearhorizon} we restricted to the geometry's near-horizon $AdS_5 \times S^5$ region and invoked holographic duality, wherein type IIB SUGRA in $AdS_5 \times S^5$ is dual to $d=4$ $SU(N_c)$ $\N=4$ SYM at large $N_c$ and large $\lambda$ while the probe $N_f \ll N_c$ 4ND D7-branes are dual to probe hypermultiplets in the fundemental representation of $SU(N_c)$. Our holomorphic or antiholomorphic $y$ are dual to a holomorphic or antiholomorphic hypermultiplet mass $m$, and in sec.~\ref{sec:fieldtheory} we presented two logically distinct proofs, entirely within SYM theory (with no use of SUGRA or holography), that such $m$ preserve $d=2$ $\N=(4,0)$ or $(0,4)$ SUSY, respectively, and that such $m$ also describe holonomies of a background $U(1)_R$ gauge field. Finally, in sec.~\ref{sec:holodual} we performed a holographic calculation to show that holomorphic or antiholomorphic $m$ produces zero VEV of the mass operator, $\langle \mathcal{O} \rangle = 0$, and that zeroes of $y$ describe D7-brane worldvolumes that interpolate between the 4ND D7-branes' worldvolume, $AdS_5 \times S^3$, and the 8ND D7-branes' worldvolume, $AdS_3 \times S^5$, which obviously involves a topology change from a trivial cycle $S^3 \subset S^5$ to the non-trivial cycle $S^5$.

As suggested in sec.~\ref{sec:intro}, via holography our solutions should be useful for studying many questions about the effects of translational and rotational symmetry breaking, including in particular the effect on transport coefficients such as electrical conductivity or charge diffusion. More generally, our solutions could be useful for studying many things, such as RG flows from $d=4$ fields (the hypermultiplets) to $d=2$ fields ($d=2$ SUSY-preserving chiral fermions), the fusion of conformal defects or ``quantum wires'', entanglement between impurities, and more. For example, a solution like $y(z) = \frac{1}{k} \sin(kz)$ is holographically dual to mass with zeroes periodically spaced by $1/k$, and hence describes a one-dimensional lattice of conformal defects/quantum wires, while a solution like $y(z) = 1/(k \sin(kz))$ describes a one-dimensional lattice of infinitely massive scattering centres. The holographically dual probe D7-brane solutions should allow for calculations of transport coefficients as well as meson spectra, entanglement entropy, and more in the presence of such $m$.

Looking farther afield, other intersections of D-branes should admit solutions with similar properties, namely BPS and SUSY-preserving solutions that depend holomorphically or antiholomorphically on spatial coordinates of the SYM theory on the worldvolume on one stack of D-branes. Many such intersections should also have holographic duals. Can we find, and possibly classify, all such intersections and their holomorphic or antiholomorphic solutions? The companion paper ref.~\cite{Ratcliffe:2026exv} will address many such questions. However, many more general questions also deserve further research. For example, in our probe D-brane solutions the only non-trivial worldvolume fields were scalars---what about solutions with non-trivial worldvolume gauge fields or fermions? Many D-brane systems admit a moduli space---how do holomorphic or antiholomorphic solutions depend on such moduli, and what effect would a singular locus or anomaly in the moduli space parameters have? What about other coordinate systems besides Cartesian---do holomorphic or antiholomorphic solutions exist for other slicings of flat space? In fact, what if the field theory is coupled to gravity, i.e. to a dynamical supergravity multiplet, rather than to a rigid supergravity background, as in our case---how do our solutions change then? For all of these questions, can we go beyond the probe limit, and find fully back-reacted solutions?

We hope that our solutions provide the first steps towards addressing these and many other important questions about systems with broken translational and rotational symmetry.

\section*{Acknowledgements}

We thank A.~Chalabi, J.~Gomis, Z.~Komargodski, C.~Kristjansen, V.~Mishnyakov, I. Yaakov, and K.~Zarembo for useful discussions. The work of P.C. is supported by a Mayflower studentship from the University of Southampton. A.~O'B. gratefully acknowledges support from the Simons Center for Geometry and Physics, Stony Brook University where some research for this paper was performed. The work of J.R is supported by the STFC consolidated grant ST/X000583/1. The work of R.R. was supported by the European Union’s Horizon Europe research and innovation program under Marie Sklodowska-Curie Grant Agreement No.~101104286. Nordita is supported in part by Nordforsk. The work of B.S. was supported in part by the STFC consolidated grant ST/T000775/1. A.~O'B. and J.~R. would like to thank the Isaac Newton Institute for Mathematical Sciences, Cambridge, for support and hospitality during the programme ``Quantum field theory with boundaries, impurities, and defects'', where work on this paper was undertaken. This work was supported by EPSRC grant EP/Z000580/1.

\appendix

%%%%%%%%%%%%%%%%%%%%%%%%%%%%%%%%%%%%%%%%%%%%%%%%%%%%%%%%%%%%%%%%%%%%%%%%%%%%%%%%%%%%%%%%%%%%%%%%%%%%%%%%%%%%%%%%%%
%%%%%%%%%%%%%%%%%%%%%%%%%%%%%%%%%%%%%%%%%%%%%%%%%%%%%%%%%%%%%%%%%%%%%%%%%%%%%%%%%%%%%%%%%%%%%%%%%%%%%%%%%%%%%%%%%%
\section*{Appendix: holographic renormalisation}
\addcontentsline{toc}{section}{Appendix: holographic renormalisation}
%%%%%%%%%%%%%%%%%%%%%%%%%%%%%%%%%%%%%%%%%%%%%%%%%%%%%%%%%%%%%%%%%%%%%%%%%%%%%%%%%%%%%%%%%%%%%%%%%%%%%%%%%%%%%%%%%%
%%%%%%%%%%%%%%%%%%%%%%%%%%%%%%%%%%%%%%%%%%%%%%%%%%%%%%%%%%%%%%%%%%%%%%%%%%%%%%%%%%%%%%%%%%%%%%%%%%%%%%%%%%%%%%%%%%

\setcounter{equation}{0}
\renewcommand{\theequation}{A.\arabic{equation}}

In this appendix we carry out the holographic renormalisation of our probe D7-brane solutions. In this appendix we work exclusively in the near-horizon limit \(r \ll L\) of the D3-brane background~\eqref{eq:metric_complex_coords}, such that the metric becomes that of \(AdS_5 \times S^5\). We then invoke the AdS/CFT correspondence, in which type IIB SUGRA in $AdS_5 \times S^5$ is dual to $\N=4$ SYM with large $N_c$ and large $\lambda$, and the probe D7-branes are dual to $N_f \ll N_c$ $\N=2$ hypermultiplets. More specifically, $y$ is dual to the hypermultiplet's complex mass operator, $\mathcal{O}$. For our holomorphic solutions, $y(z)$, we will use holographic renormalisation~\cite{deHaro:2000vlm,Bianchi:2001kw,Skenderis:2002wp} to compute $\langle \mathcal{O} \rangle$ and also the hypermultiplets' contribution to the energy. We will find that both of these quantities vanish, presumably because of SUSY, as we discuss below.

In AdS/CFT, the bulk on-shell action in AdS is dual to the generating functional of the SYM theory~\cite{Gubser:1998bc,Witten:1998qj}. However, the bulk on-shell action typically diverges due to integration to the AdS boundary, which in the coordinates of eq.~\eqref{eq:ads5s5metric}, means divergences due to integration all the way to $r \to \infty$. As mentioned below eq.~\eqref{eq:ads5s5metric}, $r$ is dual to the SYM energy scale~\cite{Susskind:1998dq,Peet:1998wn}, where the near-boundary region $r\to\infty$ is dual to the UV. The bulk large-$r$ divergences are thus dual to the expected UV divergences of SYM theory, which we must renormalise. Holographic renormalisation translates the steps of SYM renormalisation to AdS: we regulate the divergences via a cutoff at large $r$, we add to the action covariant counterterms on the cutoff surface to cancel divergent terms, and we then remove the regulator, obtaining a finite, cutoff-independent, and unambiguous result.

Ref.~\cite{Karch:2005ms} initiated the holographic renormalisation of probe D-branes, in which we add counterterms only to the D-branes' action. Refs.~\cite{Karch:2005ms,Karch:2006bv,Hoyos:2011us} determined the counterterms for 4ND D7-branes that asymptotically approach $AdS_5 \times S^3$ inside $AdS_5 \times S^5$ with worldvolume scalars that have arbitrary dependence on the SYM coordinates $(x_0,x_1,x_2,x_3)$.

For convenience we will change our $AdS_5$ coordinates from those of eq.~\eqref{eq:metric_complex_coords} to the Fefferman-Graham (FG) gauge used in refs.~\cite{Karch:2005ms,Karch:2006bv,Hoyos:2011us}. We thus replace the coordinates \((y,\yb,\rho)\) of eq.~\eqref{eq:metric_complex_coords} with the coordinates \((u,\q,\y)\) via
\begin{subequations}
\label{eq:fefferman_graham_transformation_1}
\begin{align}
\r &= \frac{L^2 \cos \q}{u},\\
y &= \frac{L^2 \sin \q}{\sqrt{2} \, u} e^{i\y},
\end{align}
\end{subequations}
where $u \in (0,\infty)$, $\theta \in [0,\pi/2]$, and $\psi \in [0,2\pi]$. The inverse of eq.~\eqref{eq:fefferman_graham_transformation_1} is 
\begin{subequations}
\label{eq:fefferman_graham_transformation_2}
\begin{align}
    u &= \frac{L^2}{r} = \frac{L^2}{\sqrt{\r^2 + 2 |y|^2}},\\
    \q &= \tan^{-1} \le(\frac{\sqrt{2} \, |y|}{\r} \ri),\\
    \y &= \frac{1}{2i} \log \le(\frac{y}{\yb}\ri).
\end{align}
\end{subequations}
In the coordinates \((u,\q,\y)\) the metric of \(AdS_5 \times S^5\) is
\beq
\label{eq:metric_fefferman_graham}
    d s^2 = \frac{L^2}{u^2} du^2 + \frac{L^2}{u^2} \left(-  dx_0^2 + dx_1^2 + 2 dz \, d\zb\right)
    + L^2 \le(d \q^2 + \sin^2 \q \, d \y^2 + \cos^2 \q \, d s_{S^3}^2 \ri).
\eeq
The $AdS_5$ boundary is now at $u \to 0$ while the Poincar\'e horizon is at $u \to \infty$.

In the coordinates \((u,\q,\y)\) the D7-branes' worldvolume scalars are $\theta$ and $\psi$. Our ansatz for these is $\q(z,\zb,u)\) and \(\y(z,\zb,u)\), whose derivatives we denote
\beq
\label{eq:derivnotation2}
\partial \theta \equiv \frac{\partial \theta}{\partial z}, \qquad \pb \theta \equiv \frac{\partial \theta}{\partial \zb}, \qquad \dot{\theta} \equiv \frac{\partial \theta}{\partial u},
\eeq
and similarly for $\partial \psi$, $\pb \psi$, and $\dot{\psi}$. If we plug our ansatz into the D7-branes' action, eq.~\eqref{eq:D7-action}, integrate over the $S^3$ to give an overall factor of $2 \pi^2$, and define the normalisation factor
\beq
\N_{D7} \equiv 2 \pi^2 L^8 N_f T_\mathrm{D7} = \frac{\l }{16 \pi^4}N_f N_c,
\eeq
then the D7-branes' action becomes
\begin{align}
    S = - \N_{D7} \int d  u \int d^4 x \frac{\cos^3\q}{u^4} \biggl[&
        \frac{1}{u^2} + \dot{\q}^2 + \sin^2 \q \, \dot{\y}^2
        + 2  \le( \p \q \, \pb \q  +  \sin^2 \q  \, \p \y \, \pb \y \ri)
        \nonumber\\
        & - u^2 \sin^2 \q \le(\p\y \, \pb \q - \pb \y \, \p \q \ri)^2
        \label{eq:action_fefferman_graham_app}
        \\ 
        &+ 2 u^2  \sin^2 \q \, \le(\dot{\q} \, \p \y - \dot{\y} \, \p \q \ri)\le(\dot{\q}\, \pb \y - \dot{\y}\, \pb \q \ri)
    \biggr]^{1/2}.
    \nonumber
\end{align}

Near the $AdS_5$ boundary at \(u \to 0\), the worldvolume scalars's equations of motion give a FG expansion of the form
\begin{subequations}
\label{eq:fg_expansion_app}
\begin{align}
    \q(z,\zb,u)& = \q_1(z,\zb) u + \q_L(z,\zb) u^3 \log (u/L)  + \q_3(z,\zb) u^3+ \cO(u^4 \log u),
    \\
    \y(z,\zb,u) &= \y_0(z,\zb) + \y_L(z,\zb) u^2 \log (u/L) + \y_2(z,\zb) u^2 + \cO(u^3 \log u),
\end{align}
\end{subequations}
where \(\q_1\), $\theta_3$, $\psi_0$, and $\psi_2$ must be fixed by boundary conditions, while \(\q_L\) and \(\y_L\) are determined in terms of \(\q_1\) and \(\y_0\) by the equations of motion:
\begin{subequations}
\label{eq:log_terms_app}
\begin{align}
    \q_L &= \q_1  \, \p \y_0 \, \pb \y_0 - \p \pb \,\q_1,\\
    \y_L &= - \frac{1}{\theta_1} \left(\p \q_1 \, \pb \y_0 + \pb \q_1 \, \p \y_0 + \q_1 \, \p \pb \,\y_0\right).
\end{align}
\end{subequations}

Plugging a holomorphic or antiholomorphic solution $y$ into the coordinate transformations of eq.~\eqref{eq:fefferman_graham_transformation_1} or~\eqref{eq:fefferman_graham_transformation_2} gives worldvolume scalars $\theta(z,\zb,u)$ and $\psi(z,\zb,u)$ that depend on all of $(z,\zb,u)$. In particular, for a holomorphic $y(z)$ the FG expansion of eq.~\eqref{eq:fg_expansion_app} is
\begin{subequations}
\label{eq:fg_expansion_holomorphic_app}
\begin{align}
    \q(z,\zb,u) &= \frac{\sqrt{2} \, |y(z)|}{L^2} \, u + \frac{\sqrt{2} \, |y(z)|^3}{3 L^6} \, u^3 + \cO(u^5),\\
    \y(z,\zb,u) &= \frac{1}{2i} \log \le(\frac{y(z)}{\yb (\zb)}\ri),
\end{align}
\end{subequations}
which clearly depend on all of $(z,\zb,u)$. However, these FG expansions do not have $\log(u/L)$ terms, so that $\theta_L=0$ and $\psi_L=0$. To be explicit: from eq.~\eqref{eq:fg_expansion_holomorphic_app} we identify
\begin{subequations}
\label{eq:fg_coefficients_holomorphic}
\begin{align}
    \q_1(z,\zb) &= \frac{\sqrt{2} |y(z)|}{L^2},\\
    \y_0(z,\zb) &=  \frac{1}{2i} \log \le(\frac{y(z)}{\yb (\zb)}\ri),
\end{align}
\end{subequations}
and plugging these into eq.~\eqref{eq:log_terms_app} gives $\theta_L=0$ and $\psi_L=0$. Analogous statements apply for antiholomorphic solutions.

In the AdS/CFT correspondence we identify the hypermultiplet mass $m$ from the boundary values of the D7-branes' worldvolume scalar via
\beq
\label{eq:massdef}
m = \frac{1}{2\pi \alpha'}\lim_{\rho \to \infty} \sqrt{2} \, y,
\eeq
which in the original 4ND D3/D7 intersection of tab.~\ref{tab:4nd} is simply the string tension, $1/(2 \pi \alpha')$, times the asymptotic separation between the D3- and D7-branes. Plugging in a holomorphic solution $y(z)$ obviously gives
\beq
\label{eq:massdefholo}
m(z) = \frac{\sqrt{2}}{2\pi \a'}\,y(z),
\eeq
as stated in eq.~\eqref{eq:mofz}. An analogous statement applies for antiholomorphic $y$. To find $m$ in terms of $\theta(z,\zb,u)$ and $\psi(z,\zb,u)$, we plug the change of coordinates in eq.~\eqref{eq:fefferman_graham_transformation_1} or~\eqref{eq:fefferman_graham_transformation_2} into eq.~\eqref{eq:massdef}, which gives
\beq
\label{eq:complex_mass_1}
m = \frac{L^2}{2\pi \a'} \lim_{u \to 0} \frac{\q(z,\zb,u) }{u} e^{i \y(z,\zb,u)}.
\eeq
Plugging the generic FG expansions of eq.~\eqref{eq:fg_expansion_app} into eq.~\eqref{eq:complex_mass_1} gives
\beq
m = \frac{L^2}{2\pi \a'} \, \theta_1 \, e^{i \psi_0}.
\eeq
For holomorphic solution $y(z)$, plugging the FG expansions of eq.~\eqref{eq:fg_expansion_holomorphic_app} into eq.~\eqref{eq:complex_mass_1}, or equivalently plugging eq.~\eqref{eq:fg_coefficients_holomorphic} into eq.~\eqref{eq:complex_mass_1}, reproduces eq.~\eqref{eq:massdefholo}, as expected.

We subsequently compute correlation functions of the mass operator, $\mathcal{O}$, by taking variational derivatives of the on-shell D7-brane action with respect to $m$. We thus plug the generic FG expansions in eq.~\eqref{eq:fg_expansion_app} into the action eq.~\eqref{eq:action_fefferman_graham_app} and perform the integrations. Of course, the integral over $u$ diverges near the $AdS_5$ boundary $u \to 0$. We regulate this divergence by integrating only over \(u \geq \e\), for some short distance cutoff \(\e\). We denote the resulting regulated action $S_{\textrm{reg}}$. Explicitly, the divergent terms in $S_{\textrm{reg}}$, obtained by expanding the integrand for small \(u\) and then performing the integral over \(u\geq \e\), are
\beq
\label{eq:d7_action_fg_app}
    S_{\textrm{reg}} = \N_{D7} \int d^4 x \, \le[- \frac{1}{4\e^4} + \frac{\q_1^2}{2 \e^2}
        + \le(\p \q_1 \, \pb \q_1  - \q_1 \, \p \pb \q_1 + 2\q_1^2 \, \p \y_0 \, \pb \y_0 \ri) \log (\e/L)
    \ri] + O(\e^0).
\eeq
Following the holographic renormalisation recipe, we add counterterms at $u=\e$ to remove these divergences, render the variational problem well-posed, and produce finite, unambiguous results for observables. The counterterms we need appear in ref.~\cite{Hoyos:2011us}: if we define
\beq
\Theta \equiv \q \, e^{i \y},
\eeq
and \(\g_{\m\n}\) as the induced metric on the intersection of the D7-branes with the cutoff surface \(u=\e\), then the counterterms we need are
\begin{subequations}
\label{eq:counterterms_general}
\begin{align}
    S_1 &= \frac{1}{4 L^4} \, \N_{D7}\,\int d^4x \, \sqrt{|\g|},\\
    S_2 &= - \frac{1}{2 L^4}\,\N_{D7}\, \int d^4x \, \sqrt{|\g|} \, |\Theta|^2,\\
    S_3 &= \frac{5 }{12 L^4} \,\N_{D7}\,\int d^4x \, \sqrt{|\g|} \, |\Theta|^4,\\
    S_4 &= - \frac{1}{2 L^2} \,\N_{D7}\,\int d^4x \, \sqrt{|\g|} \, \g^{\m\n} \p_\m \Theta \, \p_\n \Theta^* \log|\Theta|,\\
    S_5 &=  - \frac{1}{4 L^2} \, \N_{D7}\,\int d^4x \, \sqrt{|\g|} \, \g^{\m\n} \p_\m \Theta \, \p_\n \Theta^*.
\end{align}
\end{subequations}
Plugging the generic FG expansions in eq.~\eqref{eq:fg_expansion_app} into eq.~\eqref{eq:counterterms_general} and dropping terms that vanish as $\e \to 0$ then gives
\begin{subequations}
\label{eq:counterterms_app}
\begin{align}
    S_1 &= \frac{1}{4 \e^4}\, \N_{D7}\, \int d^4 x,
    \\
    S_2 &= - \N_{D7}  \int d^4 x \, \le[\frac{\q_1^2}{2 \e^2} + \q_1 \q_3 + \le( \q_1^2 \,  \p \y_0 \, \pb \y_0 - \q_1 \, \p \pb \q_1 \ri) \log(\e/L) \ri] ,
    \\
    S_3 &= \frac{5}{12}\,\N_{D7}\, \int d^4 x \,  \q_1^4,
    \\
    S_4 &= - \N_{D7} \int d^4 x\, \le(\p\q_1 \, \pb \q_1 + \q_1^2 \, \p\y_0 \, \pb\y_0 \ri) \log (\e \q_1),
    \\
    S_5 &= - \frac{1}{2} \,\N_{D7}\,\int d^4 x \, \le(\p\q_1 \, \pb \q_1 + \q_1^2 \, \p\y_0 \, \pb\y_0\ri).
\end{align}
\end{subequations}
Comparing eq.~\eqref{eq:counterterms_app} to eq.~\eqref{eq:d7_action_fg_app} shows that the counterterms will cancel all $\e \to 0$ divergences of the on-shell action. The resulting renormalised action, $S_{\textrm{ren}}$, defined as
\beq
\label{eq:srendef}
S_{\textrm{ren}} \equiv \lim_{\e \to 0} \left(S_{\textrm{reg}}+ \sum_{i=1}^5 S_i\right),
\eeq
will thus be finite and unambiguous.

%%%%%%%%%%%%%%%%%%%%%%%%%%%%%%%%%%%%%%%%%%%%%%%%%%
%%%%%%%%%%%%%%%%%%%%%%%%%%%%%%%%%%%%%%%%%%%%%%%%%%
\subsection*{A.1 Renormalised on-shell action}
%%%%%%%%%%%%%%%%%%%%%%%%%%%%%%%%%%%%%%%%%%%%%%%%%%
%%%%%%%%%%%%%%%%%%%%%%%%%%%%%%%%%%%%%%%%%%%%%%%%%%

To obtain the value of \(S_\mathrm{ren}\) we need the \(O(\e^0)\) term in $S_{\textrm{reg}}$ in eq.~\eqref{eq:d7_action_fg_app}, which requires performing the $u$ integral in eq.~\eqref{eq:action_fefferman_graham_app}. For generic \(\q(z,\zb,u)\) and \(\y(z,\zb,u)\) performing that integral explicitly is difficult. However, for holomorphic or antiholomorphic solutions $y$ the integrand simplifies, allowing us to perform the integral explicitly. For holomorphic or antiholomorphic $y$, and dropping terms that vanish as $\e \to 0$, we find
\begin{subequations}
\label{eq:SD7_holomorphic}
\begin{align}
    S_{\textrm{reg}} &= - \frac{\N_{D7}}{L^8} \int d^4 x \,  \int_{\e}^{L^2/\sqrt{2} \, |y|} \frac{du}{u^5} \le(L^4 - 2 u^2 |y|^2 \ri) \le[L^4 + u^4 (|\p y|^2 + |\pb y|^2)\ri]
    \\
    &= \N_{D7} \int d^4 x \le[-\frac{1}{4 \e^4} + \frac{|y|^2}{L^4 \e^2} + \frac{ |\p y|^2 + |\pb y|^2}{2 L^4} \le(1 + \log\le(\frac{2 \e^2 |y|^2}{L^4} \ri)\ri) - \frac{|y|^4}{L^8} \ri].
\end{align}
\end{subequations}
The upper limit on the \(u\) integral comes from the coordinate transformation in eq.~\eqref{eq:fefferman_graham_transformation_1} or~\eqref{eq:fefferman_graham_transformation_2}, which implies that \(u\) is bounded from above by \(L^2/ \sqrt{2} \, |y|\). Substituting the FG coefficients for a holomorphic or antiholomorphic solution $y$ from eq.~\eqref{eq:fg_coefficients_holomorphic} into the counterterms eq.~\eqref{eq:counterterms_app} gives
\begin{subequations}
\label{eq:holoct}
\begin{align}
    S_1 &= \frac{1}{4\e^4} \,\N_{D7}\,\int d^4 x ,
    \\
    S_2 &= -  \N_{D7}\, \int d^4x \, \le( \frac{|y|^2}{\e^2 L^4} + \frac{2|y|^4}{3 L^8} \ri),
    \\
    S_3 &=  \frac{5}{3 L^8}  \,\N_{D7}\,\int d^4 x \,  |y|^4,
    \\
    S_4 &= - \frac{1}{2L^4}  \,\N_{D7}\,\int d^4 x \, \le(|\p y|^2 + |\pb y|^2 \ri) \log \le(\frac{2 \e^2 |y|^2}{L^4}\ri),
    \\
    S_5 &= - \frac{1}{2 L^4}  \,\N_{D7}\,\int d^4 x \,  \le(|\p y|^2 + |\pb y|^2 \ri).
\end{align}
\end{subequations}

Plugging eq.~\eqref{eq:SD7_holomorphic} for $S_{\textrm{reg}}$ and eq.~\eqref{eq:holoct} for the counterterms of holomorphic or antiholomorphic $y$ into eq.~\eqref{eq:srendef} for the renormalised action then gives
\beq
\label{eq:srenresult}
S_{\textrm{ren}} = 0.
\eeq
Such a result is not entirely surprising: as mentioned below eq.~\eqref{eq:holomass}, for our holomorphic or antiholomorphic solutions $y$ the on-shell D7-brane action is a sum of 4ND and 8ND D7-brane actions, each of which vanishes independently after holographic renormalisation, due to SUSY. Moreover, as explained above eq.~\eqref{eq:D7-energy}, because $S \equiv - \int d x_0 \, E$ the result in eq.~\eqref{eq:srenresult} implies that the total energy $E$ vanishes as well.

%%%%%%%%%%%%%%%%%%%%%%%%%%%%%%%%%%%%%%%%%%%%%%%%%%
%%%%%%%%%%%%%%%%%%%%%%%%%%%%%%%%%%%%%%%%%%%%%%%%%%
\subsection*{A.2 Renormalised one-point functions}
%%%%%%%%%%%%%%%%%%%%%%%%%%%%%%%%%%%%%%%%%%%%%%%%%%
%%%%%%%%%%%%%%%%%%%%%%%%%%%%%%%%%%%%%%%%%%%%%%%%%%

We denote the operator dual to $\theta$ as \(\cO_m\) and the operator dual to $\psi$ as \(\cO_\y\). The one-point functions $\vev{\cO_m}$ and $\vev{\cO_\y}$ are given by
\begin{subequations}
\label{eq:scalar_vev_formulas}
\begin{align}
    \vev{\cO_m} &= - \frac{2 \pi \a'}{L^2} \frac{\d S_\mathrm{ren}}{\d \q_1},\\
    \vev{\cO_\y} &= - \frac{\d S_\mathrm{ren}}{\d \y_0}.
    \end{align}
\end{subequations}

To compute the variation of the renormalised action, $\delta S_{\textrm{ren}}$, we need the variations of the regulated action, $\delta S_{\textrm{reg}}$, and of the counterterms. To compute $\delta S_{\textrm{reg}}$, we define a Lagrangian $\cL$ from the regulated D7-brane action $S_{\textrm{reg}}$ from eq.~\eqref{eq:action_fefferman_graham_app} via
\beq
    S_{\textrm{reg}} = - \N_{D7} \int_{\e} d u \int d^4 x \, \cL,
\eeq
or in other words $\cL$ is the integrand in eq.~\eqref{eq:action_fefferman_graham_app}. The variations \(\q \to \q + \d \q\) and \(\y \to \y + \d\y\) then produce a variation of the regulated action,
\beq
\label{eq:sregvariation}
    \d S_{\textrm{reg}} = \N_{D7} \int d^4 x  \le[\frac{\p \cL}{\p \dot{\q}} \d\q + \frac{\p \cL}{\p \dot{\y}} \d\y \ri]_{u=\e}.
\eeq
If we plug the generic FG expansions for \(\q(z,\zb,u)\) and \(\y(z,\zb,u)\) of eq.~\eqref{eq:fg_expansion_app} into eq.~\eqref{eq:sregvariation} and integrate by parts to remove and \(z\) and \(\zb\) derivatives from the variations, and drop terms that vanish as $\e \to 0$, then we find
\beq
\begin{aligned}
\label{eq:bulk_action_variation_app}
    \d S_{\textrm{reg}} = \N_{D7} \int d^4 x \, \biggl[&
        \frac{\q_1 \, \d\q_1}{\e^2} + 4 (\q_L \, \d\q_1 + \q_1^2 \y_L \, \d\y_0) \log(\e/L) + \q_1 \, \d\q_3
        \\ & \hspace{2cm}
        +\le(3 \q_3 - 2 \q_1^3 + \q_L \ri)  \d \q_1  + \q_1^2  \le(2 \y_2 + \y_L \ri)  \d \y_0
    \biggr].
\end{aligned}
\eeq
We next need the variations of the counterterms from eq.~\eqref{eq:counterterms_app},
\begin{subequations}
\begin{align}
    \d S_1 &= 0,
     \\
    \d S_2 &= \N_{D7} \int d^4x \,\le[
        - \frac{\q_1 \, \d\q_1}{\e^2}
        - 2 \le(\q_L \, \d \q_1 + \q_1^2 \y_L \, \d \y_0\ri) \log(\e/L) 
        -\q_1 \, \d \q_3 - \q_3 \, \d \q_1
    \ri],
     \\
    \d S_3 &= \frac{5}{3} \N_{D7} \int d^4 x \, \q_1^3 \, \d \q_1,
    \\
    \d S_4 &= \N_{D7} \int d^4 x \, \biggl[
        - 2 (\q_L \, \d \q_1 + \q_1^2 \y_L \, \d \y_0) \log(\e \q_1)\nonumber
   \\ & \phantom{= \N_{D7} \int d^4 x \, h^3 c^2 \biggl[}
        + \frac{\p \q_1 \pb \q_1 - \q_1^2 \p \y_0 \pb \y_0}{\q_1} \d \q_1 
        + \le(\p \q_1 \pb \y_0 + \pb \q_1 \p \y_0 \ri) \q_1 \, \d\y_0
    \biggr],
    \\
    \d S_5 &= - \N_{D7} \int d^4 x \,  \le(\q_L \, \d\q_1 + \q_1^2 \, \y_L \, \d \y_0 \ri).
\end{align}
\end{subequations}
Following eq.~\eqref{eq:srendef}, we add these to $\delta S_{\textrm{reg}}$ to obtain $\delta S_{\textrm{ren}}$:
\beq
\begin{aligned}
\label{eq:srenvariation}
    \d S_\mathrm{ren} = 2\,\N_{D7} \int d^4 x \, \biggl[&
            \biggl(\q_3 - \frac{\q_1^3}{6} - \q_L \log(\q_1 L) + \frac{\p \q_1 \, \pb \q_1 - \q_1^2 \, \p \y_0 \, \pb \y_0}{2 \q_1}\biggr)\d\q_1
        \\
        & \hspace{1cm}
        + \q_1^2 \biggl( \y_2 - \y_L \log(\q_1 L) + \frac{\p \q_1 \, \pb \y_0 + \pb \q_1 \, \p \y_0}{2 \q_1}\biggr)\d\y_0 
    \biggr]. 
\end{aligned}\eeq
As required by the variational principle, \(\d S_\mathrm{ren}\) is finite in the limit \(\e\to 0\), and is independent of the variation of the normalisable coefficients \(\d\q_3\) and \(\d\y_2\).

Finally, plugging $\d S_\mathrm{ren}$ from eq.~\eqref{eq:srenvariation} into eq.~\eqref{eq:scalar_vev_formulas} gives the one-point functions,
\begin{subequations}
\label{eq:genericonepointfunctions}
\begin{align}
    \vev{\cO_m} &= -\frac{\sqrt{\l} N_f N_c}{4 \pi^3}\biggl[\q_3 - \frac{\q_1^3}{6} - \q_L \log(\q_1 L) + \frac{\p \q_1 \, \pb \q_1 - \q_1^2 \, \p \y_0 \, \pb \y_0}{2 \q_1}\biggr],
    \\
    \vev{\cO_\y} &= - \frac{\l N_f N_c}{8 \pi^4} \, \q_1^2 \biggl( \y_2 - \y_L \log(\q_1 L) + \frac{\p \q_1 \, \pb \y_0 + \pb \q_1 \, \p \y_0}{2 \q_1}\biggr).
\end{align}
\end{subequations}
For a holomorphic or antiholomorphic solution $y$, plugging the FG expansions of eq.~\eqref{eq:fg_expansion_holomorphic_app} into eq.~\eqref{eq:genericonepointfunctions}, we find that all terms cancel, so that
\begin{subequations}
\label{eq:onepointfunctionsvanish}
\begin{align}
    \vev{\cO_m} & = 0,\\
    \vev{\cO_\y} &= 0.
\end{align}
\end{subequations}
As mentioned in secs.~\ref{sec:intro} and~\ref{sec:holomass}, we suspect that the result in eq.~\eqref{eq:onepointfunctionsvanish} follows from SUSY, although we have not proven so.

%%%%%%%%%%%%%%%%%%%%%%%%%%%%%%%%%%%%%%%%%%%%%%%%%%
%%%%%%%%%%%%%%%%%%%%%%%%%%%%%%%%%%%%%%%%%%%%%%%%%%
\bibliographystyle{JHEP}
\bibliography{holoprobe}
%%%%%%%%%%%%%%%%%%%%%%%%%%%%%%%%%%%%%%%%%%%%%%%%%%
%%%%%%%%%%%%%%%%%%%%%%%%%%%%%%%%%%%%%%%%%%%%%%%%%%

\end{document}